\documentclass{article}
\usepackage{jcappub}
\usepackage[utf8]{inputenc}
\pdfoutput=1  % togliere il commento se le figure sono pdf, Blasciarlo se sono eps
\usepackage{color}
\usepackage{graphicx}
\usepackage{amsmath,amssymb}
\graphicspath{{./figures/}}
\usepackage{appendix}
\usepackage{placeins}
\usepackage[normalem]{ulem}
\usepackage[utf8]{inputenc}

\usepackage{booktabs} %per le tabelle
\usepackage{tablefootnote}
\usepackage{graphicx}
\usepackage{caption}
\usepackage{subcaption}
\usepackage{float}

\hypersetup{     
	urlcolor=black
}

\input epsf
\usepackage{mathtools}
\usepackage{amssymb}
\usepackage{amsmath}
\usepackage{amsfonts}
\usepackage{upgreek}
\usepackage{latexsym}
\usepackage{stfloats}
\usepackage{afterpage}

\newcommand{\be}{\begin{equation}}
	\newcommand{\ee}{\end{equation}}
\newcommand{\beq}{\begin{equation}}
	\newcommand{\eeq}{\end{equation}}
\newcommand{\bea}{\begin{eqnarray}}
	\newcommand{\eea}{\end{eqnarray}}

\newcommand{\e}[1]{\times 10^{#1}}

\newcommand{\bn}{{\mathbf  n}}

\newcommand{\bv}{{\mathbf v}}

\newcommand{\HH}{{\cal H}}

\newcommand{\De}{\Delta}

\newcommand{\de}{\delta}

\newcommand{\La}{\Lambda}
\newcommand{\si}{\sigma}

\newcommand{\Om}{\Omega}

\newcommand{\cd}{\cdot}

\definecolor{dgreen}{rgb}{0,0.6,0.0}

\definecolor{dblue}{rgb}{0,0,0.6}

\begin{document}
	\title{The Dipole of the Pantheon+SH0ES Data}
	
	\date{\today}

	\author[]{Francesco Sorrenti,}
	\author[]{Ruth Durrer and}
	\author[]{Martin Kunz}
	\affiliation[]{D\'epartement de Physique Th\'eorique and Center for Astroparticle Physics,\\
		Universit\'e de Gen\`eve, 24 quai Ernest  Ansermet, 1211 Gen\`eve 4, Switzerland}
	
	\emailAdd{francesco.sorrenti@unige.ch}
	\emailAdd{ruth.durrer@unige.ch}
	\emailAdd{martin.kunz@unige.ch}
	
	\abstract{In this paper we determine the dipole in the distance redshift relation from the Pantheon+ data. We find that, while its amplitude roughly agrees with the dipole found in the cosmic microwave background which is attributed to the motion of the solar system with respect to the cosmic rest frame,  the direction is different with a significance of slightly more than $3\si$. While the amplitude depends on the lower redshift cutoff, the direction is quite stable. For redshift cuts of order $z_{\rm cut} \simeq 0.05$ and higher, the dipole is no longer detected with high statistical significance. An important rôle seems to be played by the redshift corrections for peculiar velocities.}
	
	\maketitle
	
	\section{Introduction}
	The cosmological principle states that the Universe is statistically homogeneous and isotropic, and that on sufficiently large scales fluctuations are small. Isotropy of the Universe is best observed in the cosmic microwave background (CMB). The fluctuations of the CMB are very small with a typical amplitude of $\De T/T \sim 10^{-5}$ with one exception, the dipole which is about $10^{-3}$. It is usually assumed that most of this dipole is due to the peculiar velocity of our solar system with respect to the cosmic rest-frame. Attributing the entire dipole to our motion, one infers a velocity of~\cite{Kogut:1993ag,Planck:2013kqc,Planck:2018nkj,Saha:2021bay}
	\be
	v_0=(369\pm 0.9){\rm km/s} \,, \qquad  
	(l,b) = (263.99\pm 0.14, 48.26\pm 0.03)
	\label{e:d-Planck}
	\ee
	where $(l,b)$ denote the longitude and latitude of the velocity direction in galactic coordinates. In this work we shall use the directions wrt the baricenter of the solar system termed `right ascension' (ra) and `declination' (dec) (at J2000, i.e. January 1, 2000). In these coordinates the direction of the CMB dipole is
	\be
	({\rm ra,dec}) =(167.942\pm 0.007, -6.944\pm 0.007)\,.
	\label{e:d-Planck-solar}
	\ee
	
	Even if a small part of the dipole comes from intrinsic fluctuations on the last scattering surface, we  expect this to be of the same order as the higher multipoles and hence to contribute  not more than about 1\% (which is nevertheless significantly larger than the measurement error quoted in \eqref{e:d-Planck}).
	
	Observing a sufficiently large number of far away sources, these should in principle also define the cosmic rest-frame on average, and we therefore expect them to have a similar dipole as the CMB.
	This has been tested using radio galaxies~\cite{Ellis1984,Tiwari:2016,Bengaly:2017slg,Colin:2019opb,Siewert:2021} and also quasars or clusters~\cite{Secrest:2020has,Dam:2022wwh,Migkas:2021zdo}.  However, even though the dipole of these galaxy number counts points roughly in the same direction as the CMB dipole, the inferred velocity is typically about a 
	factor of two larger. The statistical significance of this discrepancy has been determined~\cite{Secrest:2022uvx} to be about $5\si$. These findings 
	led the authors of Ref.~\cite{Secrest:2022uvx} to challenge the cosmological principle. However, there is also the possibility that this result may be due to too simplistic model assumptions~\cite{Dalang:2021ruy,Guandalin:2022tyl}. Furthermore, in presently available data the intrinsic dipole as expected in the standard $\La$CDM model is not very much smaller than the kinematic dipole. In \cite{Atrio-Barandela:2014nda} the authors determined the bulk flow by constructing
	a statistic that evaluates the dipole moment at cluster locations. Also there they found a significant cluster bulk flow.
	In the future, using number counts e.g.\ from Euclid or SKA, it will be possible to discriminate between an intrinsic dipole and the dipole due to the observer velocity with the help of a `weighted analysis' ~\cite{Nadolny:2021hti}.

	In this work we follow another direction. We use the fact that in supernova distances, the Doppler term in the redshift fluctuation enhances the dipole due to our peculiar velocity at low redshift. This allows one to see a significant dipole already with a rather limited number of sources.
	
	This dipole has been recovered from supernova data for the first time nearly 20 years ago from the data of about 100 Supernovae Type Ia (SNe 1a), and it was found to be largely consistent with the CMB data~\cite{Bonvin:2006en}. Later, the anisotropy of the supernovae Ia distribution has been studied in~\cite{Javanmardi:2015sfa}. Recently, this analysis has been repeated and refined using  the Pantheon data~\cite{Horstmann:2021jjg}. In a newer paper~\cite{Dhawan:2022lze} both, the dipole and the quadrupole of the Pantheon and the JLA compilation of supernovae have been analysed with similar results for the dipole. Recently, in~\cite{Kalbouneh:2022tfw}, the authors have determined the first 3 multipoles of the  Cosmicflows-3 dataset~\cite{Tully:2016ppz} and the Pantheon supernovae
	catalogue.
	In the present paper we use the significantly larger and improved data set Pantheon+. This catalogue contains many more low redshift supernovae and it is these that contribute most significantly to the dipole, as we shall see. We study the dipole of the luminosity distance derived from 1701 lighcurves of 1550 SNe 1a published as the Pantheon+ data~\cite{Brout:2022vxf}. 
	Naively, we expect this dipole to be dominated by our peculiar velocity with respect to the Friedmann-Lema\^\i tre background. Therefore, the inferred velocity should agree with the  CMB dipole~\eqref{e:d-Planck}.  In the previous  Pantheon data~\cite{Pan-STARRS1:2017jku}, reasonable agreement albeit with a somewhat too small velocity amplitude was indeed found after applying corrections to the redshifts published by Pantheon~\cite{Horstmann:2021jjg}.
	
	Here we show that while the dipole in the Pantheon+ data set agrees well with the one of the Pantheon data, due to its better statistics and therefore smaller error bars, it significantly disagrees (at slightly more than $3\si$) with the velocity inferred from the Planck data. Especially the direction disagrees by about $40^o$. We also study the dependence of the dipole on the imposed redshift cut. We find that our result is stable for sufficiently low redshift cuts such that the dipole is detected with high significance.
	\vspace{0.2cm}\\
	{\bf Notation :} We consider a spatially flat Friedmann-Lema\^\i tre universe with linear scalar perturbations in longitudinal gauge,
	\be
	ds^2 = a^2(t)[-(1+2\Psi)dt^2 + (1-2\Phi)\de_{ij}dx^idx^j] \,.
	\ee
	The functions $\Phi$ and $\Psi$ are the Bardeen potentials. Einstein's summation convention is assumed. Spatial 3d vectors are denoted in bold face. A dot denotes the derivative with respect to conformal time $t$. The comoving Hubble parameter is $\HH=\dot a/a$ while the physical Hubble parameter is $H=\dot a/a^2$. The speed of light is set to unity.

	\section{Theoretical Model}\label{s:theo}
	Within linear perturbation theory the luminosity distance becomes direction dependent. The full expression in a spatially flat Universe is given by~\cite{Sasaki:1987ad,Bonvin:2005ps,Durrer:2020fza}
	\bea
	D_L(z,\bn) &=& \bar D_L(z)\left\{1 +\frac{1}{\HH(z)r(z)}(\bv_0\cd\bn)-\left(\frac{1}{\HH(z)r(z)}-1\right)\left[(\bv\cd\bn)-\Psi- \int_0^{r(z)}dr(\dot\Psi+\dot\Phi)\right] -\Phi
	\right. \nonumber \\
	&& \hspace*{1.6cm} \left.
	+\int_0^{r(z)}\frac{dr}{r}\left[1-\frac{r(z)-r}{2r(z)}\De_{\bn} \right](\Phi+\Psi)\right\} \,.
	\label{e:fluct}
	\eea
	
	Here $\bv$ is the velocity field in longitudinal gauge at the source position and $\bv_0$ denotes  the velocity of  the observer\footnote{There are also other `observer terms' which contribute to the monopole and which are neglected in \eqref{e:fluct}, see~\cite{Biern:2016kys} for the full expression including all observer terms}. The value at the observer contributes the kinematic dipole. The symbol $\De_{\bn} $ denotes the Laplacian on the sphere of directions, $r(z)$ is the comoving distance out to redshift $z$ and $\HH(z)$ is the comoving Hubble parameter. In a flat $\La$CDM Universe the background luminosity distance $\bar D_L$ is given by
	\be
	\bar D_L(z)=\frac{1+z}{H_0}\int_0^z \frac{dz'}{\sqrt{\Om_m(1+z')^3+1-\Om_m}} \,, \label{ansatz_monopole}
	\ee
	it depends on the cosmological parameters $h=H_0/(100$km/s/Mpc$)$ and on $\Om_m$.
	
	At small redshifts, $z\ll 1$, where $\HH(z)r(z) \simeq z \ll 1$, the dipole term $\bv_0\cd\bn$ largely dominates these anisotropies,
	\bea
	D_L(z,\bn) &\simeq& \bar D_L(z)\left(1 + \frac{1}{\HH(z)r(z)}\bv_0\cd\bn\right)  \,, \qquad z\ll 1\,, \label{e:ansatz_dipole}
	\eea
	where $\bv_0$ is the observer velocity and $c$ is the speed of light. More correctly this term would be\\ $\left(\bv_0-\bv(\bn,r(z))\right)\cd\bn$, but we have neglected the source velocity $\bv(\bn,r(z))$, assuming that it averages out and does not contribute significantly to the dipole (for a detailed analysis, see Sec.~\ref{sec:peculiar_velocities}). This is correct, if there is not a large, direction independent `bulk velocity'  which is common to all supernovae, independent of direction. Such a common source velocity would  have to be subtracted from the observer velocity. But since it cannot be modelled independently from the data, we understand $\bv_0$ as this difference between  the observer velocity and a common bulk velocity.
	At high redshift, $z\gtrsim 1$, the lensing term, the second term in the integral on the second line of \eqref{e:fluct}, dominates.
	
	Eq.~\eqref{e:ansatz_dipole} is simply the dominant term at low redshift of the first order Taylor expansion of 
	
	\be\label{e:vexact}
	D_L(z) \simeq \bar D_L(\bar z) = \bar D_L(z-\de z) =\bar D_L \left(z + \bv_0\cd\bn \right)\,.
	\ee 
	In this expression, only the peculiar redshift contribution to the fluctuation of the luminosity distance, which dominates at low redshifts, is considered.
	We have also used Eq.~\eqref{e:vexact} for our analysis (as is actually  done in~\cite{Horstmann:2021jjg}) and found that the results do not differ appreciably in cases with well detected dipole. Therefore this difference does not affect our conclusions.
	
	We model the data with the ansatz \eqref{e:ansatz_dipole}, leaving the velocity $\bv_0$ as well as $h$ and $\Om_m$ as free parameters which are estimated from the data with a Markov Chain Monte Carlo (MCMC) routine.
	
	%In addition to this physically motivated ansatz, we shall also fit the data in a model independent approach via
	%\be
	%D_L(z,\bn) = M(z) + \bd(z)\cd\bn + \frac{1}{2}
	%Q_{ij}(z)n^jn^j
	%\ee
	%where $M$ is the monopole, $\bd$ is the dipole and $(Q_{ij})$ is the tracefree, symmetric quadrupole. We model the redshift dependence of these 9 functions with a second order Padé approximation,
	%\be
	%X(z) = \frac{\al_X+\beta_Xz+\ga_Xz^2}{1+\de_X z + \ep_X z^2}
	%\ee
	%We fit the resulting 45 parameters $\al_M$ to $\ep_{Q_{23}}$ to the data using MCMC.
	
	At very low redshift, we expect the velocities of the supernovae to be correlated with our velocity which leads to a bias that may underestimate the true dipole wrt the cosmic flow. On the other hand, for higher redshifts, the pre-factor $1/r(z)\HH(z)$ becomes smaller and the dipole is correspondingly reduced. For this reason we also study the dependence of the dipole signal on the redshift cut which we impose.
	
	\section{The data}
	
	We use the Pantheon+ data which provides distance moduli $\mu$ for 1550 SNe,
	\be
	\mu = 5\log_{10}(D_L/10{\rm pc}) =  5\log_{10}(D_L/1{\rm Mpc}) +25 =  5\log_{10}D_L  +M \,.
	\label{e:distance_modulus}
	\ee
	After the last equal sign $D_L$ can be expressed in some arbitrary (fixed!) units which then determine the offset $M$. This offset is perfectly degenerate with the Hubble constant $H_0$ to which $D_L$ is inversely proportional, see \eqref{ansatz_monopole}. Note that also the dipole term, $\propto (\HH(z)r(z))^{-1}$ does not break this degeneracy as its is independent of $H_0$ since $\HH(z)\propto H_0$ and $r(z) \propto  H_0^{-1}$. Nevertheless, the ansatz can be used to fit $\Om_m$ or, equivalently $\Om_\La=1-\Om_m$ or other models for dark energy or models that include curvature. The degeneracy between $H_0$ and $M$ is lifted by the SNe for which also Cepheid distances are known.
	
\begin{figure}[!ht]
		\centering
		\includegraphics[width=0.8\linewidth]{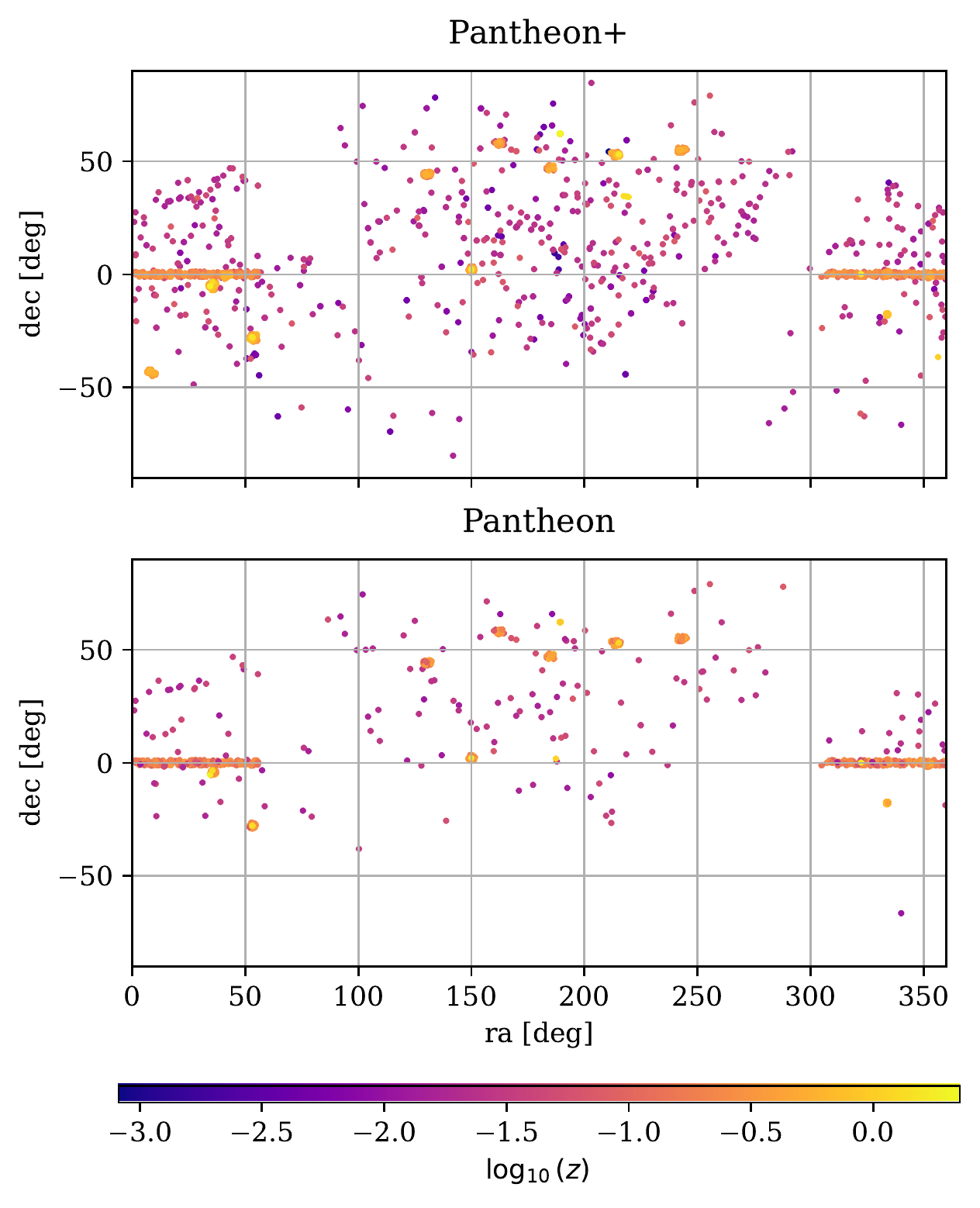}
		\caption{Positions and redshifts of 
			the supernovae in the  Pantheon+ (top panel) and Pantheon (low panel) data. We see that on the one hand, the Pantheon+ data has much more low redshift supernovae (blue and dark red dots) and on the other hand it is more uniformly distributed in the sky.} 
		\label{fig:test}
	\end{figure}

	We used the data and modified versions of codes from \url{https://github.com/PantheonPlusSH0ES/DataRelease}, where both the Pantheon+ data set and covariance matrix are available.
	We have tested our code using the Pantheon data for which this analysis has already been performed as reported in~\cite{Horstmann:2021jjg}. The Pantheon data, including the corrected redshifts and covariance matrix, was kindly provided to us by N.~Horstmann.
	In Fig.~\ref{fig:test} we show the angular distribution of the two data sets, with their redshifts indicated as colors. The number of supernovae above a given redshift cut are indicated in table~\ref{t1}.

	\begin{table}[!ht]
		\centering
		\begin{tabular}{ c c c c }
			\toprule
			$z_{\rm cut}$ & Pantheon+ & Cepheid hosts & Horstmann \\
			& without Cepheids & & \\
			\midrule
			No cut & 1624 & 77& 1048 \\
			0.005 & 1615 & 50 & 1048 \\  
			0.01 & 1576 &7 & 1046  \\
			0.0175  & 1468 & 2 &  1010  \\
			0.025  & 1312 &  0& 976  \\
			0.0375  & 1126 & 0 & 915  \\ 
			0.05  & 1054 & 0 & 890  \\
			0.1  & 960 & 0 & 837  \\
			\bottomrule
		\end{tabular}
		\caption{Number of supernova lightcurves for each data set changing the lower cut in redshift $z_{\rm cut}$. Clearly, the Pantheon sample used in Horstmann et al. has much fewer very low redshift SNe.\label{t:cut}. In the third column we also report the number of cepheid hosted supernovae inside the Pantheon+ catalogue. \label{t1} }
	\end{table}

	We consider different redshift cuts to the data, including only supernovae above a certain redshift $z_{\rm cut}$. The higher the cut, the less supernovae remain in the sample and the smaller their contribution to the dipole which is proportional to $(\HH(z)r(z))^{-1}$. On the other hand, very low redshift supernovae have velocities which are correlated to the observer velocity which will typically reduce the dipole amplitude.

\begin{figure}[!ht]
		\centering
		\includegraphics[width=1
		\linewidth]{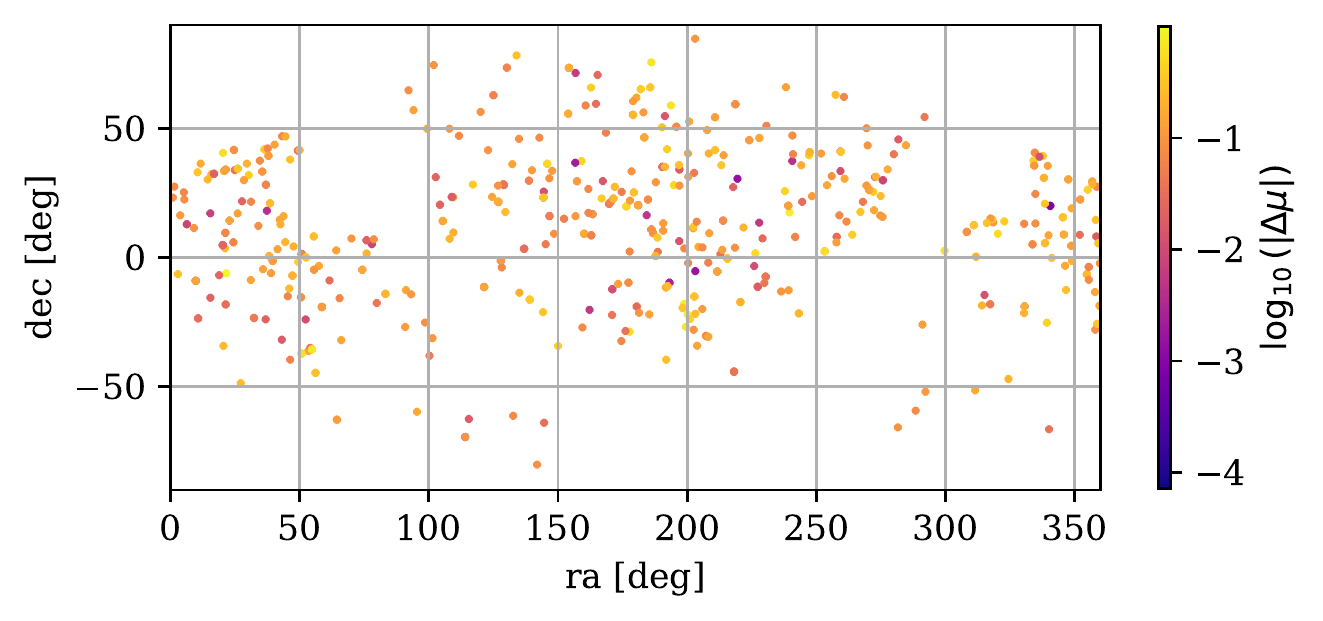} 
		\label{fig:position_snae2}
		\caption{Positions and absolute value of the residuals of the supernovae distance moduli in the  Pantheon+ data for redshift $z < 0.0375 $ with a dipole given by the Planck results. We choose this redshift limit as supernovae at higher redshifts lead to significantly larger errors in the dipole, see Table~\ref{tab:params_all_range}.}
		\label{fig:contribution_low_snae}
	\end{figure}

	From a first analysis, assuming the the order of magnitude of the dipole as in Planck (see Eqs.~\eqref{e:d-Planck}), we see that all the supernovae at $z<0.0375$, that we expect to be most relevant, are fairly uniformly distributed in the sky (see Fig. \ref{fig:contribution_low_snae}). This ensures that the detection of a ``dipole'' is not simply due to a selection effect. 
	
	We may therefore hope to find a `sweet spot' where both effects add to a minimum. The number of supernovae above a given redshift cutoff  for both samples are given in table~\ref{t:cut}. In this table one also sees that while above redshift 0.1, the number of supernovae in the two catalogs only differs by 123, at lower redshifts, $z<0.1$, Pantheon+ contains 741 light curves while Pantheon  contains only 211. This is the main reason why the dipole is much better determined by Pantheon+ than by Pantheon alone. Note also that above $z=0.0175$ only two galaxies also host a cepheid. The highest redshift galaxy with  cepheid  data is $z=0.01765$.
	
\begin{figure}[!ht]
		\centering
		\includegraphics [scale=0.8]{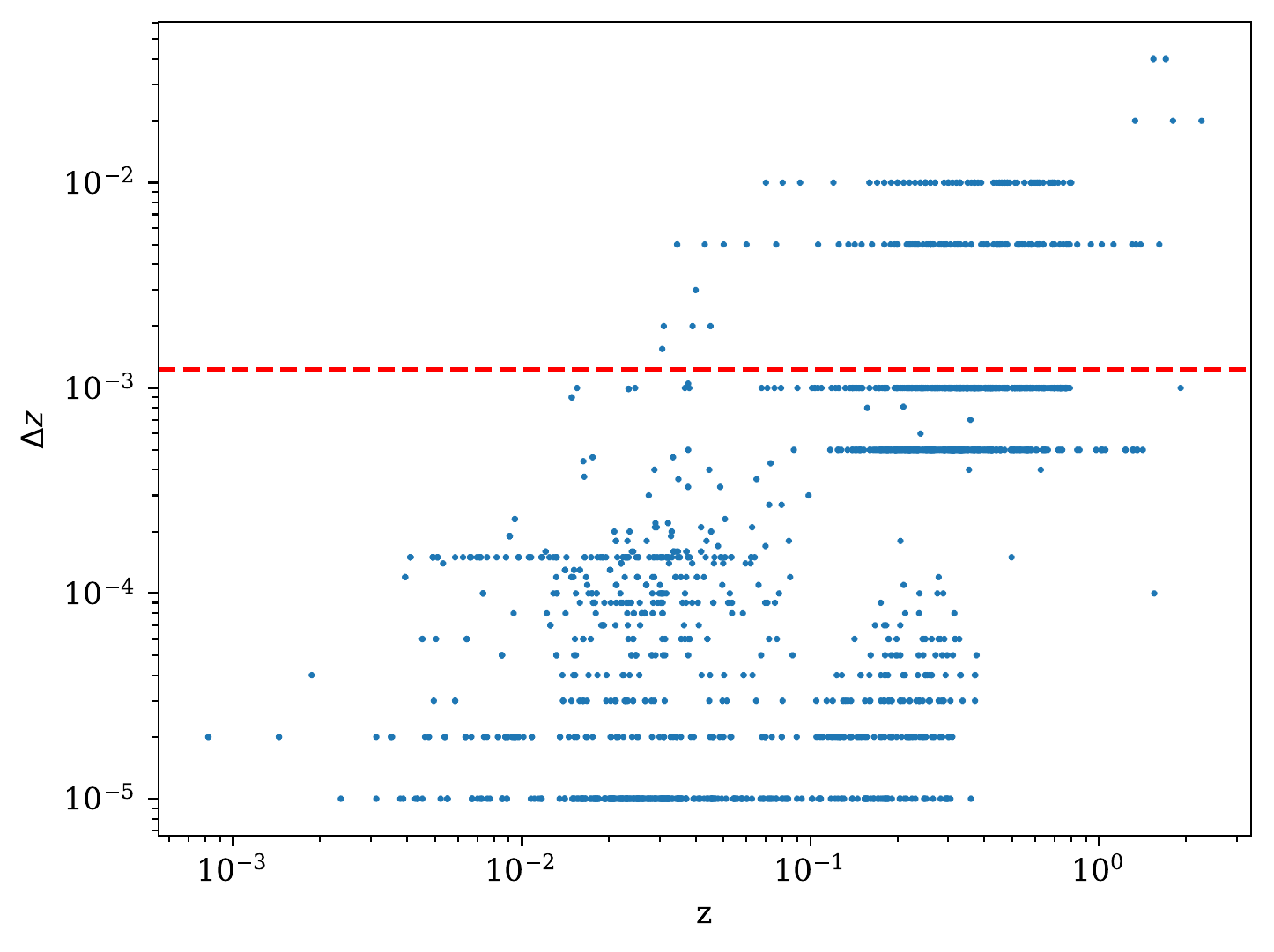}
		\caption{Scatter plot of redshift uncertainties $\Delta z$ of the Pantheon+ data set. The horizontal dashed red line, given by $v_{\rm Planck}/c$, shows us roughly the expected amplitude of the dipole.} 
		\label{fig:error_z}
	\end{figure}

	The redshifts used are the heliocentric J2000 redshifts derived in~\cite{Carr:2021lcj}. An accurate redshift determination is very important for our results.
	As we see in Fig. \ref{fig:error_z}, at low redshifts most uncertainties are significantly smaller than the dipole correction that, according to Eq.\eqref{e:vexact}, is of order $ v_{\rm Planck}/c$. For $z<0.03$ virtually all redshift uncertainties are smaller than the one induced by the CMB dipole. At higher redshifts this is no longer then case. For $z>0.1$ about half the uncertainties are larger than the amplitude of the CMB dipole.
	
	\section{Results and Discussion}
	\subsection{The Dipole of the Pantheon+ data set}
	~~~
	In the first part of our analysis, we follow the  steps prescribed by the Pantheon+ collaboration  in~\cite{Brout:2022vxf}, and maximize the following likelihood:
	\be \label{eq:likelihood}
	\log(\mathcal{L}) = - \frac{1}{2} \Delta \boldsymbol{\mu}^T C^{-1} \Delta \boldsymbol{\mu},
	\ee
	where $C$ is the covariance matrix and $\Delta \boldsymbol{\mu}$ is a vector whose elements refer to the data set and are given by:
	\be
	\Delta \mu^i =
	\begin{cases}
		\mu^i + \delta M -\mu_\mathrm{ceph}^i,  \quad i \in  \text{Cepheid hosts}\\
		\mu^i + \delta M -\mu_\mathrm{model}^i, \quad \text{otherwise.} \\
	\end{cases}
	\ee 
	The quantity $\delta M$ is a nuisance parameter which is determined by the supernovae in galaxies which also host observed Cepheids on which supernova distances are calibrated via the expression above (Cepheid hosts), where $\mu_\mathrm{ceph}$ is the Cepheid calibrated host-galaxy distance modulus. The supernovae from galaxies which do not host Cepheids are used to constrain our model. The model magnitude, $\mu_\mathrm{model}$, is
	\be
	\mu^i_\mathrm{model} = 5 \log \biggl( \frac{D_L(z_i, \bn_i)}{\rm Mpc} \biggr) + 25 
	\ee 
	where $D_L(z, \bn) $ is given by \eqref{e:ansatz_dipole}.
	
	We perform a parallelized Markov Chain Monte Carlo (MCMC) routine fitting the parameters $H_0$, $\Om_m$ as well as the nuisance parameter $\delta M$ and the velocity amplitude, $v_0$ and its direction in solar bari-center coordinates (ra,dec).
	
	For transformations of the position data we used the package \texttt{astropy}~\cite{astropy:2013,astropy:2018,astropy:2022}.
	We use the python packages \texttt{emcee}~\cite{emcee} for the MCMC and \texttt{chainconsumer} for the chain analysis~\cite{chainconsumer}. The MCMC is also run in parallel using the Python package \texttt{schwimmbad}~\citep{schwimmbad}. Our sampler is composed by 32 walkers using the ``stretch move" ensemble method described in~\cite{autocorr}. 
	Following the \texttt{emcee} documentation (see especially \\ \url{https://emcee.readthedocs.io/en/stable/tutorials/autocorr/}), we studied the integrated auto-cor\-re\-lation time $\tau$ as convergence diagnostics.\footnote{For a formal definition of the auto-correlation time $\tau$, the interested reader is referred  to \cite{autocorr}.}
	$\tau$ can interpreted as the number of steps after which the chain forgets its starting point. The chain is then assumed to be converged when $N/50 > \tau$, where $N$ is the number of steps made in the chain. For our analysis, in particular, we have seen that each walker generates a chain with a number of step of order $10^5$ before converging.
	Finally, we discard as burn-in the number of points given by twice the maximal integrated auto-correlation time of all parameters.
	
	For each parameter we use a wide uniform prior as reported in Table ~\ref{t:prior}.
	
	\begin{table}[!ht]
		\centering
		\begin{tabular}{ c c }
			\toprule
			Parameter & Prior range  \\
			\midrule
			$v_0$ & [0, 1200] km/s\\
			$\delta M$ & [-100, 100] \\  
			$H_0$ & [30, 100] km/s/Mpc  \\  
			$\Om_m$ & [0, 1]  \\
			\text{ra} & [0\textdegree, 360\textdegree] \\
			$\sin(\text{dec})$ & [-1, 1]  \\
			\bottomrule
		\end{tabular}
		\caption{Amplitude of the uniform priors applied to the parameters used in the MCMC routine. Instead of dec, we vary $\sin(\text{dec})$ in order to evenly sample surface elements on the celestial sphere. We then re-parametrize the posterior for dec using the $\arcsin$ function (note that $\sin(\text{dec})=\cos\vartheta$ for the standard polar coordinate $\vartheta$). \label{t:prior}}
	\end{table}
	
\begin{figure}[!ht]
		\centering
		\includegraphics [scale=0.75]{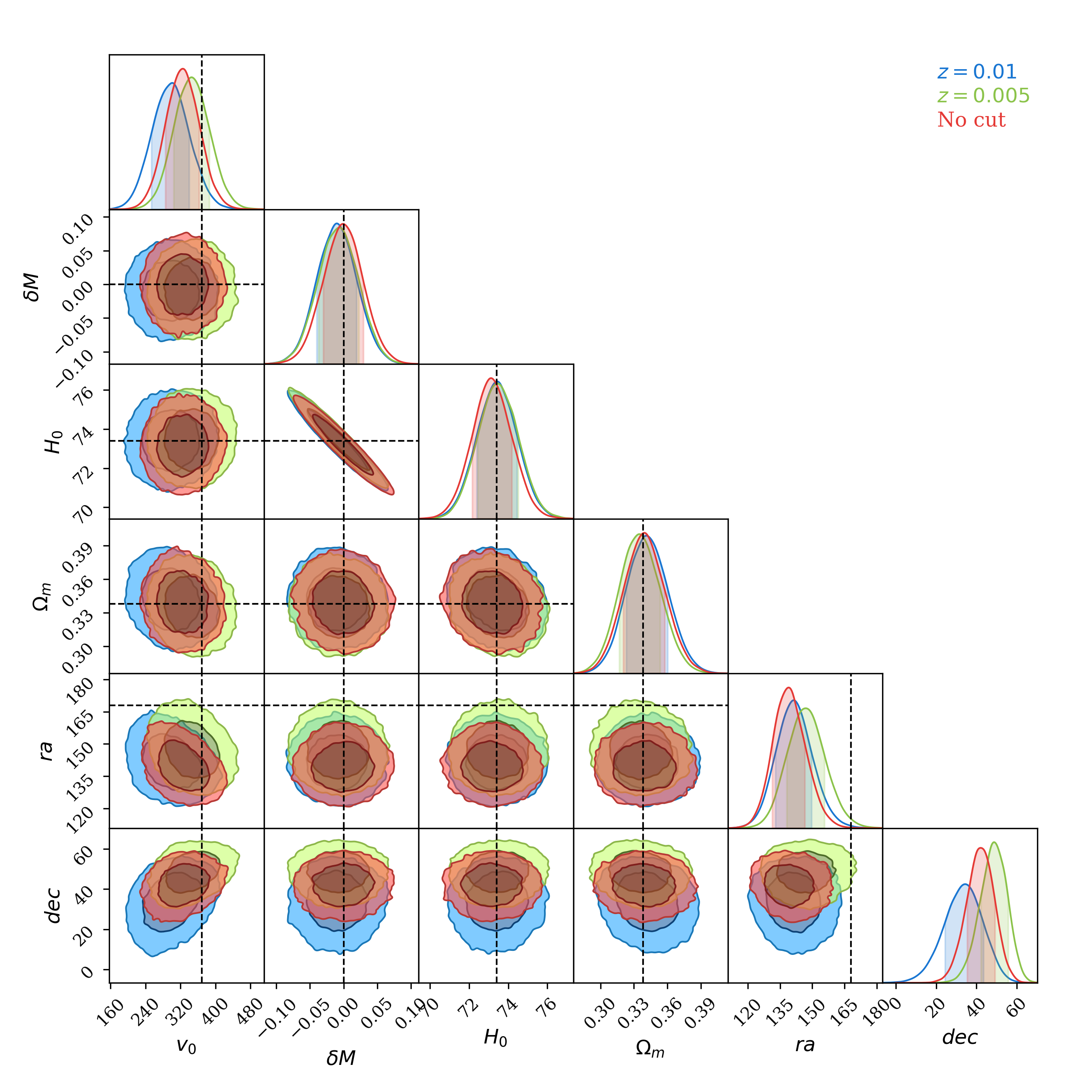}
		\caption{Contour plots for the Pantheon+ data set with different cuts in the redshift of the Supernovae. Darker shades include 65\% confidence regions while the lighter shades include the 95\% confidence regions. Here and in all the other plots, the dashed lines show the reference values $v_0$, \textit{ra, dec} from Planck given in \eqref{e:d-Planck-solar} and $\Omega_m=0.338$, $H_0=73.4$ given by~\cite{Brout:2022vxf}.} 
		\label{pic_marginalized1}
	\end{figure}

	The results of this MCMC analysis are shown in Figs.~\ref{pic_marginalized1} and \ref{pic_marginalization_more} and summarized in Table~\ref{tab:params_all_range}.

	\begin{table}[!ht]
		\centering
		\begin{tabular}{cccccccc}
			\toprule
			$z_{\rm cut}$ & N. of data &$v_0$ 
			[km/s] & $\delta M$ &  $H_0$ [km/s/Mpc]& $\Omega_m$ & ra [\textdegree] & dec [\textdegree] \\ 
			\midrule
			No cut & 1701 & $328^{+35}_{-42}$ & $-0.001^{+0.031}_{-0.029}$ & $73.11^{+1.07}_{-0.96}$ & $0.339^{+0.018}_{-0.019}$ & $139.4^{+7.2}_{-8.0}$ & $42.0^{+7.2}_{-6.6}$ \vspace{6 pt} \\
			0.005 & 1692 & $344^{+42}_{-40}$ & $-0.005^{+0.028}_{-0.032}$ & $73.5\pm 1.0$ & $0.335^{+0.019}_{-0.018}$ & $147.6^{+8.0}_{-9.5}$ & $48.9^{+6.9}_{-6.7}$ 
			\vspace{6 pt} \\
			0.01 & 1653  & $302^{+38}_{-49}$ & $-0.010\pm 0.030$ & $73.47^{+0.97}_{-1.09}$ & $0.340^{+0.020}_{-0.017}$ & $141.1^{+8.6}_{-8.2}$ & $34.4^{+9.1}_{-10.1}$
			\vspace{6 pt} \\ 
			0.0175 & 1545 & $377^{+57}_{-62}$ & $-0.012^{+0.031}_{-0.028}$ & $73.46^{+1.10}_{-0.97}$ & $0.342^{+0.016}_{-0.020}$ & $132.4^{+10.3}_{-8.2}$ & $45.2^{+8.3}_{-9.4}$ \vspace{6 pt} \\ 
			0.025 & 1389 &  $434^{+91}_{-77}$ & $-0.010^{+0.030}_{-0.028}$ & $73.38^{+1.10}_{-0.95}$ & $0.341^{+0.020}_{-0.017}$ & $137.1^{+11.9}_{-9.6}$ & $42.1^{+9.9}_{-10.6}$\vspace{6 pt} \\ 
			0.0375 & 1203 &  $490^{+110}_{-130}$ & $-0.010^{+0.030}_{-0.029}$ & $73.6^{+1.1}_{-1.0}$ & $0.338^{+0.018}_{-0.021}$ & $141^{+18}_{-15}$ & $33^{+17}_{-18}$ \vspace{6 pt} \\ 
			0.05 & 1131 & $370^{+150}_{-160}$ & $-0.011^{+0.031}_{-0.029}$ & $73.55^{+1.17}_{-0.99}$ & $0.333^{+0.022}_{-0.019}$ & $167^{+37}_{-30}$ & $21^{+34}_{-28}$ \vspace{6 pt} \\ 
			0.1 & 1037 &   $620^{+250}_{-310}$ & $-0.006^{+0.028}_{-0.031}$ & $73.5^{+1.0}_{-1.2}$ & $0.338^{+0.025}_{-0.026}$ & $211^{+29}_{-31}$ & $-2^{+46}_{-24}$ \vspace{6 pt} \\ 
			\bottomrule
		\end{tabular}
		\\
		
		\caption{Constraints on parameters for the dipole inferred in the Pantheon+ data set for different cuts in the redshift of the Supernovae. Note that the errors given here are the 68\% confidence errors from the MCMC routine. These are purely statistical errors and therefore should be interpreted with a grain of salt. We always use all SNe containing Cepheids since otherwise any constraints on $H_0$ and $\delta M$ are lost at higher $z_{\rm cut}$.  \label{tab:params_all_range}}
	\end{table}
	
	The supernovae in galaxies hosting also Cepheids are  included for each redshift cut, as they are just used to determine $\de M$ and do not depend on our model.  Not including them would lead to a perfect degeneracy between $\de M$ and $H_0$. The vertical dashed lines in Fig.~\ref{pic_marginalized1} indicate the values of the cosmological parameters obtained in the Pantheon+ analysis and the velocity from the Planck data (the value of the declination, $-6.9^\circ$, is outside the limits of the plot). 
	
	\begin{table}[!ht]
		\centering
		
		\begin{tabular}{cccc}
			\toprule
			\multicolumn{4}{c}{$\bv_{\rm bulk}=\bv_{\rm Planck}-\bv_0$} \\ \midrule
			$z_{\rm cut}$ & Amplitude [km/s] & ra & dec  \\ 
			\midrule
			No cut & $326 \pm 37 $ & $205.4$\textdegree$\pm11.3$\textdegree  &  -54.0\textdegree$\pm6.2$\textdegree \vspace{6 pt} \\
			0.005 & $349 \pm 40$  & 194.9\textdegree$\pm12.6$\textdegree  &  -60.3\textdegree$\pm6.3$\textdegree 
			\vspace{6 pt} \\
			0.01 & $282 \pm 46$  & 206.0\textdegree$\pm14.0$\textdegree  &  -49.7\textdegree$\pm8.4$\textdegree 
			\vspace{6 pt} \\ 
			0.0175 & $379 \pm 55$  & 213.7\textdegree$\pm15.7$\textdegree  &  -55.4\textdegree$\pm7.4$\textdegree   \vspace{6 pt} \\ 
			0.025 & $385 \pm 77$  & 229.4\textdegree$\pm24.8$\textdegree  &  -60.7\textdegree$\pm9.8$\textdegree \vspace{6 pt} \\ 
			0.0375 & $362 \pm 133$  & 258.0\textdegree$\pm39.8$\textdegree  &  -59.1\textdegree$\pm21.5$\textdegree  \vspace{6 pt} \\ 
			0.05 & $178 \pm 198$ & 183.1\textdegree$\pm533.9$\textdegree$\ast$  &  -83.0\textdegree$\pm52.0$\textdegree  \vspace{6 pt} \\ 
			0.1 & $432\pm 296$ & 66.4\textdegree$\pm41.2$\textdegree  &  -3.0\textdegree$\pm50.1$\textdegree 
			\vspace{6 pt} \\ 
			\bottomrule
		\end{tabular}
		\\
		\caption{Differences between the best fit value for the CMB dipole by Planck and the value for the velocity dipole from Table \ref{tab:params_all_range}. We expect this quantity to be the velocity bulk flow {outside a ball of radius $3000z_{\rm cut}h^{-1}$Mpc}. The errors are computed with the help of the python package \texttt{uncertainties}~\cite{uncertainties}, assuming a negligible error for the Planck dipole. ($\ast$: The large uncertainty of the right ascension for $z_{\rm cut} = 0.05$ should be considered as a formal value, it just tells us that the direction cannot be determined, which is not surprising as the amplitude is compatible with $0$.)}
		\label{table_bulk_flow}
	\end{table}
	
	\begin{figure}[!ht]
		\centering
		\includegraphics {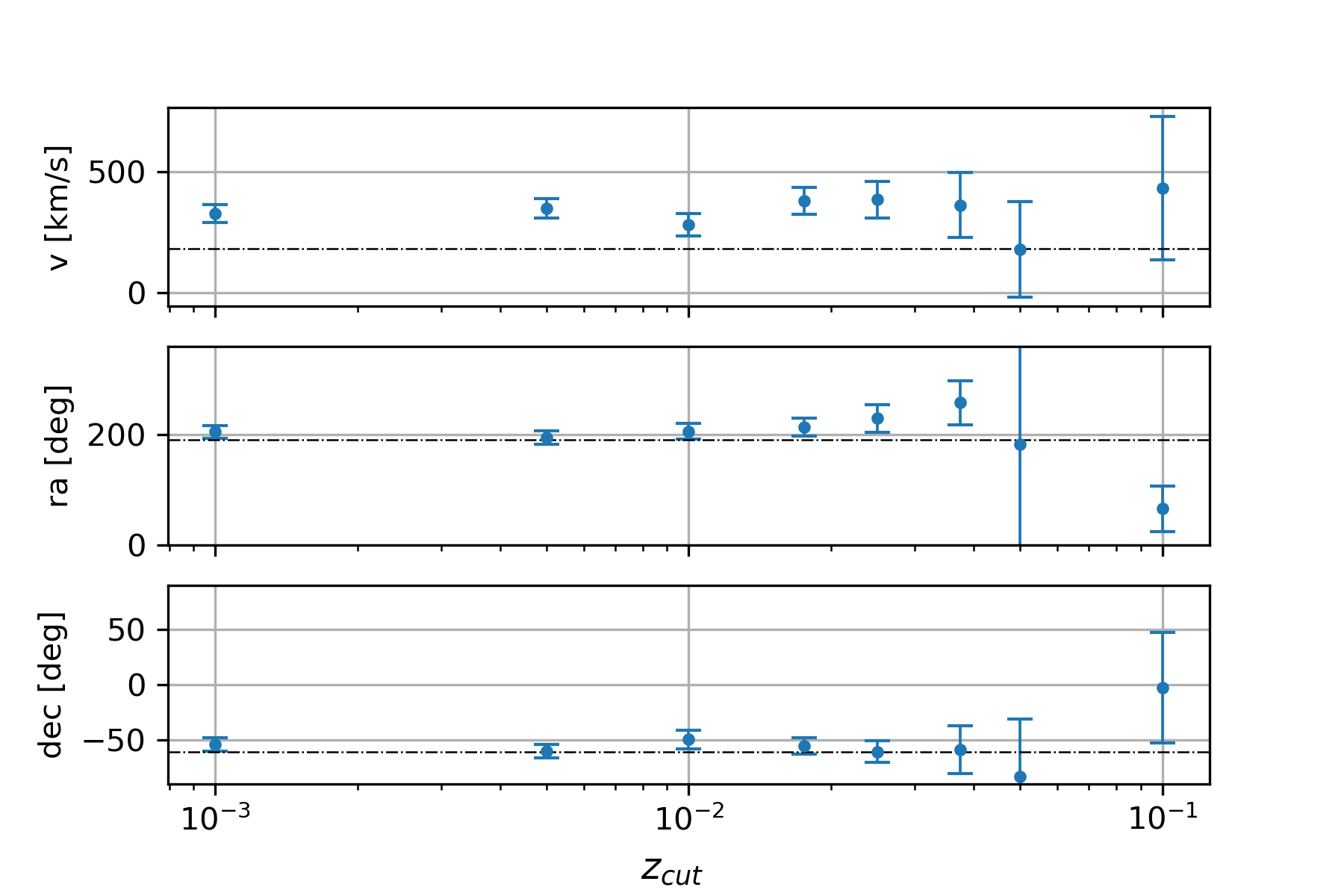}
		\caption{Here we show graphically the differences $\bv_{\rm bulk}=\bv_{\rm Planck}-\bv_0$ reported in Table \ref{table_bulk_flow} for increasing $z_{\rm cut}$. The black horizontal dash-dotted lines indicate the values $|\bv_{\rm bulk}|= 182\,$km/s, (ra, dec) = (191\textdegree ,-61\textdegree) applied as `bulk velocity corrections' in~\cite{Carr:2021lcj}.} 
		\label{pic_|table}
	\end{figure}
	
	As a first result, we find that the Hubble parameter and matter density agree well with the best fit values published in the original analysis~\cite{Brout:2022vxf} and the errors are surprisingly insensitive to the redshift cut. This is very different for the velocity. While the amplitude remains within about 2$\si$ of the Planck value, the direction for the  cutoffs shown in the plot but also for other cutoffs $z_{\rm cut}\leq 0.025$ is  significantly different. For the cutoffs shown in Fig.~\ref{pic_marginalized1} the Planck direction is excluded at more than $3\si$ (note that the Planck value for the declination ($-6.9$) is actually outside the plot range). Only for a cutoff at $z_{\rm cut} \geq 0.05$, the direction is within $1\si$ of the one measured by Planck. But this is due to the fact that the errors significantly increase, while the best fit value actually does not change very much. This is also explained by the fact that for $z_{\rm cut} \geq 0.05$, a vanishing dipole, $v_0=0$, is within the 95\% confidence contour. As anticipated, at very low $z_{\rm cut}$ the velocity amplitude is smaller than the Planck value and it reaches this value at about $z_{\rm cut}=0.0175$. For larger $z_{\rm cut}$ the errors increase substantially but the Planck velocity amplitude remains with $1\sigma$. However, the direction, as long as it is well determined, is very different.
	
	In Table~\ref{table_bulk_flow} we show the difference of the best fit dipole velocity $\bv_0$  and our peculiar velocity as inferred by Planck, $\bv_{\rm Planck}$. This can be interpreted as the total 'bulk velocity' of the galaxies at $z>z_{\rm cut}$. For $z_{\rm cut}<0.0375$ this difference is stable (within error bars) and of the same order as $\bv_{\rm Planck}$. At higher redshifts the errors become considerable, but the difference vector remains stable within errorbars. The simplest interpretation of this is that all SNe out to $z\simeq 0.0375$, corresponding to a ball of comoving radius  $R= 112h^{-1}$Mpc, move with a coherent bulk velocity given by $\bv_0-\bv_{\rm Planck}$. In Fig.~\ref{pic_|table} we compare this bulk velocity with the one used in the Pantheon+ analysis~\cite{Brout:2022vxf}. While the direction agrees very well, the amplitude found with our agnostic approach is nearly double the bulk velocity adopted in~\cite{Brout:2022vxf} for redshift cuts which allow a good determination. The peculiar velocities adopted in \cite{Brout:2022vxf} are described in detail in~\cite{Carr:2021lcj}. They are inferred from an estimate of the gravitational field generated by  the nearby density field. They are not inferred from the Supernova data directly. The authors use the 2M++ velocity field as estimated in~\cite{Carrick:2015xza}. Also the bulk flow is reconstructed by the 2M++ data. It is close to the $\bv_{\rm ext}$ reported in~\cite{Carrick:2015xza} but not identical, as the peculiar velocities averaged over $200h^{-1}$Mpc are added. A final bulk velocity of $182\,$km/s in direction $(\ell,b)=(302^o,2^o)$ is found, see~\cite{Carr:2021lcj} for more details. Even though the dipole inferred from the heliocentric redshift or from $z_{\rm HD}$ used in the Pantheon+ analysis differs, the inferred cosmological parameters $H_0$ and $\Om_m$ agree very well.
	
	Furthermore, it is interesting to notice how these inferred values are within 1-2 $\sigma$ in agreement with the ones recently obtained from the analysis of \textit{CosmicFlows4} \cite{Tully_2023}, for which a bulk flow of $|\bv_{\rm bulk}|= 395\,$km/s, (ra, dec) = (178\textdegree ,-66\textdegree) within a ball of radius R=$150h^{-1}$Mpc was obtained \cite{Watkins_2023}.

	\begin{table}
		\centering
		\begin{tabular}{ c c c c }
			\toprule
			$z_{\rm cut}$ & $\chi^2 _ {\rm No-dip}$ - $\chi^2 _ {\rm best-fit}$  & $\chi^2 _ {\rm Planck}$ - $\chi^2 _ {\rm best-fit}$  & $\chi^2 _ {z_{\rm HD}}$ - $\chi^2 _ {\rm best-fit}$ \\
			\midrule
			No cut & 88.2 & 66.4 & 9.1
			\\
			0.005 & 88.5 & 68.5 & 19.1 \\
			0.01 & 62.1  & 41.4 & 15.0\\
			0.0175 & 53.6 & 42.6  & 14.4 \\
			0.025 & 41.7 & 19.2  & -2.1 \\
			0.0375 & 22.3 & 5.3  & 1.3 \\
			0.05 & 8.7 & 0.9  & -1.0 \\ 
			0.1 & 7.4 & 3.4  & 2.9 \\  
			\bottomrule
		\end{tabular}

		\caption{$\chi^2$ differences for different redshift cuts between the hypothesis for the dipole and the best fit dipole determined by our analysis. No dipole (first column), the dipole from Planck (second column) and the Hubble diagram redshifts $z_{\rm HD}$ including source velocities in the redshift as in the analysis from the Pantheon+ collaboration (third column).}
		\label{table_chisq}
	\end{table}

	In Table~\ref{table_chisq} we report the $\chi ^2$ differences between the best fit with our fitted dipole and the best fit  assuming no dipole, first column, the best fit including the  Planck dipole, second column  and the best including the dipole from the Planck data and the bulk velocity used in the Pantheon+ analysis. The best fit dipole is very strongly favored over no dipole or the Planck dipole. In the last column we also consider the bulk velocity used by the Pantheon+ collaboration. In addition to the Planck dipole from the observer velocity. This bulk flow is modeled as $182\,$km/s  out to redshift $z=0.067$ (corresponding to 200$h^{-1}$Mpc) after which it is assumed to decay according to the $\La$CDM linear growth~\cite{Carr:2021lcj}.
	Assuming a normal distribution,  which is not far from what we obtain in Fig.~\ref{pic_marginalized1}, and using a redshift cut $z_{\rm cut}\leq 0.02$, the $p$-values for no dipole or the Planck dipole are all above $1-5\times 10^{-9}$ for a normal distribution with 3 additional parameters, the 3 components of the dipole which we fit in addition to the cosmology. In other words, the probability to observe what we do while no dipole or the Planck dipole is the true underlying model is less than $5\times 10^{-9}$. For  the $\chi^2$-difference between the dipole as modeled by the Pantheon+ collaboration we still obtain a p-value of $p=97.5$\%  and more. If a larger redshift cut is imposed the Pantheon+ modelling is comparable to the `best fit' dipole modelling. For these values, however the detection of the dipole is also much less significant. For $0.025<z_{\rm cut}< 0.05$, the dipole is still significant but can be modeled as the Planck dipole, while for $z_{\rm cut}\geq 0.05$, the dipole is no longer clearly detected.

	\subsection{Comparison with the Dipole of Pantheon}\label{sec:pantheon_comparison}

	\begin{figure}[!ht]
		\centering
		\includegraphics [scale=0.8]{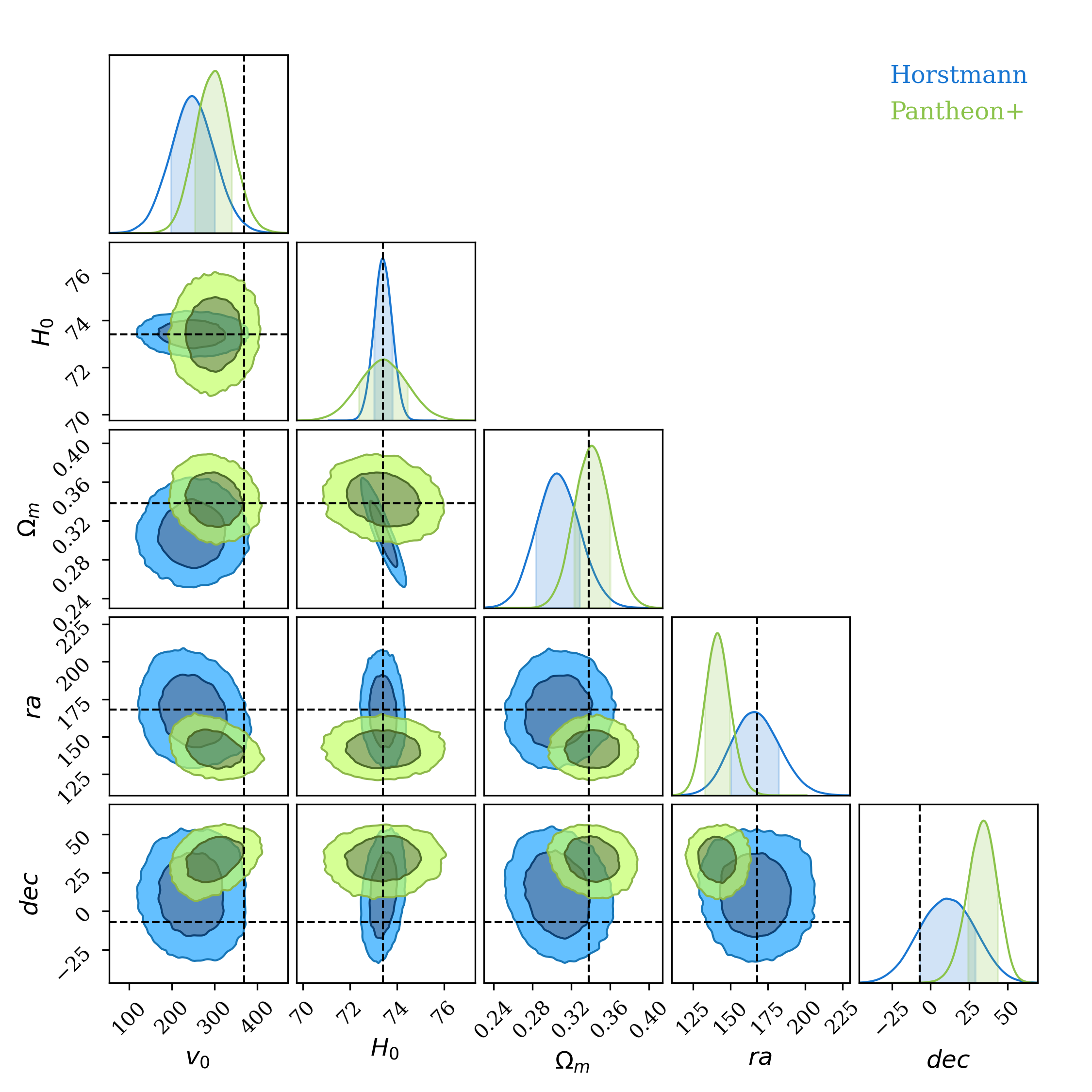}
		\caption{Contour plots for the Pantheon+ data set with lower cut in redshift z=0.01 and the Pantheon data set as provided by Horstmann. 
		}
		\label{pic3}
	\end{figure}
	
	We have repeated our analysis with the Pantheon data that was already analyzed in~\cite{Horstmann:2021jjg}. In this paper corrections to the supernova redshifts published in~\cite{Pan-STARRS1:2017jku} have been applied as found in~\cite{Steinhardt:2020kul}. Furthermore, some of the  SNe directions have been corrected, see~~\cite{Horstmann:2021jjg}. Our results shown in Figs.~\ref{pic3} and~\ref{f:compar} agree perfectly with those published in~\cite{Horstmann:2021jjg} if we impose no cut in redshift as did the authors of~\cite{Horstmann:2021jjg}. This is also a test of our code which is independent of the one by~\cite{Horstmann:2021jjg}. While the best fit velocity amplitude is rather on the low side, the dipole direction does agree with the Planck data. The full dipole agrees with Planck within about 2.4$\si$. As this removes only 2 SNe from the Pantheon data, we do not separately study the cut of $z_{\rm cut}=0.01$.  However, if we remove the supernovae with $z<z_{\rm cut}=0.05$, the dipole amplitude does peak roughly at the Planck value (it is even slightly larger), but the errors become very large. Within $2\si$ values in the full range $0\leq v_0\leq 1000\,$km/s are allowed and we do not detect the dipole with high significance, see Fig.~\ref{f:compar}. For this data set also the errors in the cosmological parameters $H_0$ and $\Om_m$ increase  when the cut is imposed. The directions of the dipole agree with the Planck dipole with and without cut, even though, especially the declination is not very stable under the cut. But this is not surprising since a very low, even vanishing dipole is not excluded at more than $2\si$.
	\begin{figure}[!ht]
		\centering
		\includegraphics[width=0.75 \linewidth]{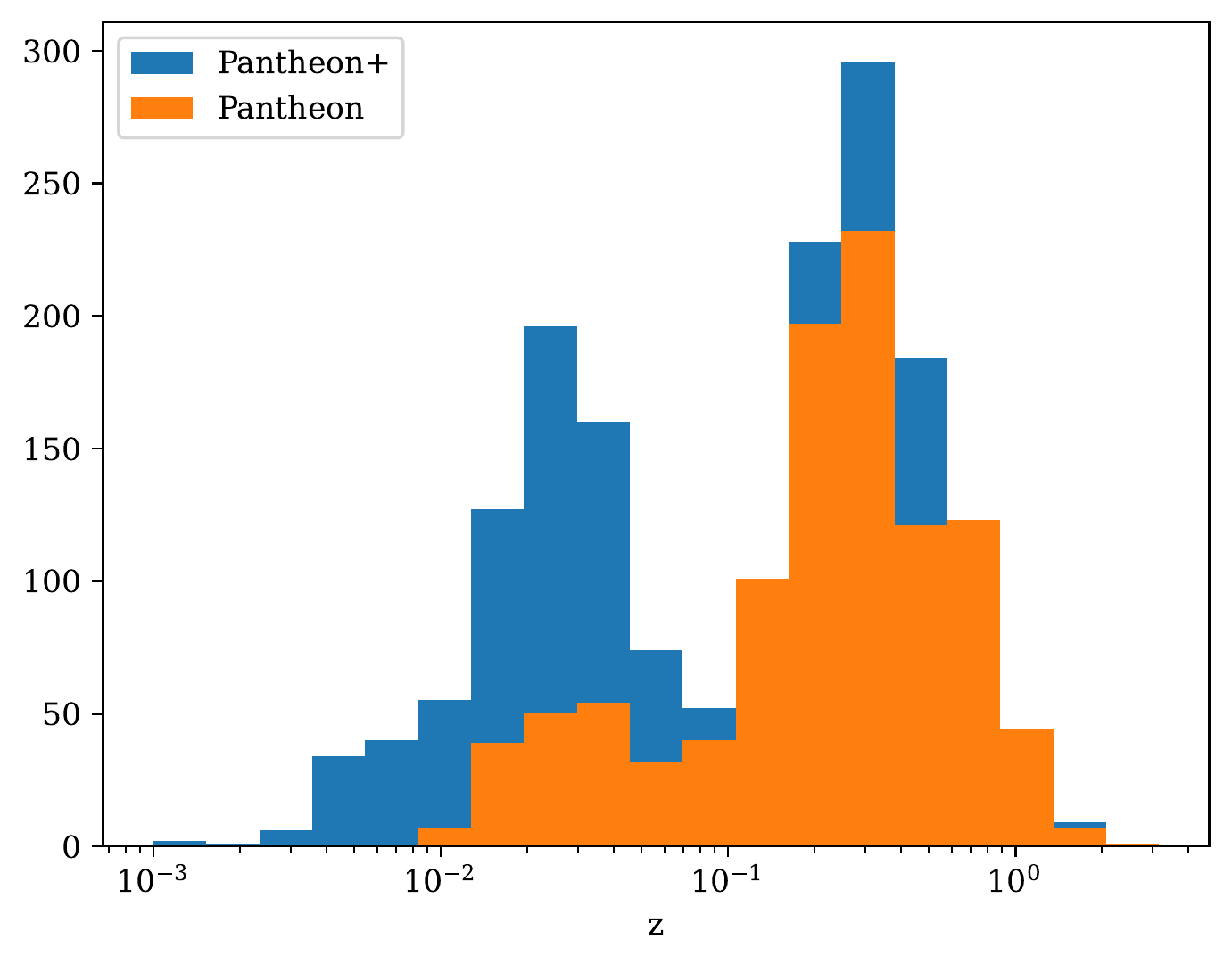} 
		
		\caption{Redshift distribution for the Pantheon+ and Pantheon datasets. The largest amount of new data in Pantheon+, with respect to Pantheon, is located  at redshifts $z<0.1$.}
		\label{fig:pantheon_pantheon_plus_distribution}
	\end{figure}
	
	It is also instructive to compare the inferred dipole from the two data sets. In Fig.~\ref{pic3} we show $H_0$, $\Om_m$ and the dipole amplitude and direction obtained from our MCMC for both data sets with a redshift cut at $z_{\rm cut}=0.01$ which corresponds to a distance of $30h^{-1}$Mpc and seems a reasonable compromise between correlated velocities and too large errors. Note also that this cutoff removes only 2 of the Pantheon lightcurves while it removes 124 light curves from Pantheon+ .
	
	Clearly, the two data sets are in good agreement, all error ellipses overlap within 1$\si$. Of course they are by no means independent as most if not all the supernovae contained in Pantheon are also in Pantheon+; i.e., the former is a subset of the latter. At the same time, it is important to underline that the two datasets are different in many ways, in particular concerning the calibration.\footnote{The main differences are summarised in Section 2 of \cite{Scolnic_2022}}
	
	The smaller data set Pantheon clearly has substantially larger errors. The fact that in our analysis the Pantheon+ errors for $H_0$ are larger than the ones in the Pantheon data is artificial since in the Pantheon analysis no Cepheids are included and no marginalisation over $\de M$ is performed, but $\de M$ is fixed to zero. 
	
	It is most interesting to note that while Pantheon still roughly agrees with both, Pantheon+ and the Planck dipole, the   Pantheon+ data, very significantly disagrees with the Planck dipole direction. This data is now good enough to see the dipole with high significance and to determine its direction with relatively small errors. It clearly does not point in the same direction as the Planck dipole. This is best quantified by the $\chi^2$ difference reported in Table~\ref{table_chisq}.
	\begin{figure}[!ht]
		\centering
		
		\includegraphics[width=1
		\linewidth]{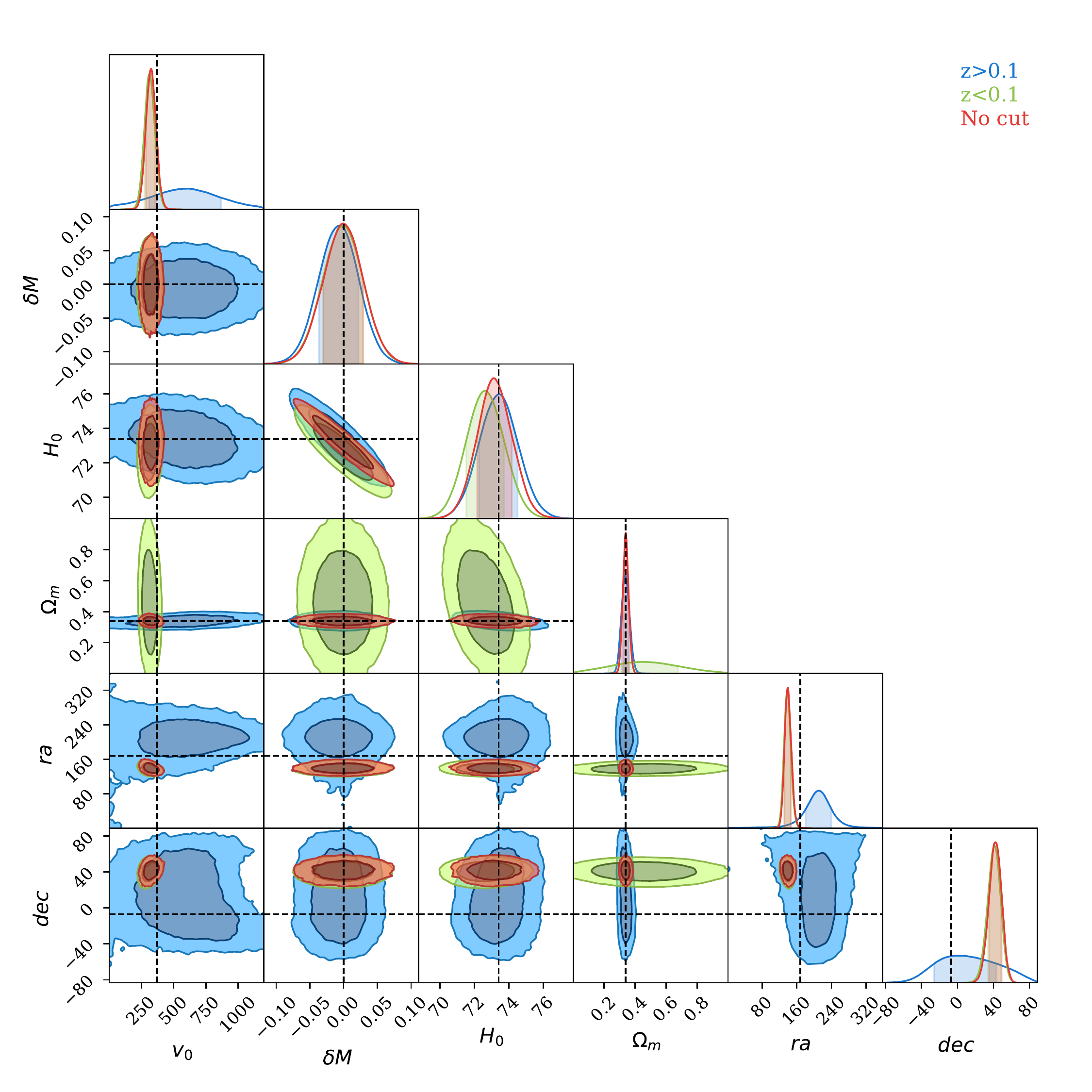} 
		
		\caption{Contour plots of the Pantheon+ dataset considering three different cuts: all the supernovae, only the supernovae with $z<0.1$ and only the supernovae with $z>0.1$. }
		\label{fig:dipole_low_high_z_comparison}
	\end{figure}
	
	As we can see from Fig. \ref{fig:pantheon_pantheon_plus_distribution}, the largest amount of new data in Pantheon+, with respect to Pantheon, are located  at redshifts $z<0.1$.  In Fig. \ref{fig:dipole_low_high_z_comparison} we compare the contour plots that we obtain considering three different cuts: all the supernovae, only the supernovae with $z<0.1$ and only the supernovae with $z>0.1$. In this way, we can see which is the major contribution coming from the new supernovae.

	The contour plots with all the supernovae and with only low redshift supernovae agree perfectly in the detection of the dipole. This is another indication of the fact that the low redshift supernovae are the ones that determine the dipole. However, the low-z supernovae do not provide strong constraints on cosmological parameters such as $\Omega_m$, as their posterior distribution is significantly broader. Conversely, high-z supernovae which constrain the cosmological parameters $\Om_m$ and $H_0$ effectively and therefore drive these constraints, while low-z supernovae determine the dipole.

	\subsection{Fixing the velocity direction}
	
	\begin{figure}[th]
		\centering
		\includegraphics[width=0.8\linewidth]{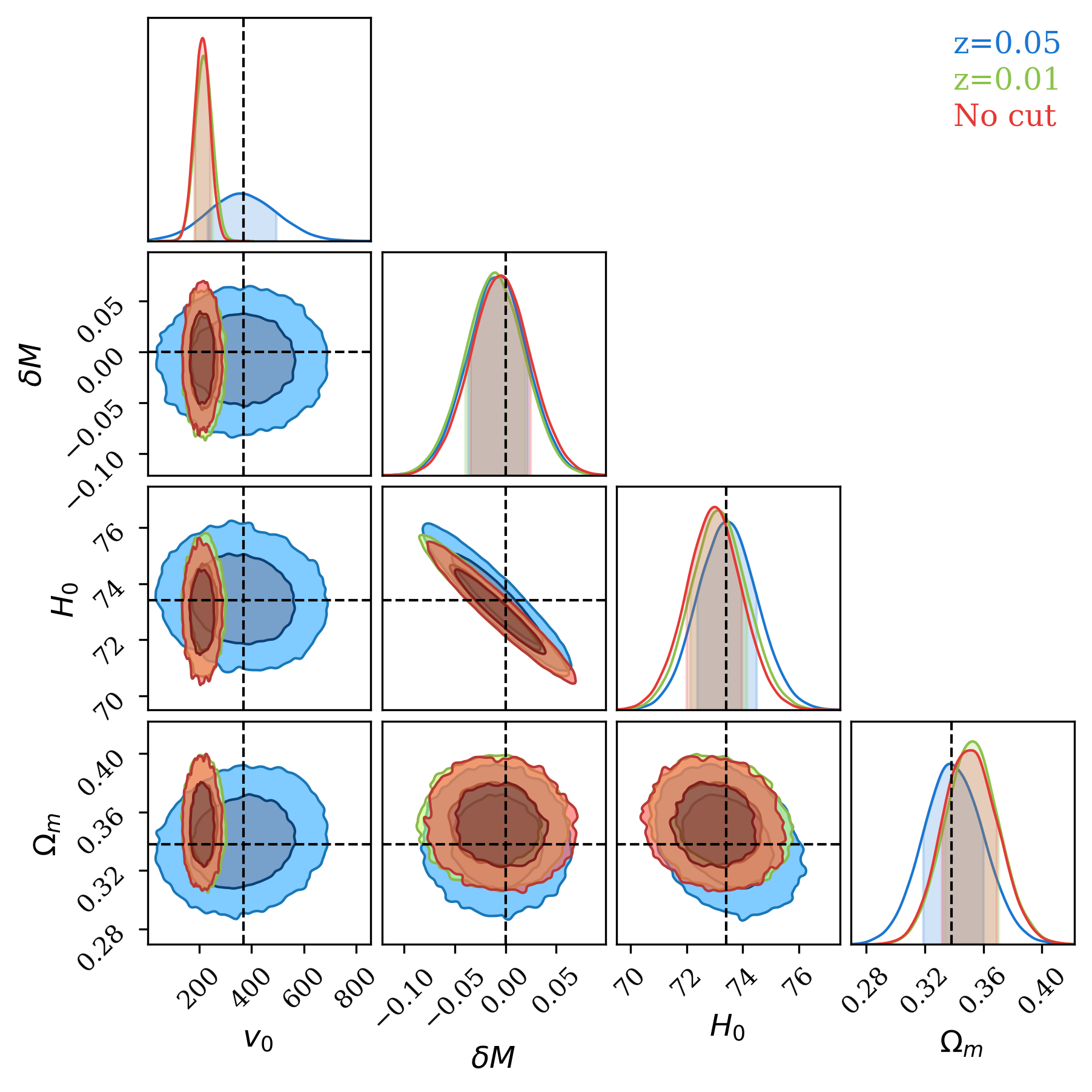}  \caption{Contour plots for the Pantheon+ data set with different cuts in the redshift of the Supernovae for an analysis with the direction of the dipole fixed to the one found by Planck.}
		\label{pic:velocity_fixed}
	\end{figure}
	
	Since the main difference of our dipole wrt. the Planck results is  the sky-position of the dipole, we also analysed the Pantheon+ data fixing the direction of the dipole to the one from Planck for different redshift cuts, see Figs. ~\ref{pic:velocity_fixed} and \ref{pic_vfixed_more}.
	
	For $z_{\rm cut}\leq 0.025$ The inferred amplitude of the dipole is smaller than the Planck value with more then 68\% confidence for $z_{\rm cut}= 0.025$ and more than 95\% for smaller cuts, see Fig. ~\ref{pic:velocity_fixed}. For larger redshift cuts, the dipole amplitude roughly agrees with the Planck value but for $z_{\rm cut}\geq 0.05$, the dipole is just marginally detected, at 95\% confidence.

	\subsection{A visual model comparison}
\begin{figure}[ht]
		\centering
		\includegraphics [scale=1]{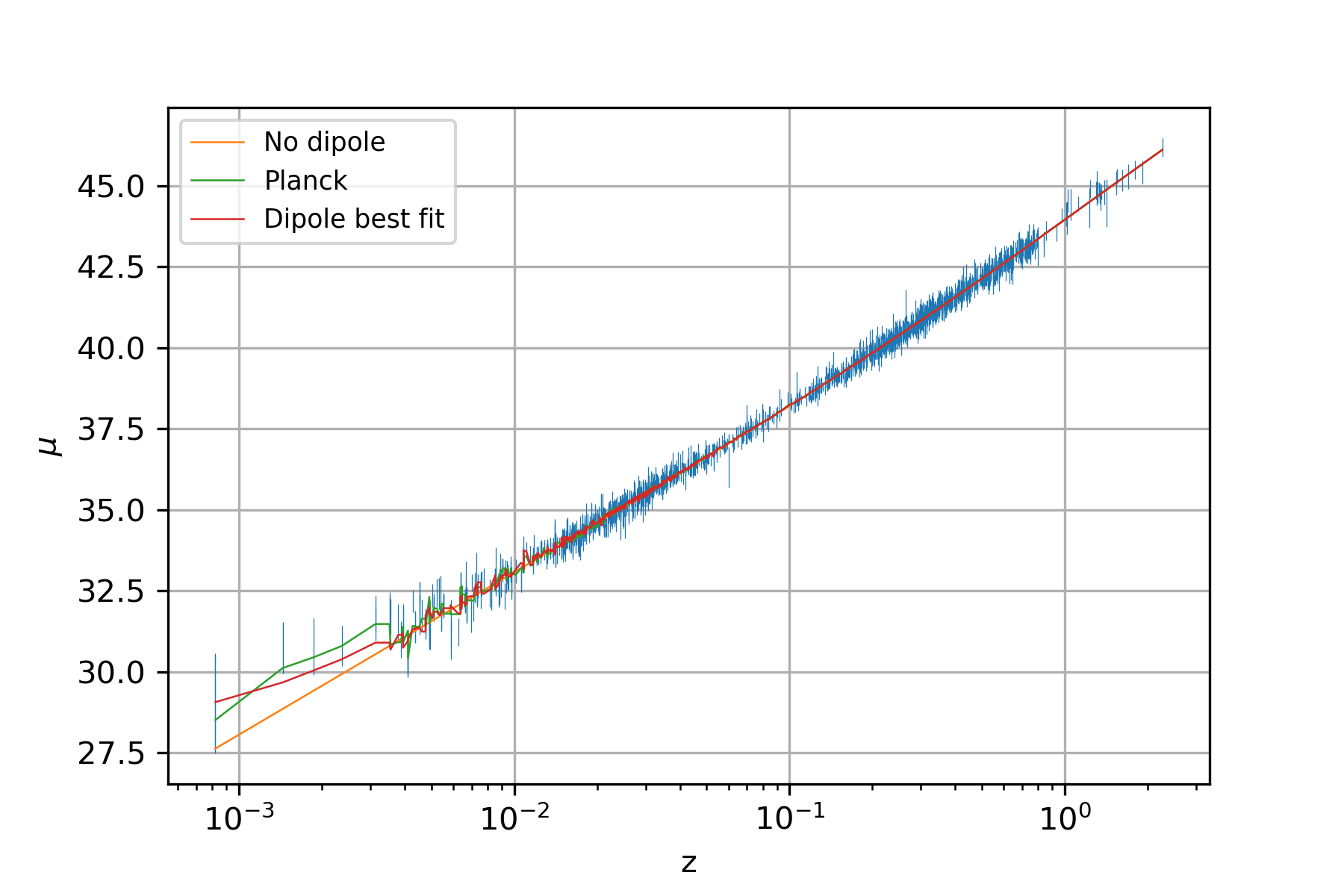}
		\caption{Plot of the distance moduli $\mu$ as a function of z (without imposing any lower cut in redshift). The blue vertical error bars are determined by the diagonal of the covariance matrix: therefore, they can be used only for visual purposes and not for fitting cosmological parameters. \label{f:mu-plot}}
	\end{figure}

\begin{figure}[ht]
		\centering
		\includegraphics [scale=0.8]{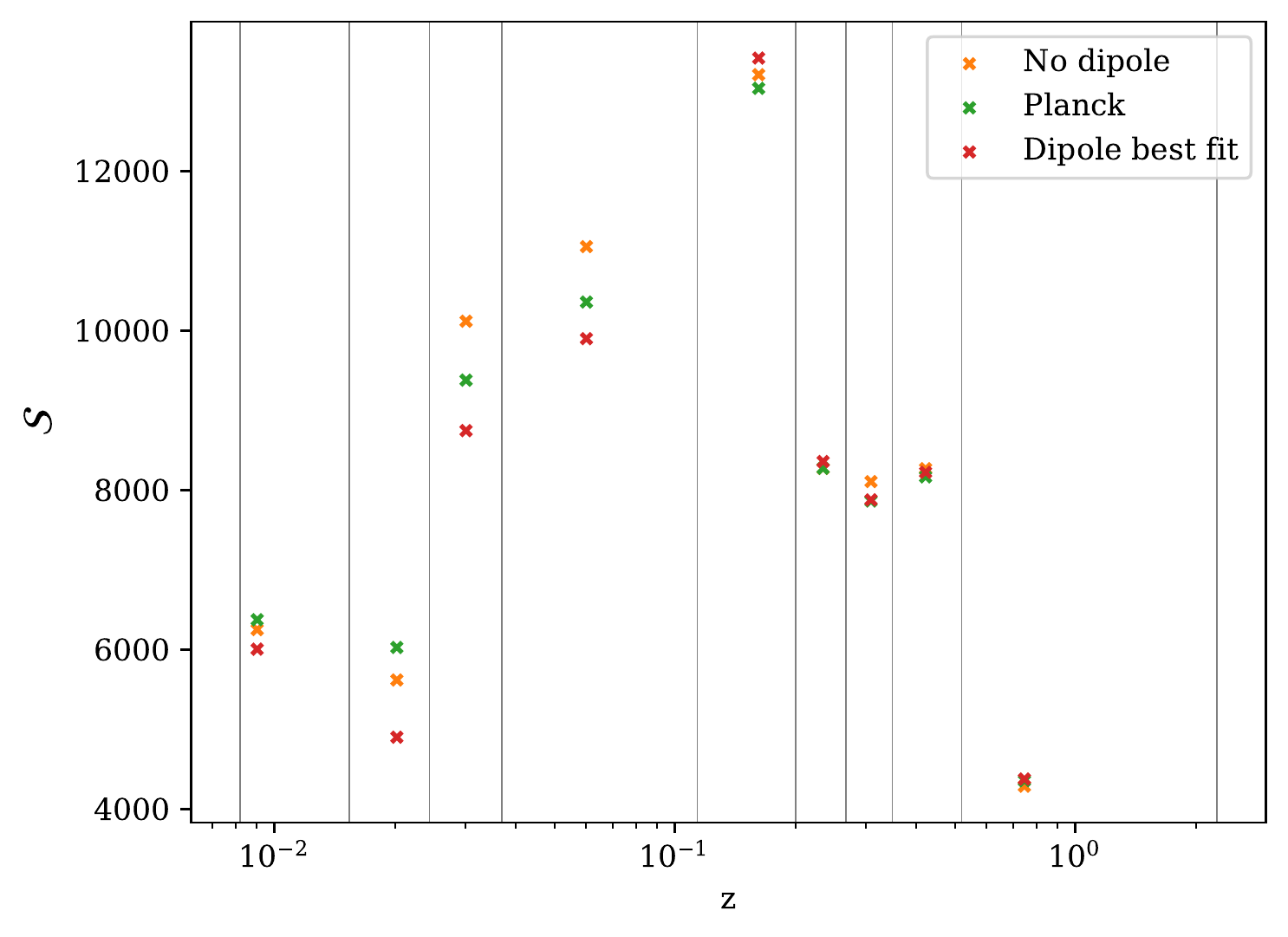}
		\caption{The binned metric $\mathcal{S}= \Sigma_i ( \Delta \mu_i )^2/(\sigma_i)^2$, with $\sigma_i$ diagonal entries of the covariance matrix. We binned the data in 9 bins containing the same number of supernovae (i.e. 189 data points). The vertical grey lines show the bin edges reported in Tab. \ref{tab:redshift-bin}. Each point is located along the horizontal axis in correspondence of the mean of the redshifts falling in that bin. }\label{f:mu-residuals}
	\end{figure}

	In Fig.~\ref{f:mu-plot} we plot the data together with the best fit values for the distance modulus $\mu$ for three different fits: the Planck dipole is assumed and only $H_0$ and $\Om_m$ are fitted (green line), the dipole is set to zero and $H_0$ and $\Om_m$ are fitted (orange line), the dipole is fitted together with $H_0$ and $\Om_m$ (red line). At first sight it looks as if the Planck dipole (green line) would fit the data better than the best fit dipole (red line). But this comes only from the first five closest supernovae which are statistically not relevant (besides that fact that we expect their redshifts to be most strongly affected by peculiar velocities). The many supernovae at slightly higher redshift, in the range $0.005<z<0.03$, are too crowded to evaluate the goodness of fit by eye. Also, the error in magnitude of each individual supernova in this range is larger than the distance between the different fits. For this reason we show in Fig.~\ref{f:mu-residuals} also the binned residuals of the different fits. More precisely, using the same colours, we plot the quantity $\mathcal{S}= \Sigma_i ( \Delta \mu_i )^2/(\sigma_i)^2$, where $\sigma_i$ are the diagonal entries of the covariance matrix.  This quantity contributes to the likelihood that we assumed to be Gaussian (see Eq.~\eqref{eq:likelihood}). We binned our data in 9 bins. We choose the bin widths such that each bin contains the same number of supernovae (i.e. 189 data points). The edges of the bins are indicated as vertical lines in Fig.~\ref{f:mu-residuals}. We also report them in  Table~\ref{tab:redshift-bin} in the Appendix. Each point is located along the horizontal axis in correspondence of the mean of the redshifts falling in that bin. As expected, our best fit analysis is significantly better (smaller value of $\mathcal{S}$, at redshifts $z < 0.1 $, i.e. at all redshift at which the dipole is detected with high significance -- as we saw at the end of Sec.~\ref{sec:pantheon_comparison}, low redshift supernovae are the ones encoding most information about the dipole.
	
	\subsection{The effects of the redshift cuts}\label{sec:redshift_cut}
	In order to test our assumptions and conclusions according to which, when increasing $z_{\rm cut}$, we reduce the detectability of the dipole, we applied our pipeline to two mock datasets.\\
	\begin{minipage}{0.48\linewidth}
		\begin{figure}[H]
			\centering
			\includegraphics [scale=0.5]{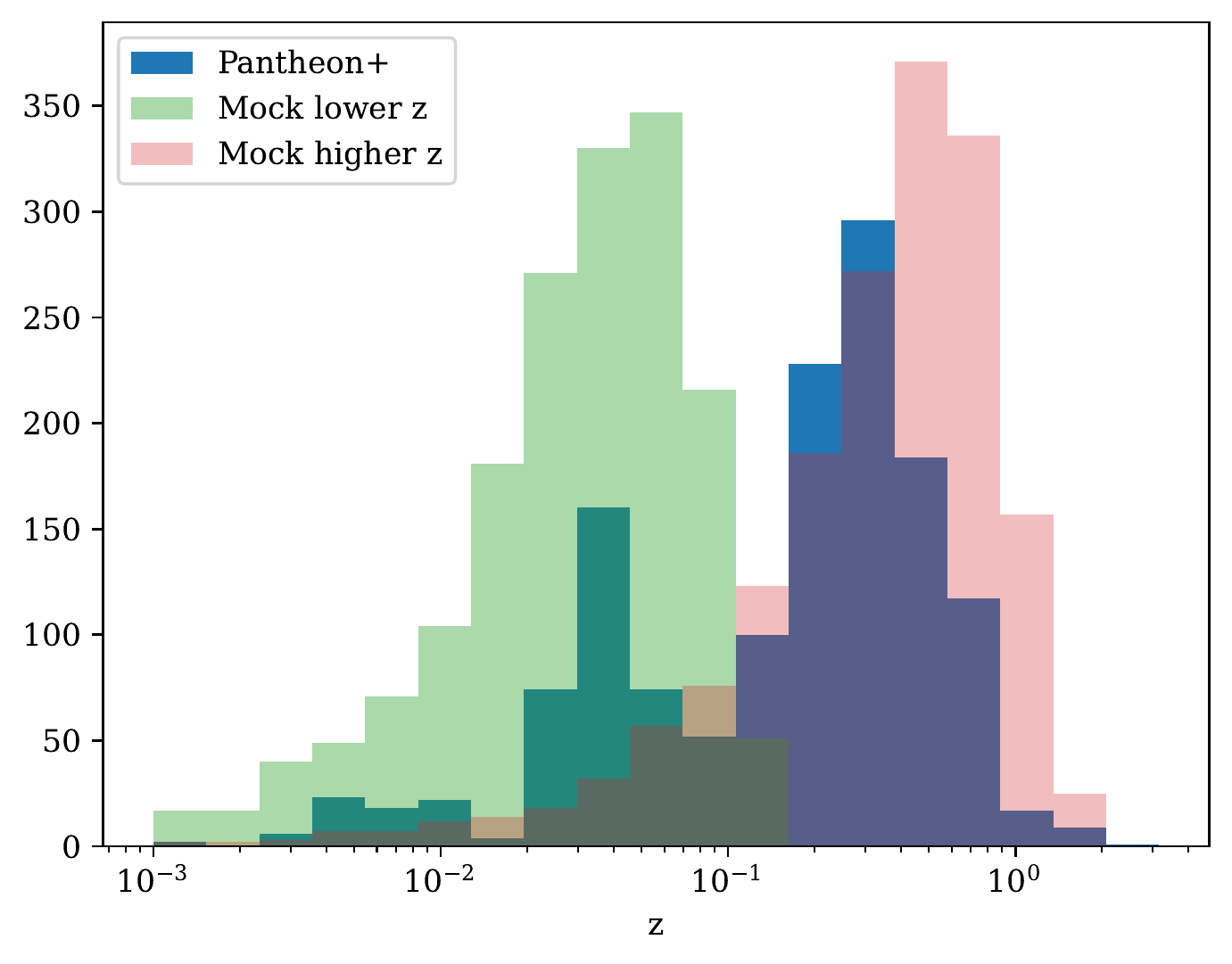}
			\caption{\small Redshift distributions of the Pantheon+ dataset and of the two mocks which have more data at \textbf{low} z, i.e. $z<0.1$ (orange histogram), or at \textbf{high} z, i.e. $z>0.1$ (green histogram). \label{f:pantheon_plus_mocks}} 
			\vspace{0.17cm}
		\end{figure}
	\end{minipage}~~
	\begin{minipage}{0.48\linewidth}
		~~ \vspace{0.27cm}\\
		\begin{table}[H]
			\centering
			\begin{tabular}{ c c c  }
				\toprule
				$z_{\rm cut}$ & Mock \textbf{low} z & Mock \textbf{high} z \\
				\midrule
				No cut & 1624 & 1624  \\
				0.005 & 1576 & 1623  \\  
				0.01 & 1459 & 1622 \\
				0.0175  & 1263 & 1612 \\
				0.025  & 1068 & 1608 \\
				0.0375  & 785 & 1592 \\ 
				0.05  & 548 & 1577 \\
				0.1  & 72 & 1479 \\
				\bottomrule
			\end{tabular}
			\caption{\label{t:mock}\small Number of supernova lightcurves for each mock data set changing $z_{\rm cut}$. In both cases, we did not consider supernovae hosted in Cepheids.} %, so that the total amount of lightcurves is 1624 instead of 1701. }
	\end{table}
\end{minipage}

Starting from the Pantheon+ data, we substitute the measured redshift  with the one from a random redshift distribution and the distance modulus with the one computed according to Eqs. \eqref{e:distance_modulus} and \eqref{e:ansatz_dipole}, assuming $\Omega_m=0.338$, $H_0=73.4$ and a known dipole of amplitude $v_0=400$km/s fixed at $ra=170$\textdegree and $dec=20$\textdegree.
The two mock redshift distributions are presented in Fig. \ref{f:pantheon_plus_mocks}. One of them has more data at \textbf{low} z, i.e. $z<0.1$, while in the other has more data at \textbf{high} z, i.e. $z>0.1$. 
We then applied our pipeline to these mock catalogues considering different $z_{\rm cut}$, as reported in Tab.~\ref{t:mock}.
For sake of simplicity, we used a diagonal covariance whose elements are taken from the diagonal of the covariance matrix of the main analysis and we also did not consider supernovae hosted in cepheids, so that we fixed $\delta M=0$. 

\begin{figure}[ht]
	\centering
	\includegraphics [scale=0.75]{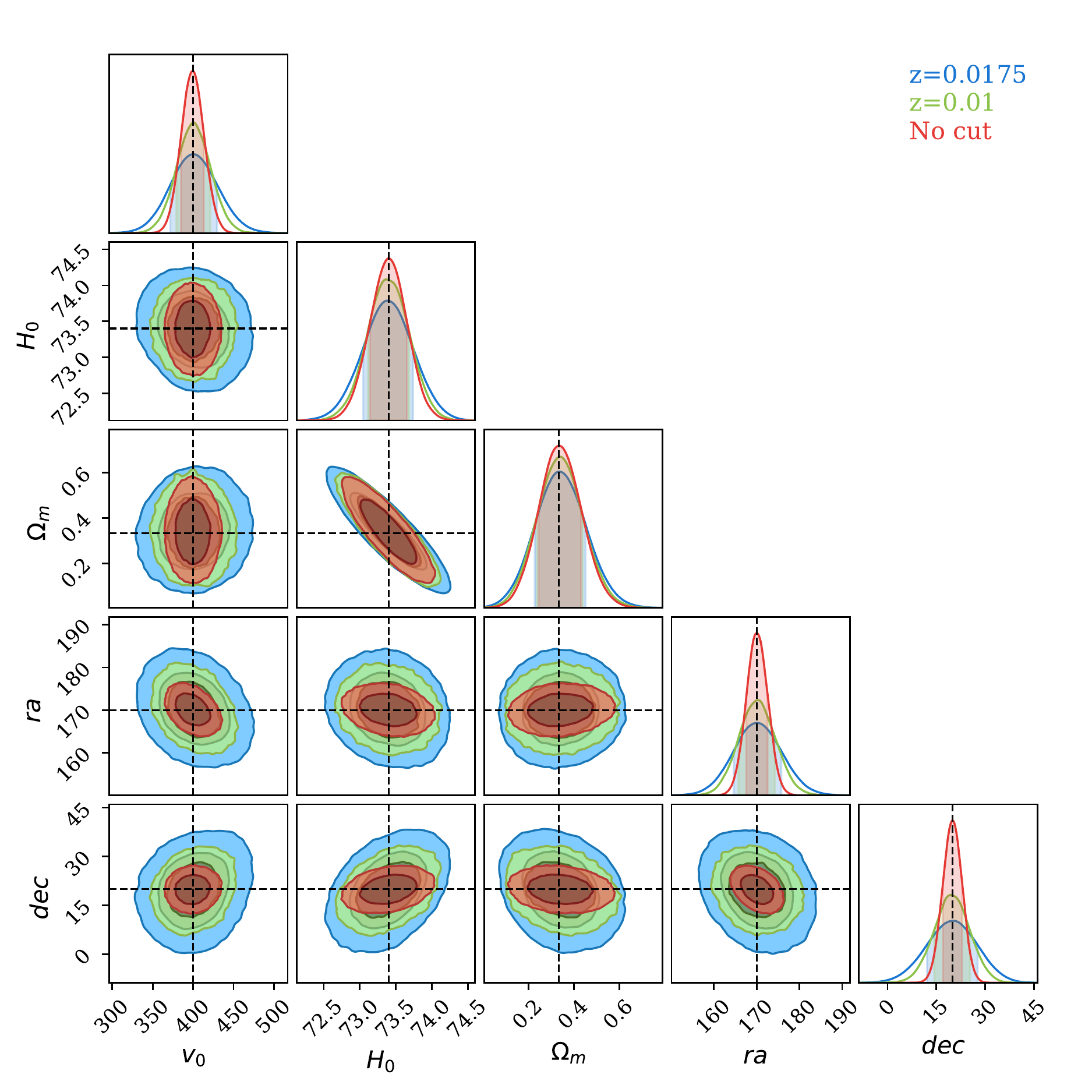}
	\caption{Contour plots for the mock data set at \textbf{lower} redshift. The dashed lines show the reference values $v_0=400$km/s, $\Omega_m=0.338$, $H_0=73.4$, $ra=170$\textdegree  and $dec=20$\textdegree .\label{pic:mock_low_z_175e2_1e2_no_filter}} 
	
\end{figure}

\begin{figure}[ht]
	\centering
	\includegraphics [scale=0.75]{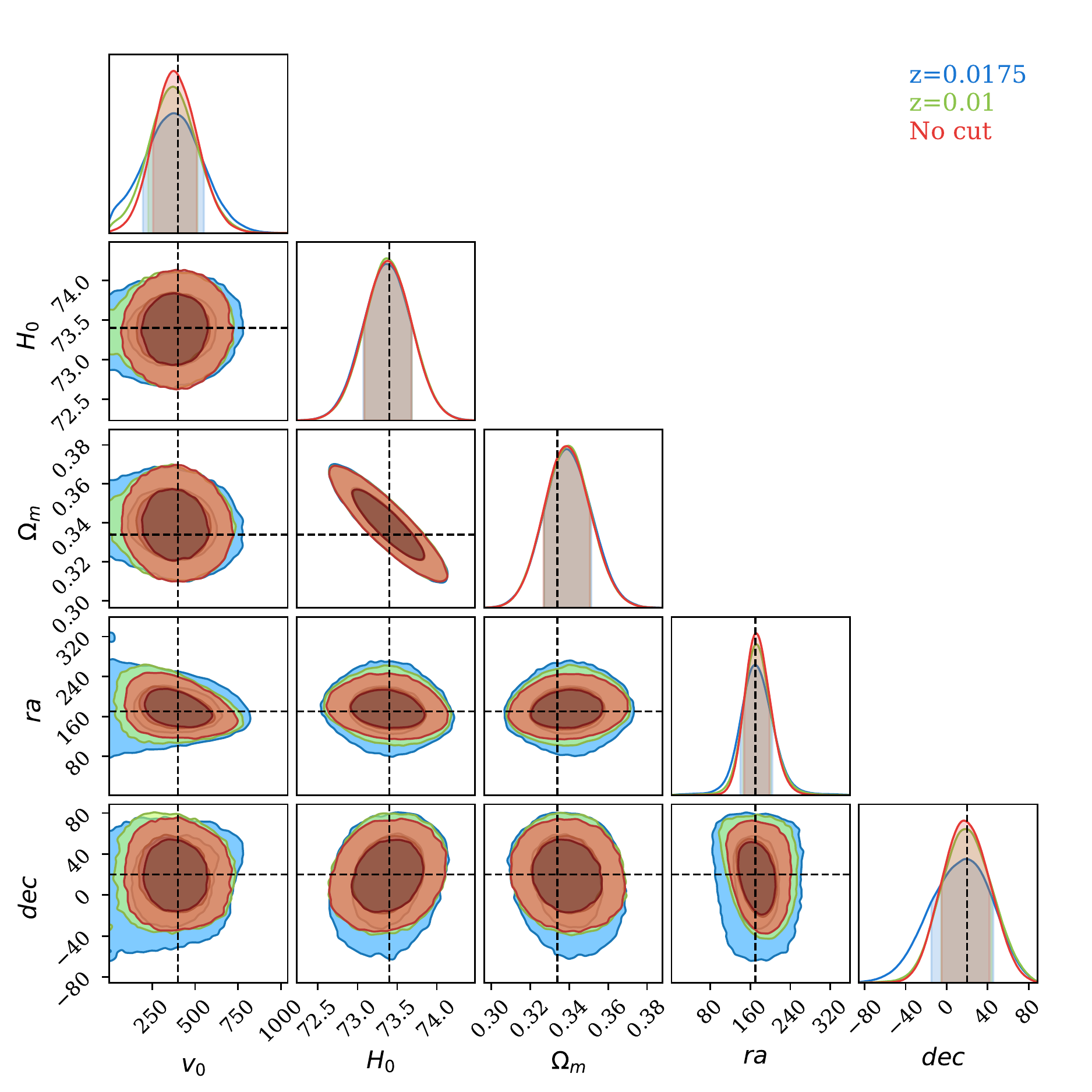}
	\caption{Contour plots for the mock data set at \textbf{higher} redshift. The dashed lines show the reference values $v_0=400km/s$, $\Omega_m=0.338$, $H_0=73.4$, $ra=170$\textdegree  and $dec=20$\textdegree .\label{pic:mock_high_z_175e2_1e2_no_filter}} 
\end{figure}

First of all, in both cases we  recover the expected values of the input parameters (this is another validity check of our pipeline). Moreover, we find that, the dipole is much better determined in the low-z sample, Fig.~\ref{pic:mock_low_z_175e2_1e2_no_filter} than in the high-z mock, Fig.~\ref{pic:mock_high_z_175e2_1e2_no_filter}. In both samples, increasing $z_{\rm cut}$ increases  the error of the dipole amplitude and direction while it has rather small effects on $\Omega_m$ and $H_0$.

More results of this study are reported in Figs.~\ref{pic:mock_extra_plots_low_z} and~\ref{pic:mock_extra_plots_high_z} shown in the Appendix.

\begin{figure}[ht]
	\centering
	\includegraphics [scale=0.75]{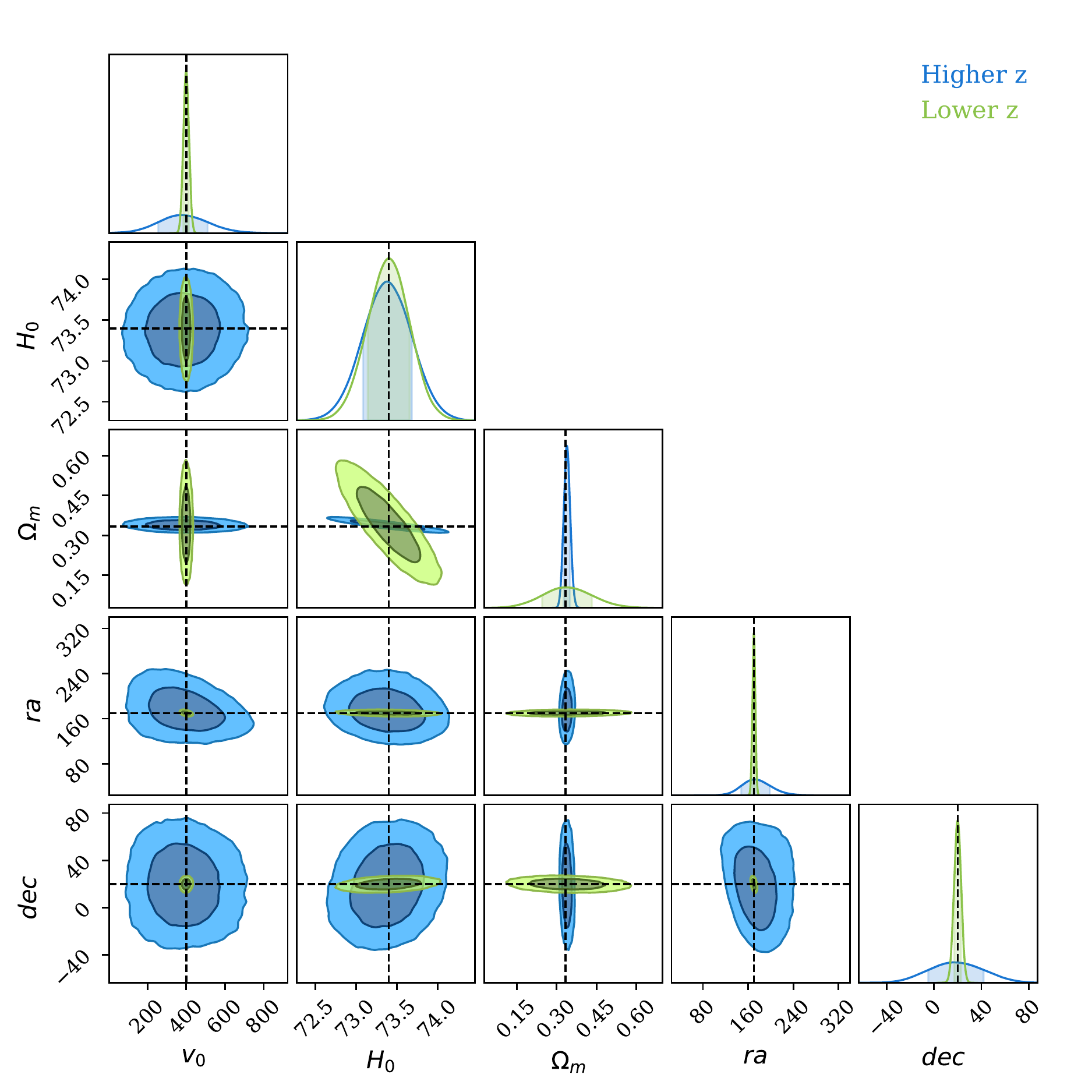}
	\caption{Contour plots for both the mock data set described in Sec.~\ref{sec:redshift_cut} without imposing any redshift cut. It is clear that lower redshift supernovae constrain the dipole while higher redshift ones constrain the cosmological parameter.\label{pic:mock_high_z_low_z_no_filter}} 
\end{figure}

\subsection{The peculiar velocities in the Pantheon+ analysis} \label{sec:peculiar_velocities}

An important difference between our treatment and  the Pantheon+ analysis~\cite{Brout:2022vxf}, lies  in the peculiar velocities of the SNe which we neglect in our analysis.  The main reason we do this is that they should not contribute significantly to the dipole which is the object of study in this work. We test this assumption here.

\begin{figure}[ht]
	\centering
	\includegraphics [scale=0.75]{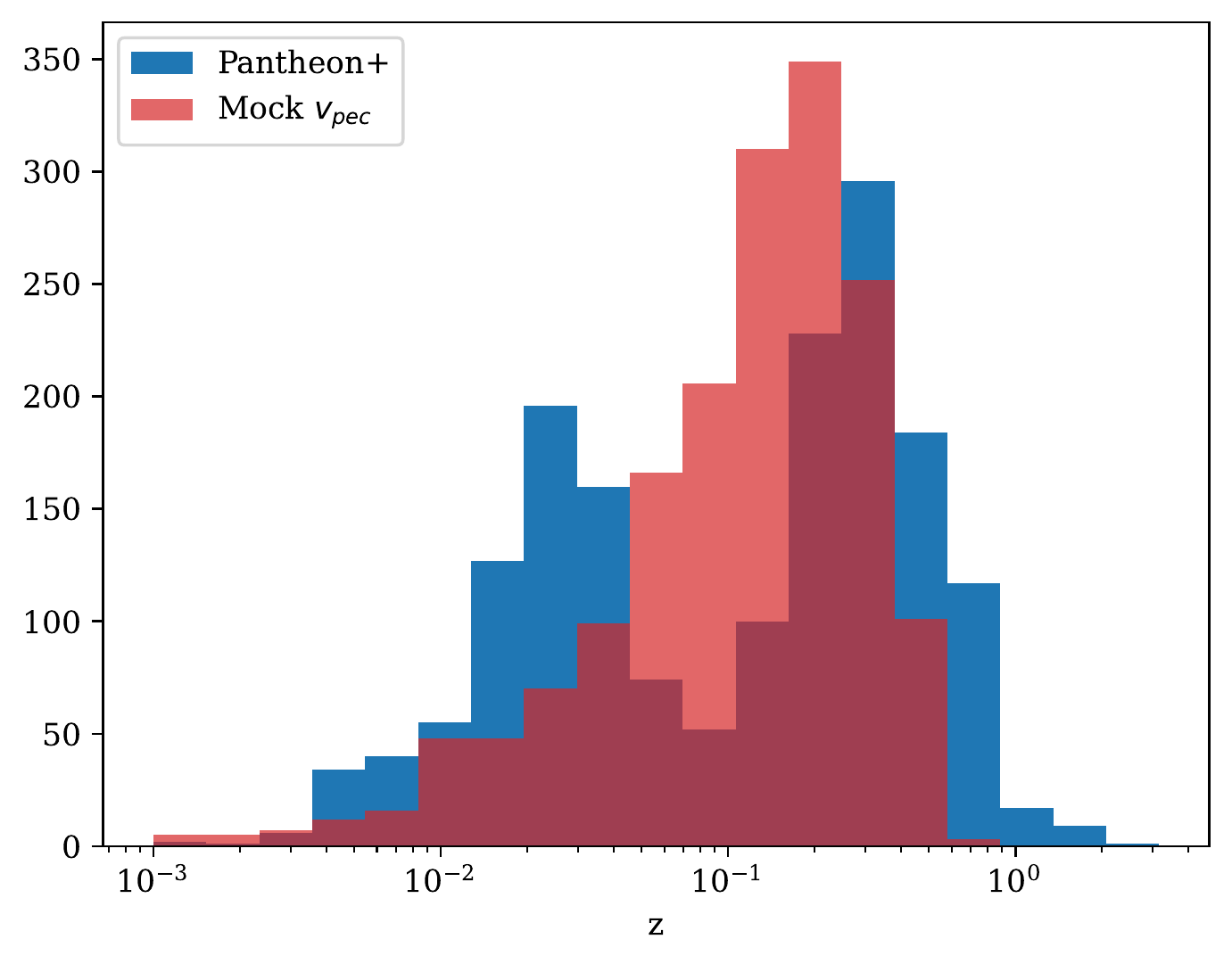}
	\caption{Redshift distributions of the Pantheon+ and mock datasets used for testing the effect of peculiar velocities in our analysis. \label{f:redshift_v_pec_distributions}} 
\end{figure}

We perform the same test as in Sec.~\ref{sec:redshift_cut}, assuming the same cosmology and covariance matrix and setting $\delta M =0$. The mock redshift distribution is shown in Fig.~\ref{f:redshift_v_pec_distributions} and the dipole, with amplitude $v_0=300$km/s, is fixed at $ra=270$\textdegree and $dec=-70$\textdegree. From this dipole, we subtract a radial peculiar velocity contribution $v_{\rm pec}$ so that Eq.~\eqref{e:ansatz_dipole} becomes:

\bea
D_L(z,\bn) &\simeq& \bar D_L(z)\left(1 + \frac{1}{\HH(z)r(z)}\left(\bv_0\cd\bn - v_{\rm pec} \right) \right)  \,. \label{e:dipole_vpec}
\eea

\begin{figure}[ht]
	\centering
	\includegraphics [scale=0.55]{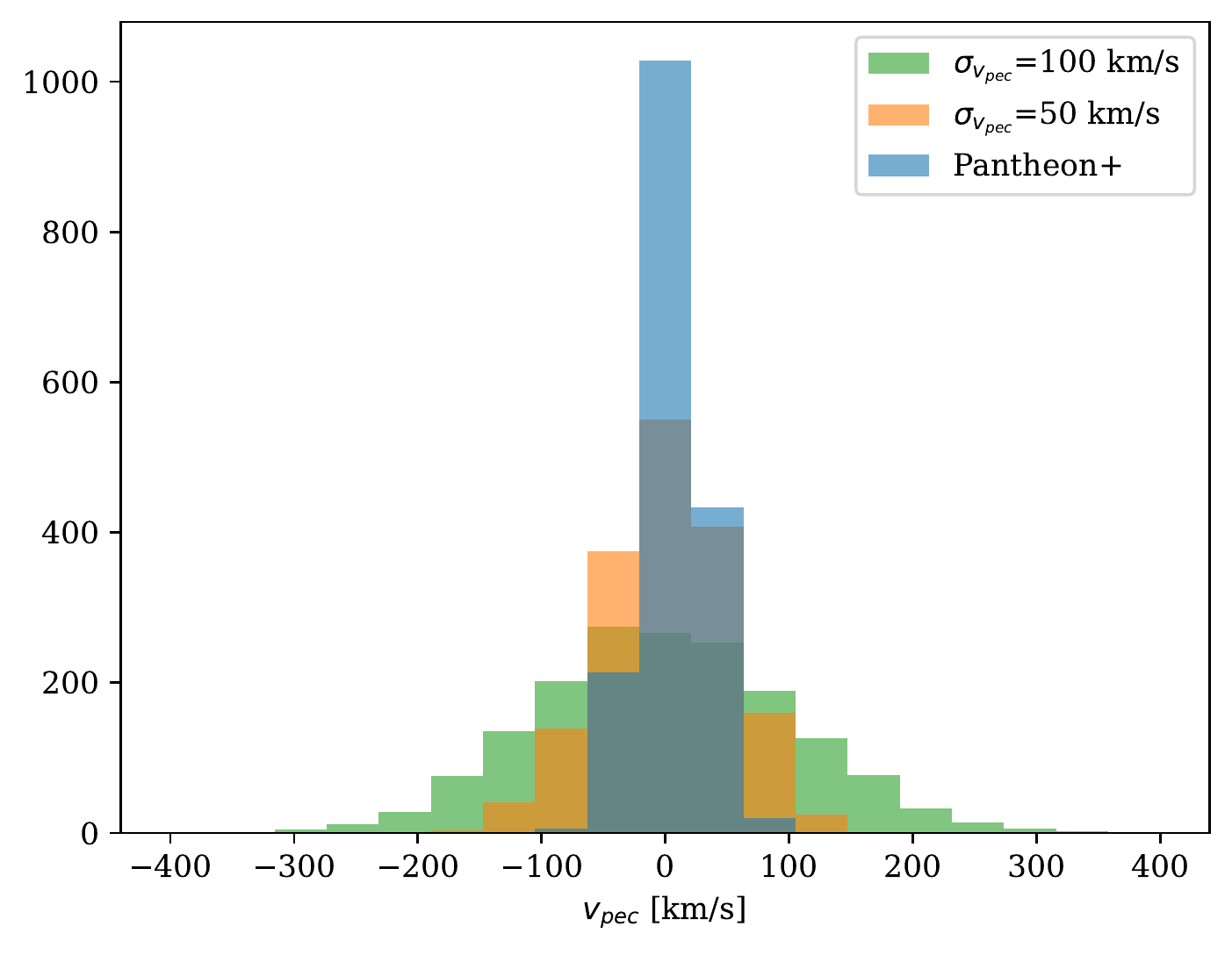}\includegraphics [scale=0.55]{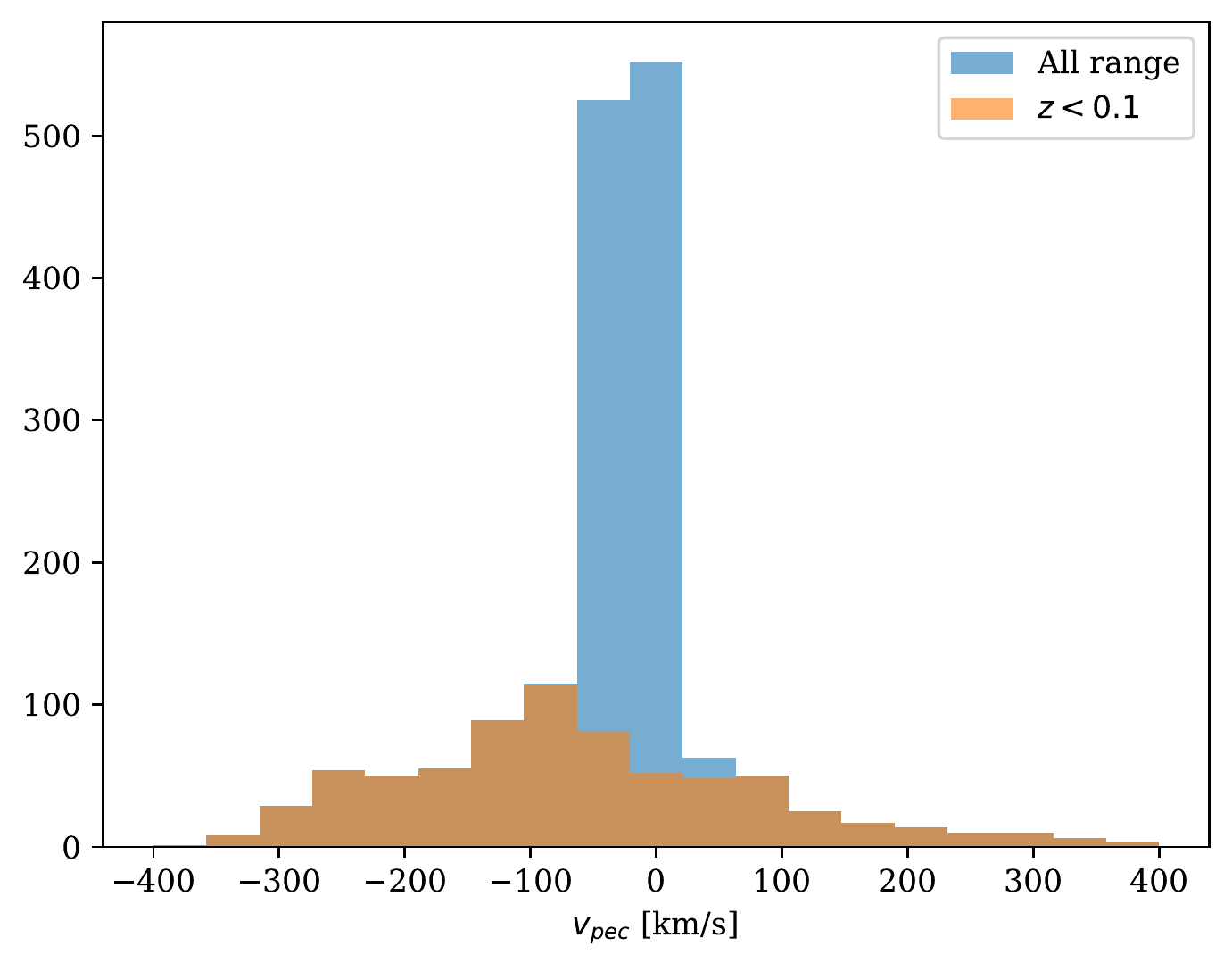}
	\caption{Left panel: Distributions of the radial peculiar velocities for different variance $\sigma_{v_{\rm pec}}$ together with the peculiar velocities values (light blue distribution) used by Pantheon+.\\
		Right panel: The peculiar velocities of the full Pantheon+ data set (blue) and  the ones of SNe with $z<0.1$ which are most relevant for the dipole. The Pantheon+ distribution is not symmetric: this is mainly due to to the presence of the bulk motion correction $\bv_{\rm bulk}\cd\bn$, with $|\bv_{\rm bulk}|= 182\,$km/s and (ra, dec) = (191\textdegree ,-61\textdegree). \label{f:v_pec_distributions}}
\end{figure}

\begin{figure}[ht]
	\centering
	\includegraphics [scale=0.75]{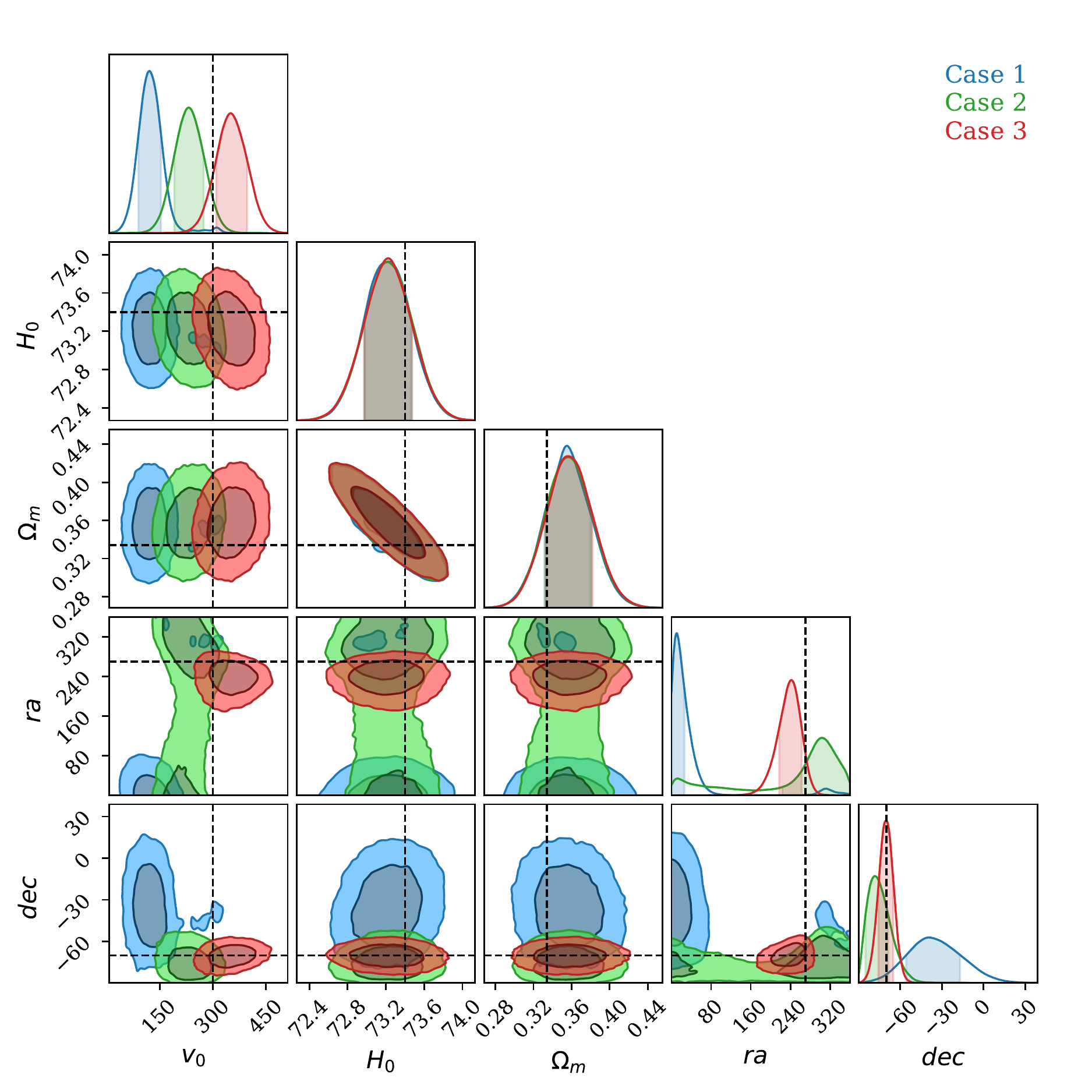}
	\caption{Contour plots for the mock data sets described in Sec.~\ref{sec:peculiar_velocities}. In \textit{Case 1} (blue contours) we have directly subtracted in Eq.\ \eqref{e:dipole_vpec} the peculiar velocity values provided by Pantheon+. In \textit{Case 2} (green contours), we subtracted the bulk motion correction, with $|\bv_{\rm bulk}|= 182\,$km/s and (ra, dec) = (191\textdegree ,-61\textdegree), according to~\cite{Carr:2021lcj}. In \textit{Case 3} (red contours), we assume for the bulk motion corrections the inferred values reported in Table~\ref{table_bulk_flow} without applying any redshift cut, i.e. $|\bv_{\rm bulk}|= 326\,$km/s and (ra, dec) = (205.4\textdegree ,-54\textdegree).The dashed lines show the reference values $v_0=300\,$km/s, $\Omega_m=0.338$, $H_0=73.4$, $ra=270$\textdegree  and $dec=-70$\textdegree.\label{f:last_test}}
\end{figure}

In Fig.~\ref{f:v_pec_distributions} we show the distribution of $v_{\rm pec}$ values applied in our test. We consider two different Gaussian distributions with vanishing mean and increasing variance $\sigma_{v_{\rm pec}}$. In the same figure we have also reported the $v_{\rm pec}$ values provided and used by the Pantheon+ collaboration and used in their analysis~\cite{Brout:2022vxf}. As we can see, the contributions we add to our mock datasets are significantly larger than the ones applied in the original Pantheon+ analysis and the corrections to the luminosity distance are correspondingly larger. As a consequence, if peculiar velocities are not  negligible,  we should see a deviation from the expected values of the dipole when applying our MCMC pipeline. 

\begin{figure}[!ht]
	\centering
	\includegraphics [scale=0.75]{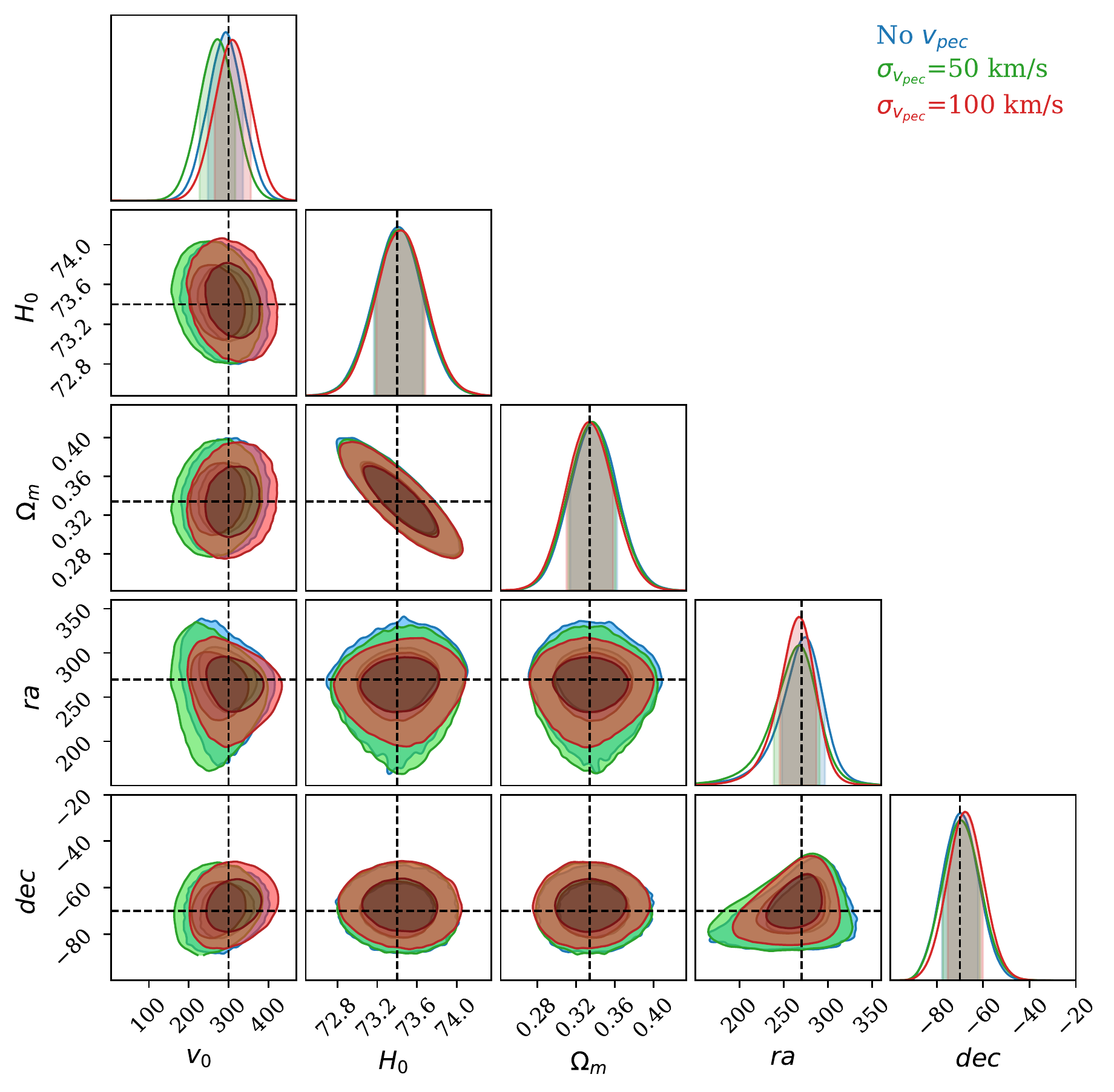} 
	\caption{\label{pic:mock_peculiar_velocities_50_100}Contour plots for the mock data sets described in Sec.~\ref{sec:peculiar_velocities}. The blue contours are obtained for the dataset without any peculiar velocity contributions, while the others are obtained applying the $v_{\rm pec}$ corrections from the distributions in Fig.~\ref{f:v_pec_distributions} with increasing variance $\sigma_{v_{\rm pec}}$. The dashed lines show the reference values $v_0=300\,$km/s, $\Omega_m=0.338$, $H_0=73.4$, $ra=270$\textdegree  and $dec=-70$\textdegree .}
\end{figure}

This is not the case. In fact, Fig~\ref{pic:mock_peculiar_velocities_50_100} shows that the shift is very small even considering peculiar velocity corrections up to five times larger than the maximum correction applied in the original Pantheon+ analysis as it happens in the case of the distribution with $\sigma_{v_{\rm pec}}=100$km/s. We have also studied $\sigma_{v_{\rm pec}}=200$km/s and have seen no significant shift in the inferred dipole. We can therefore conclude that neglecting Gaussian peculiar velocities with vanishing mean is justified and does not affect the inferred dipole. However, as we see clearly in Fig.~\ref{f:v_pec_distributions}, the peculiar velocities added to the Pantheon+ data do not have vanishing mean and hence will affect the dipole.

In Fig.~\ref{f:last_test} we show results from a study of the Pantheon+ peculiar velocities. We have added  to our mock dataset the peculiar velocities provided by Pantheon+. In \textit{Case 1} (blue contours of Fig.~\ref{f:last_test}) we have directly subtracted in Eq.\ \eqref{e:dipole_vpec} the peculiar velocity values provided by Pantheon+. As we see, doing so modifies the resulting dipole. This  is actually not surprising since the peculiar velocity correction of Pantheon+ is not simply a Gaussian distribution with vanishing mean, but includes a bulk motion i.e., a common velocity of all low redshift SNe, as we can already infer by visual inspection of Fig.~\ref{f:v_pec_distributions} where the peculiar velocity distribution is not symmetric. In order to recover the expected values, we rewrite Eq.~\eqref{e:dipole_vpec} as:
\bea
D_L(z,\bn) &\simeq& \bar D_L(z)\left(1 + \frac{1}{\HH(z)r(z)}\left(\bv_0\cd\bn - (v_{\rm pec} - \bv_{\rm bulk}\cd\bn ) \right) \right)  \,. \label{e:dipole_vpec_vbulk}
\eea
where we subtract the  bulk motion correction $\bv_{\rm bulk}\cd\bn$ from the peculiar velocdity. 

In \textit{Case 2} (green contours of Fig.~\ref{f:last_test}), we subtract the bulk motion of Pantheon+ given by  $|\bv_{\rm bulk}|= 182\,$ km/s and (ra, dec) = (191\textdegree ,-61\textdegree), according to~\cite{Carr:2021lcj}. In this case, we manage to partially recover the expected values with some minor deviations. The situation improves for \textit{Case 3} (red contours), in which we assume for the bulk motion corrections the inferred values reported in Table~\ref{table_bulk_flow} without applying any redshift cut, i.e. $|\bv_{\rm bulk}|= 326\,$km/s and (ra, dec) = (205.4\textdegree ,-54\textdegree). In this case we manage to recover the expected values within $1 \sigma$. This can be regarded as another validity check of our main analysis.

Finally, we also consider it problematic that the peculiar velocities inferred in~\cite{Brout:2022vxf} come purely from linear gravitational infall~\cite{Carr:2021lcj,Peterson:2021hel}, even though it is known from numerical simulations that at late times vorticity is as relevant as (if not larger than) the gradient flow~\cite{Jelic-Cizmek:2018gdp}. Therefore, it is not clear how trustworthy these peculiar velocities are and whether one should rather enhance the error bars in the value of $H_0$. The most important contribution to the dipole, however is the bulk flow which seems to be modeled in a rather ad hoc manner. It is assumed that all SNe out to $z=0.067$, corresponding to a ball of radius $R=200h^{-1}$Mpc
move as one with a mean velocity of $182$\,km/s.

\begin{figure}[!ht]
	\centering
	\includegraphics [scale=0.75]{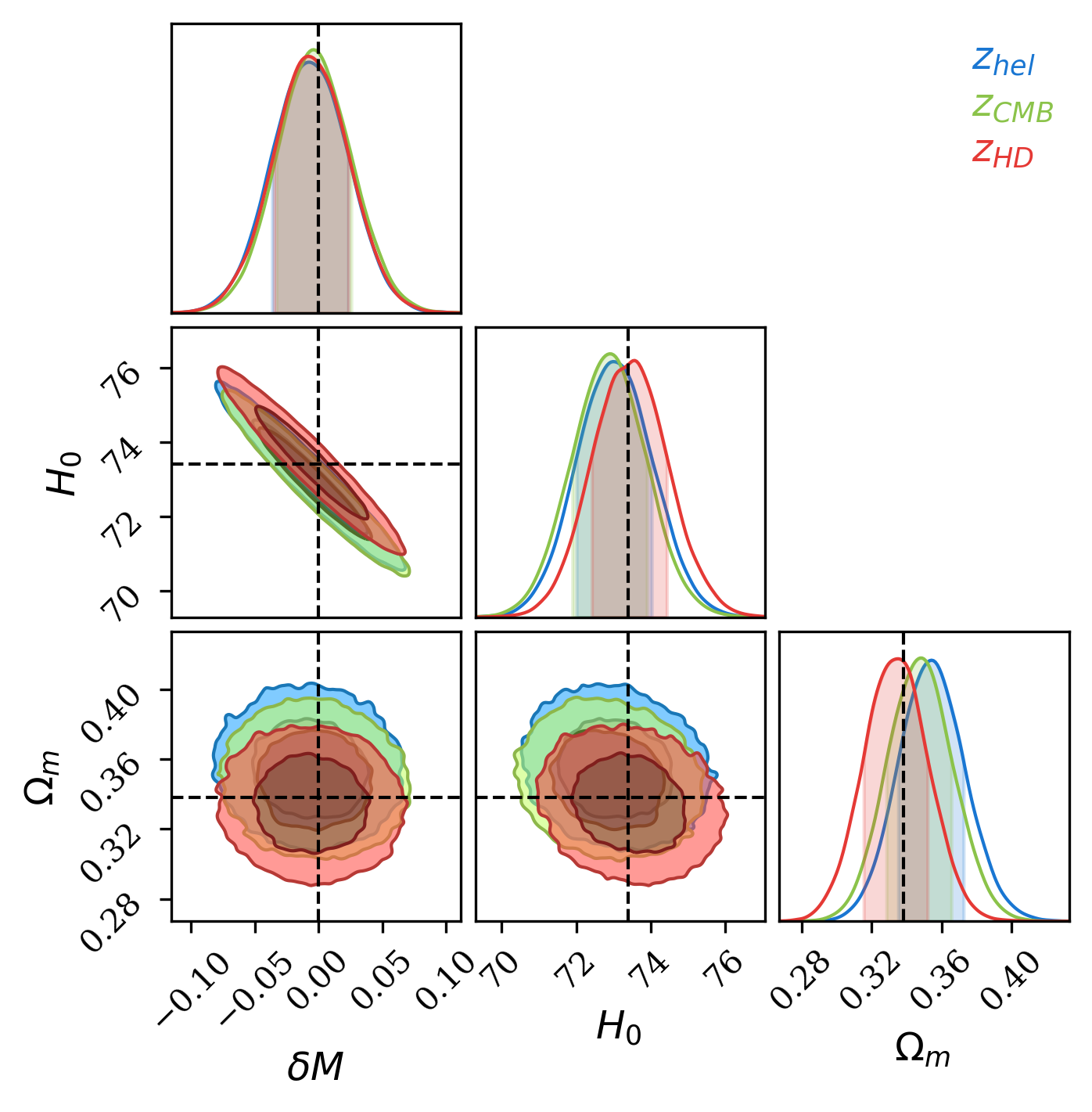}
	\caption{Contour plots for the different redshift values provided by the Pantheon+ catalogue (no lower cut in redshift). We can see there is a $\sim 0.65 \sigma$ discrepancy in $\Omega_m$ between $z_{\rm HD}$ and $z_{CMB}.$ \label{f:pancomp}}
\end{figure}

Fortunately, the effects of these redshift corrections on cosmological parameters are all within one standard deviation. Nevertheless, using CMB or heliocentric redshift, the inferred matter density increases by about $0.67\si$ with respect to the value inferred using the peculiar velocity corrected redshift $z_{\rm HD}$, see Fig.~\ref{f:pancomp}. 

\section{Conclusions}

In this work we have determined the dipole of the Pantheon+ data. We found that without imposing a cut in the redshift, this dipole is significantly different from the CMB dipole yielding a chi-square difference of about 66.
Also a reasonable lower cutoff, $z_{\rm cut} =0.01$, corresponding to 30Mpc does not substantially affect this result, the best-fit dipole still has a chi-square difference of  41 with respect to the Planck dipole. Nevertheless, the Planck dipole is always a better fit than no dipole at all, see Table~\ref{table_chisq}.
Increasing the cut to  $z_{\rm cut} =0.05$ or more, the dipole essentially disappears and is no longer clearly detectable in the data. 

The sharp decrease in the detectability of the dipole at $z\gtrsim 0.05$ is not very surprising. At $z=0.01$ we have $1/\HH(z)r(z) \simeq 1/z = 100$ so that $v_0/\HH(z)r(z) \simeq 0.1$ while at $z=0.05$ this correction is about five times smaller. This is of the same level as the white noise amplitude of the full data set, therefore with this redshift cut the dipole cannot be detected at high significance. 

One might doubt our first order approximation for the dipole in the luminosity distance used in Eq.~\eqref{e:ansatz_dipole}. However, in Ref.~\cite{Horstmann:2021jjg} the Taylor expansion of the velocity distance around the background redshift is not performed and the results obtained there are equivalent to ours. Therefore, this approximation is sufficient to determine the dipole. We have also tested this by modeling the redshift as $z=\bar z - \bv_0\cd\bn$ and obtained the same results as with our first order analysis.

While the Pantheon data still are marginally consistent with the CMB dipole, the Pantheon+ data are no longer in agreement. This might be due to the fact that the nearby supernovae $0.01\leq z\leq 0.02$ which contribute most to this dipole, have velocities which are still too significantly correlated with our own motion and cannot reliably determine our motion with respect to the cosmic restframe. In the Pantheon+ analysis this problem has been addressed by adding a bulk flow to the low redshift supernovae. Such a bulk flow of course also contributes to the dipole. As explained in Section~\ref{s:theo}, the dipole truly measures $\bv_0 =\bv_{\rm obs}-\bv_{\rm bulk}$. However, assuming an ad hoc bulk flow does not seem better motivated than simply inferring the best fit dipole as done in this work. Assuming that the CMB dipole truly measures the observer velocity, the bulk flow is then simply $\bv_{\rm bulk}=\bv_{\rm Planck}-\bv_0$. According to Table~\ref{table_bulk_flow}, our analysis yields $|\bv_{\rm bulk}|\simeq 350\,$km/s. This is the coherent velocity of a ball of radius $3000z_{\rm cut}h^{-1}$Mpc $\simeq 100h^{-1}$Mpc. This is significantly larger than the bulk flow used in the Pantheon+ analysis which is $|\bv_{\rm bulk}|= 182\,$km/s. Interestingly, however, the directions of these bulk flows agree very well.

To test the hypothesis that there is this significant bulk flow, it will be important to study also higher multipoles in the supernovae data, to which a bulk flow should in principle also contribute. We plan to investigate this in a future project.

\section*{Acknowledgement}
We thank Nick Horstmann for helpful discussions and for providing us with the corrected Pantheon data. We thank the referee for a careful review which helped us to improve this work. The computations were performed at University of Geneva on Yggdrasil HPC cluster. This work is supported by the Swiss National Science Foundation.
\vspace{2cm}

%\newpage

\section*{Appendix}
\subsection*{Pantheon+ redshift dependence }
\begin{figure}[!ht]
	
	\begin{subfigure}{.49\textwidth}
		
		\centering
		\includegraphics [scale=0.4]{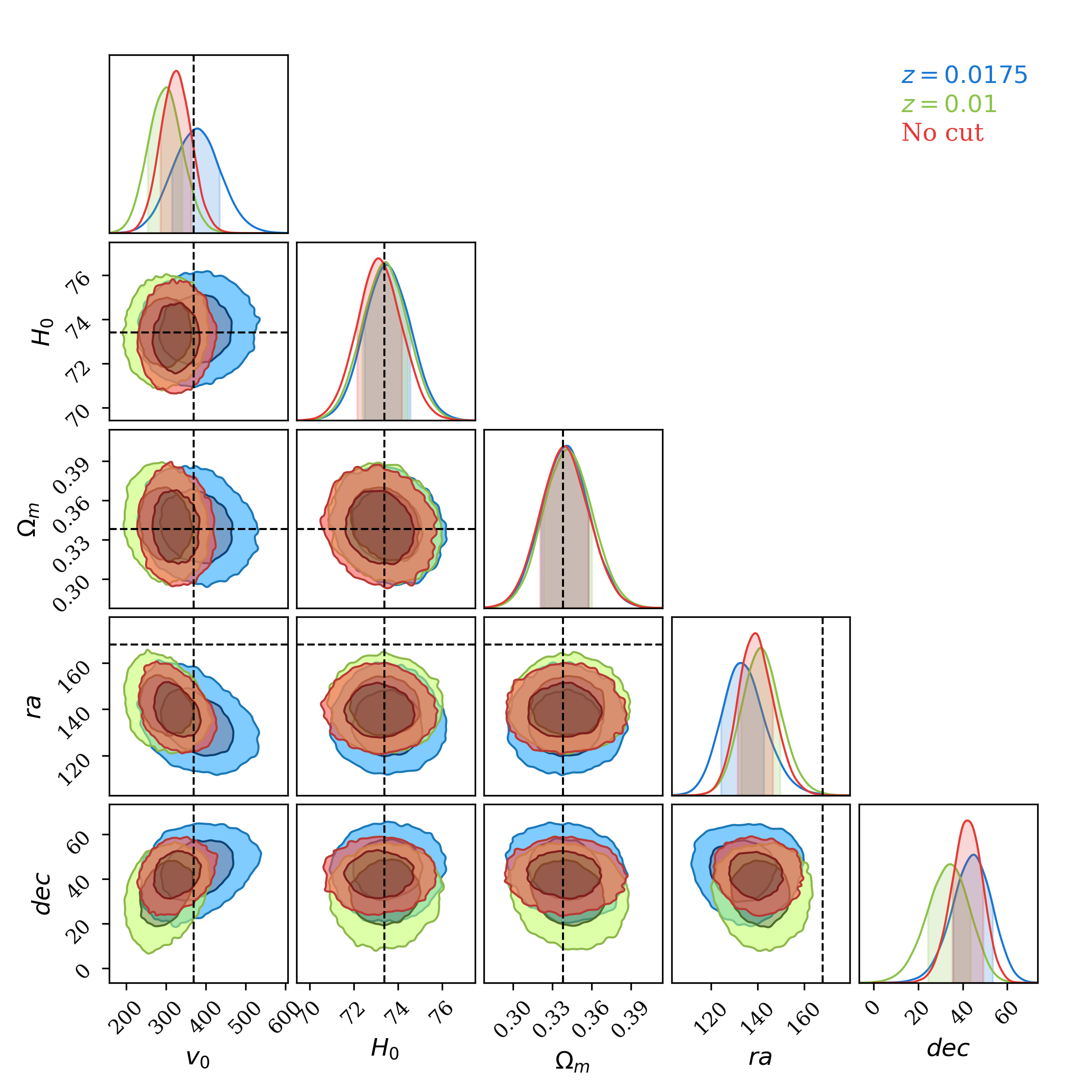}
		\caption{Low $z_{\rm cut}$.} 
		
	\end{subfigure}
	\begin{subfigure}{.49\textwidth}
		\centering
		\includegraphics [scale=0.4]{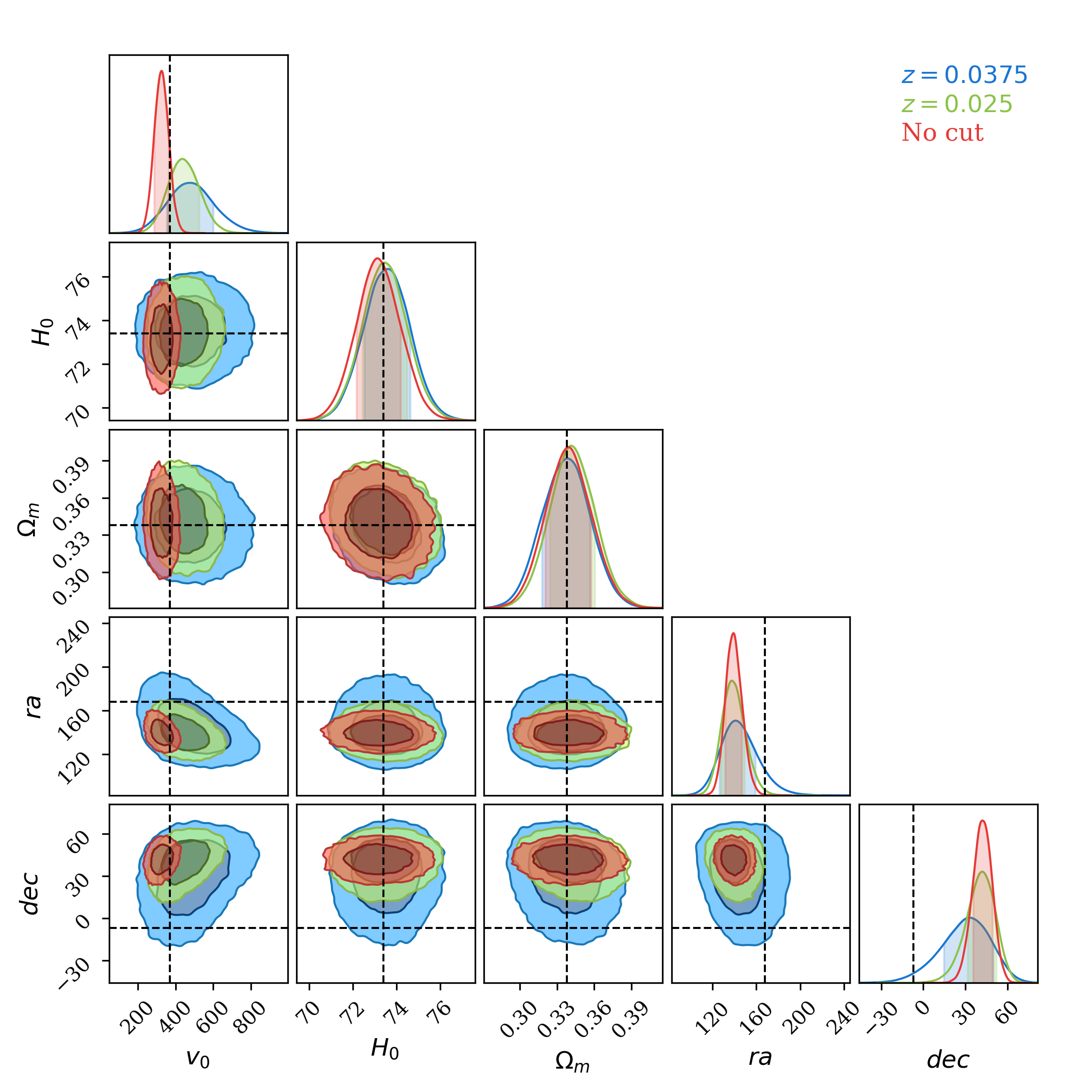}
		\caption{Medium $z_{\rm cut}$.} 
		
	\end{subfigure}
	\\
	\begin{subfigure}{.49\textwidth}
		\centering
		\includegraphics [scale=0.4]{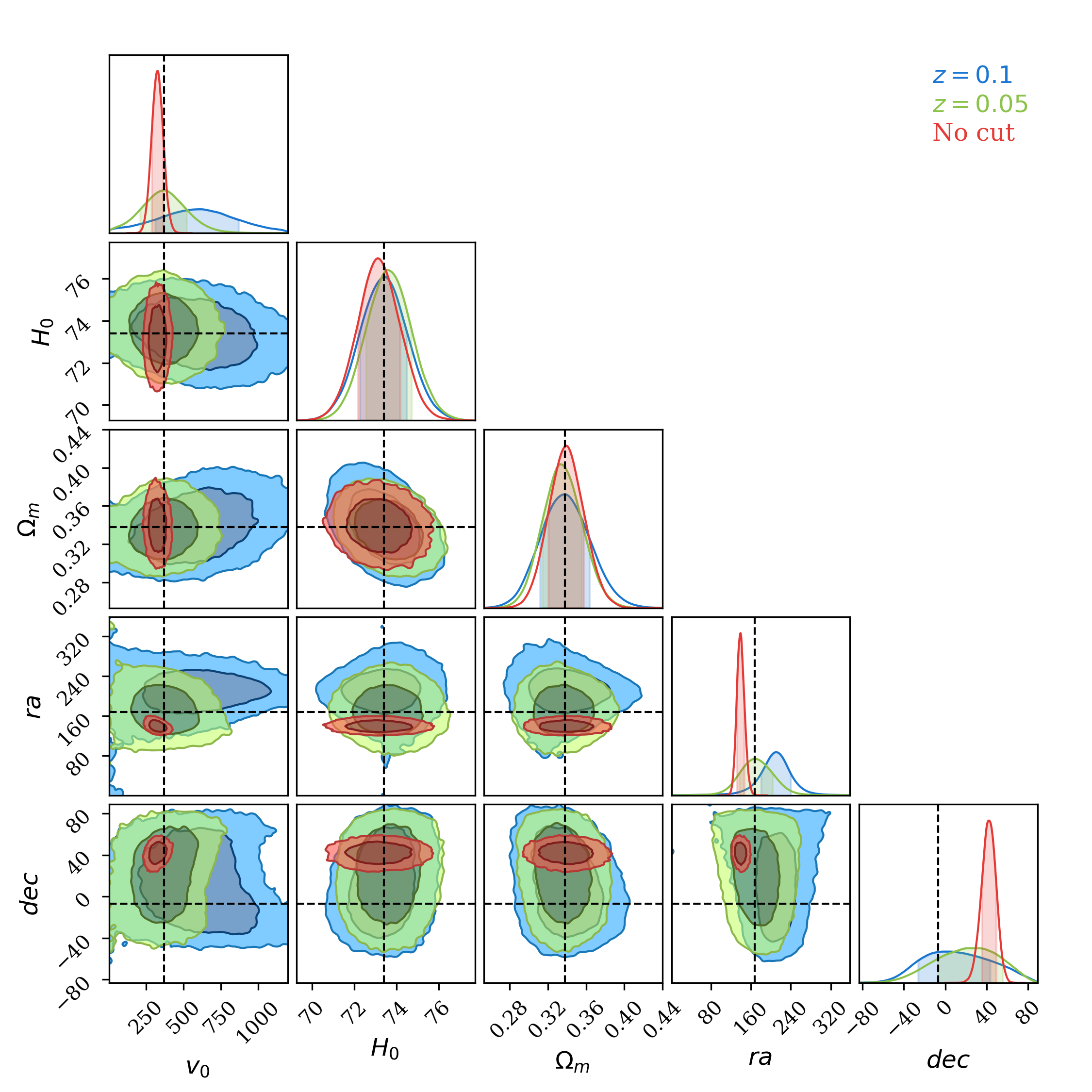}
		\caption{High $z_{\rm cut}$.} 
	\end{subfigure}
	\caption{\label{pic_marginalization_more} The results of our MCMC fitting procedures for the redshift cuts not shown in Fig.~\ref{pic_marginalized1}. To reduce the size of the figure we do not plot $\de M$ which is also marginalized over.}
	
\end{figure}

\begin{figure}[!ht]
	
	\begin{subfigure}{.5\textwidth}
		
		\centering
		\includegraphics [scale=0.45]{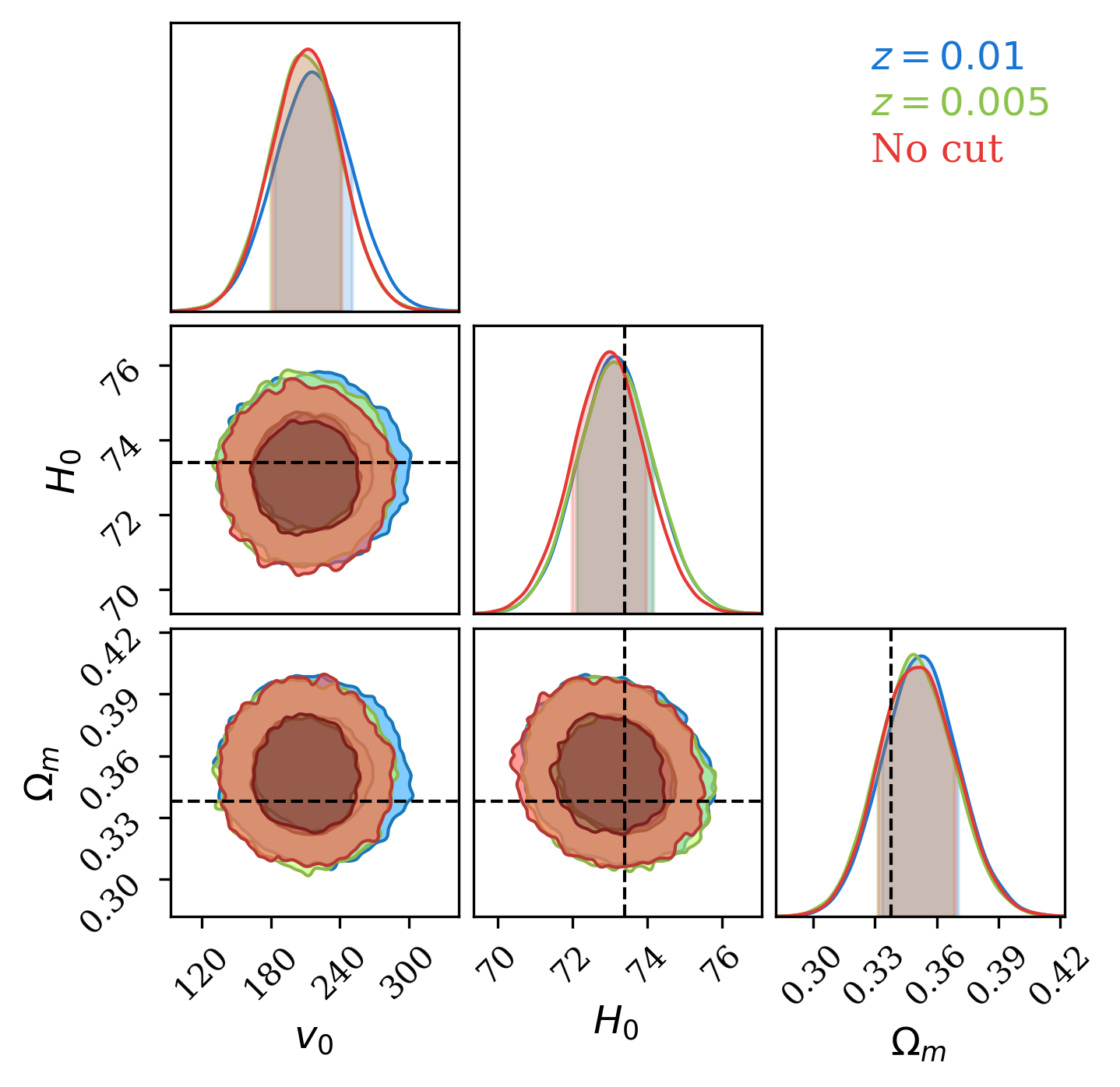}
		\caption{Lowest $z_{\rm cut}$.} 
	\end{subfigure}
	\begin{subfigure}{.5\textwidth}    
		\centering
		\includegraphics [scale=0.45]{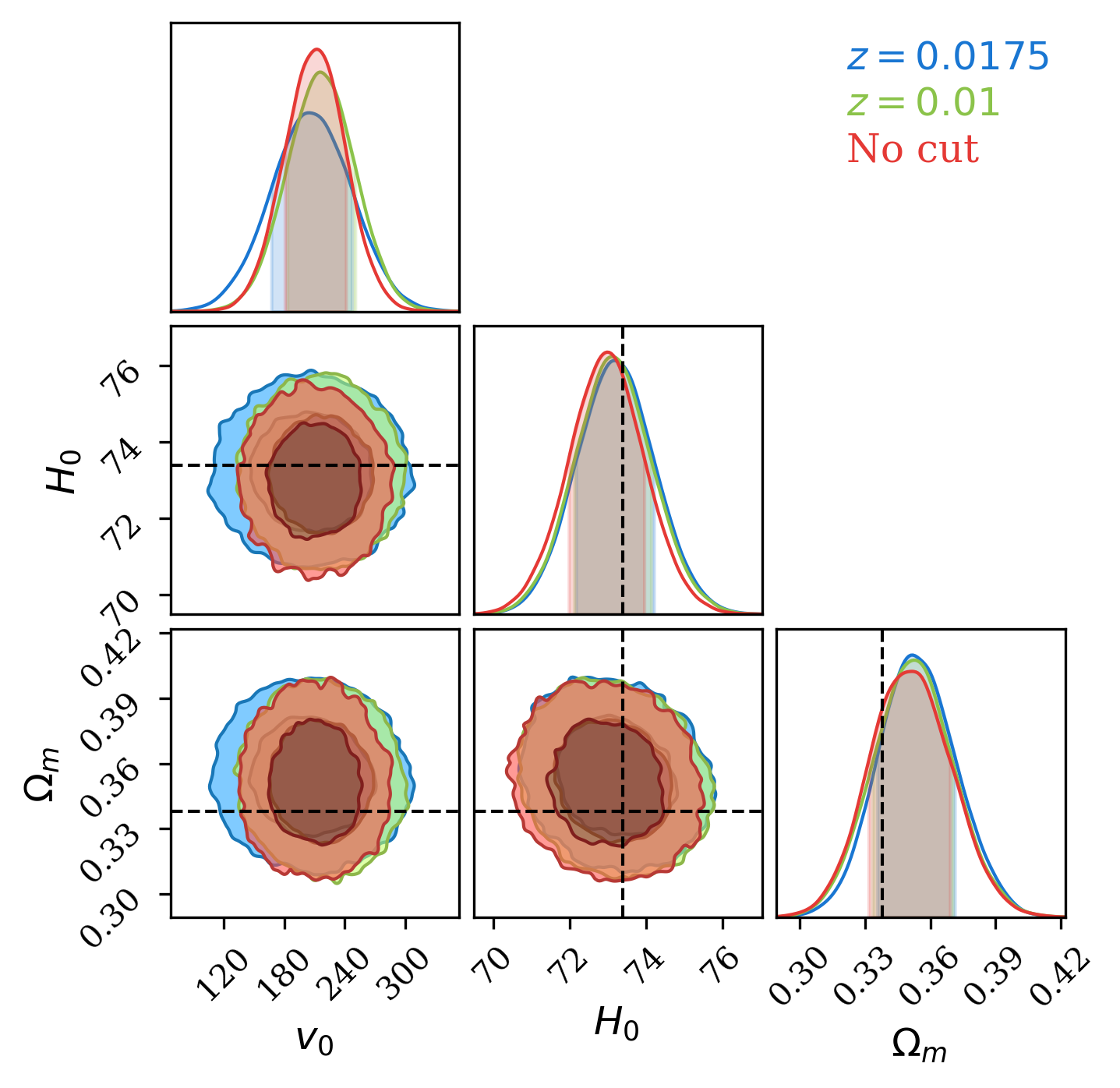}
		\caption{Low $z_{\rm cut}$.} 
		
	\end{subfigure}
	
	\begin{subfigure}{.5\textwidth}
		\centering
		\includegraphics [scale=0.45]{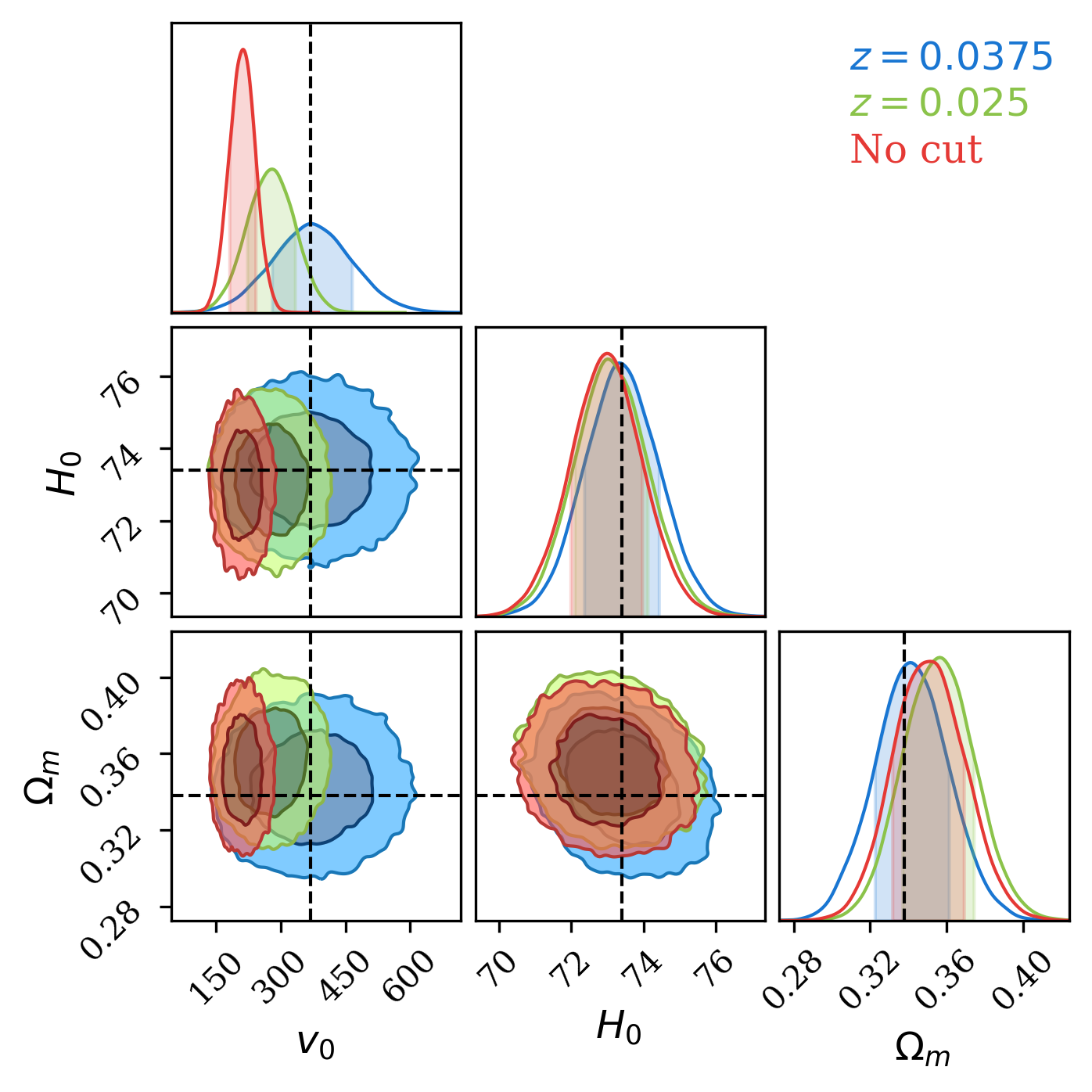}
		\caption{Medium $z_{\rm cut}$.} 
		
	\end{subfigure}
	\begin{subfigure}{.5\textwidth}
		\centering
		\includegraphics [scale=0.45]{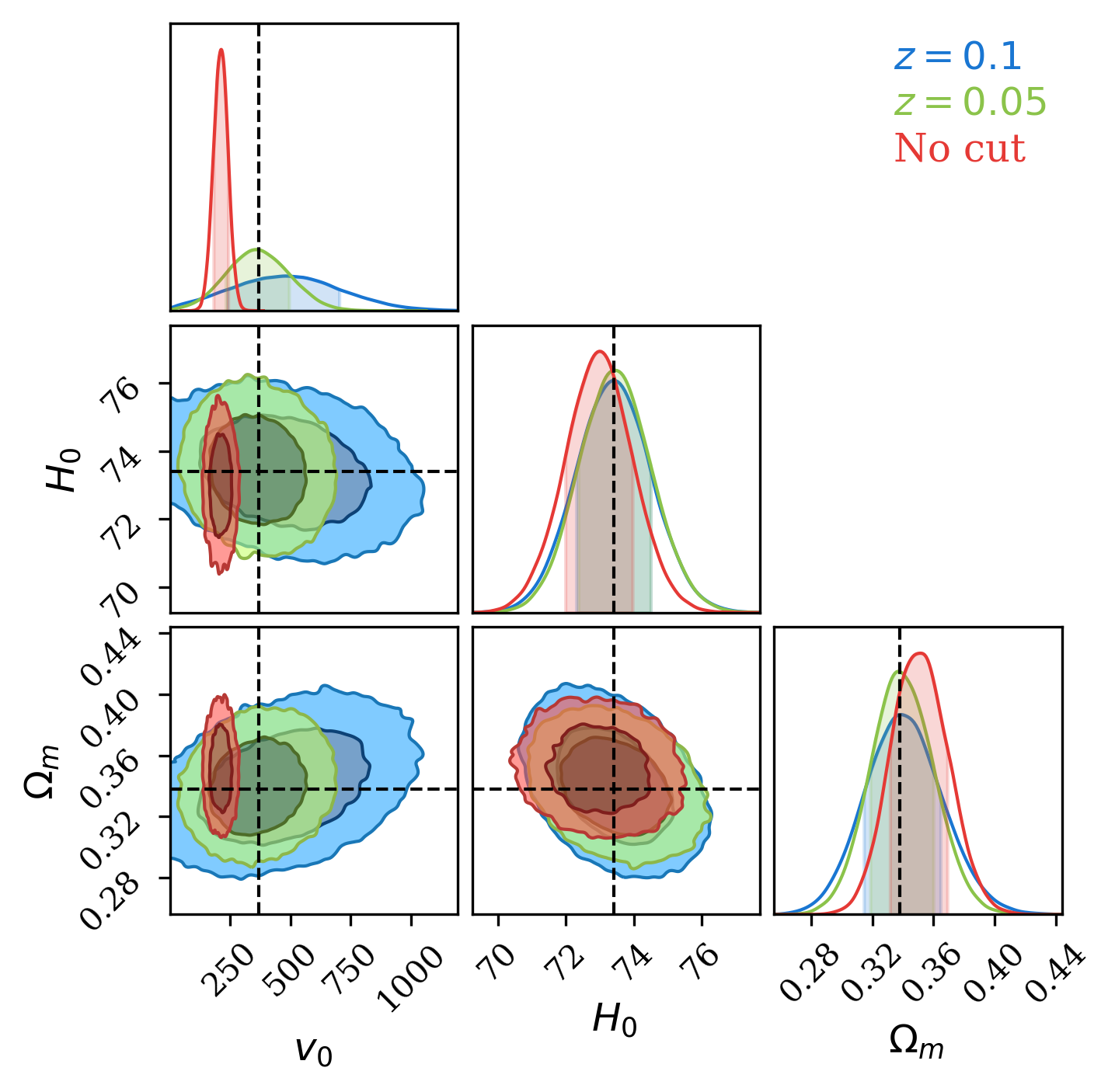}
		\caption{High $z_{\rm cut}$.} 
	\end{subfigure}
	\caption{\label{pic_vfixed_more} 
		Contour plots for the Pantheon+ data set with different cuts in the redshift of the Supernovae for an analysis with the direction of the dipole fixed to the one found by Planck.}
	
\end{figure}

In this appendix we also show the contour plots for different redshift cuts in Fig.~\ref{pic_marginalization_more}. Interestingly, for $z_{\rm cut}\geq 0.05$ the dipole is no longer detected with high significance. The  95\% confidence contour for the velocity includes $v_0=0$. In Fig.~\ref{pic_vfixed_more} we also show the corresponding plots for velocity direction fixed to the Planck value. At higher redshift cuts, $z_{\rm cut}\geq 0.0375$, the best fit velocity
approaches the Planck value, but the errors become rather large.

\subsection*{Comparison with the Dipole of Pantheon - Plots}
\begin{figure}[!ht]
	\begin{subfigure}{.5\textwidth}
		\centering
		\includegraphics[width=0.75\linewidth]{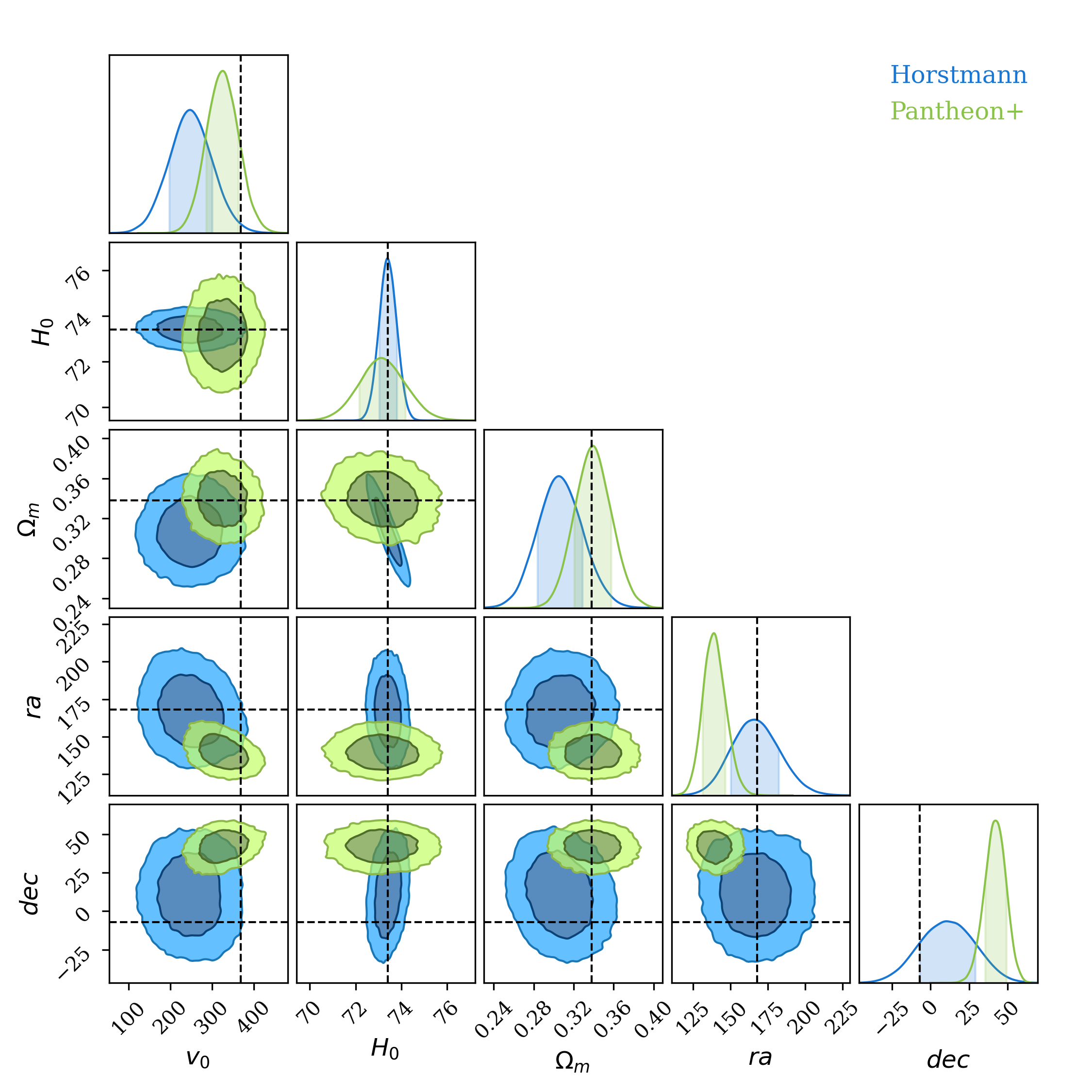}  
		\caption{No cut}
		\label{fig:sub-comparison-no}
	\end{subfigure}
	\begin{subfigure}{.5\textwidth}
		\centering
		\includegraphics[width=0.75\linewidth]{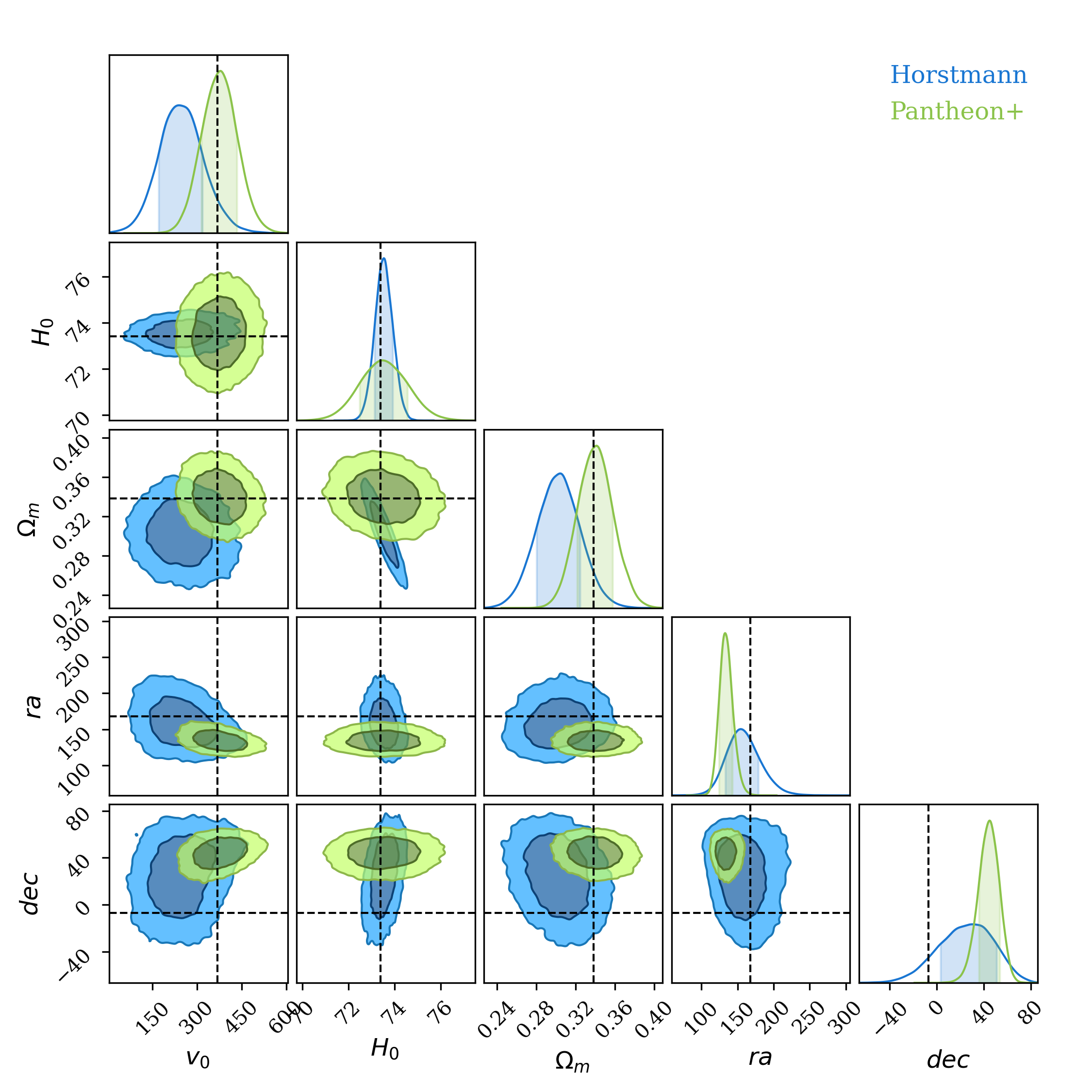} 
		\caption{$z_{\rm cut}$=0.0175}
		\label{fig:sub-comparison-0.0175}
	\end{subfigure}
	\\
	\begin{subfigure}{.5\textwidth}
		\centering
		\includegraphics[width=0.75\linewidth]{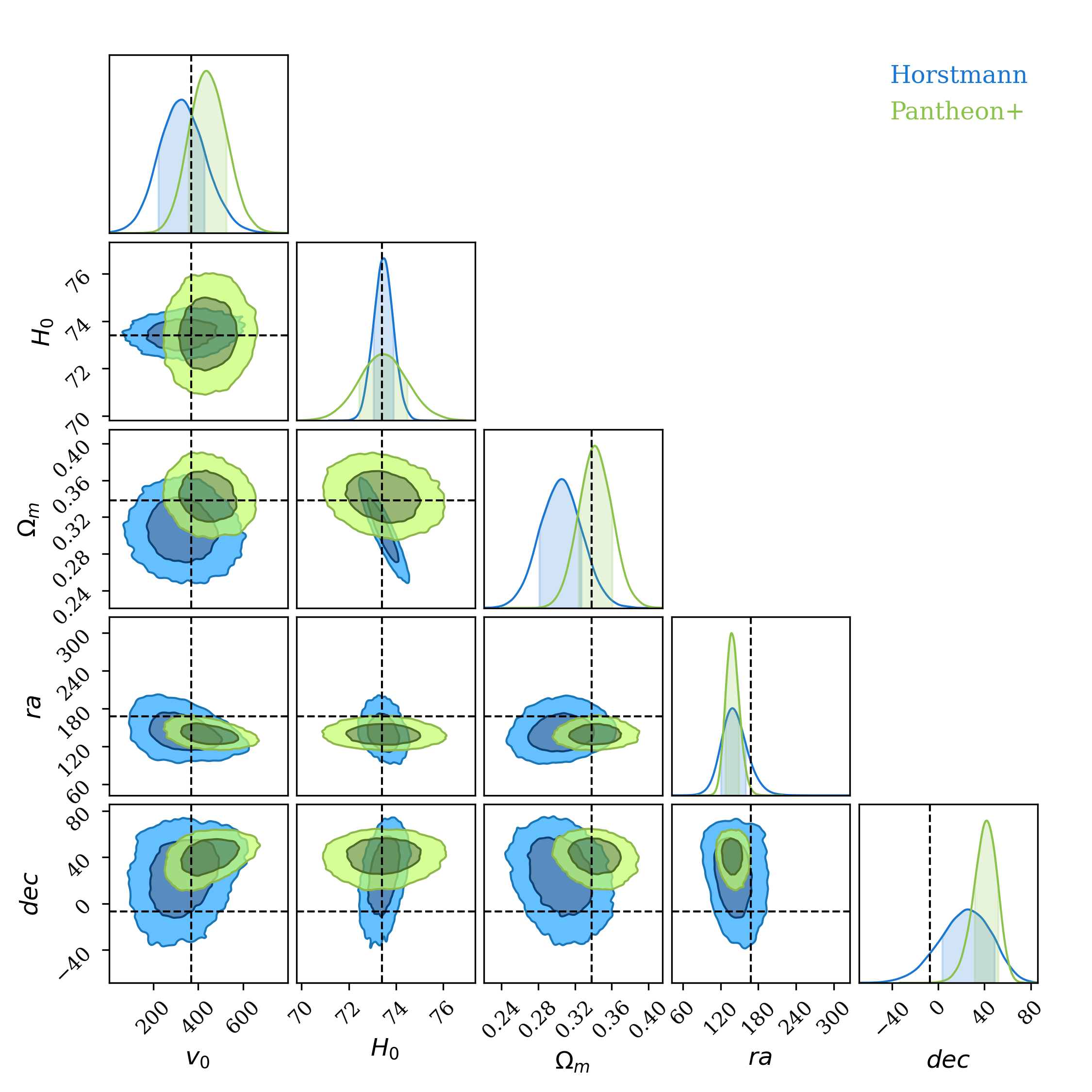}  
		\caption{$z_{\rm cut}$=0.025}
		\label{fig:sub-comparison-0.025}
	\end{subfigure}
	\begin{subfigure}{.5\textwidth}
		\centering
		\includegraphics[width=0.75\linewidth]{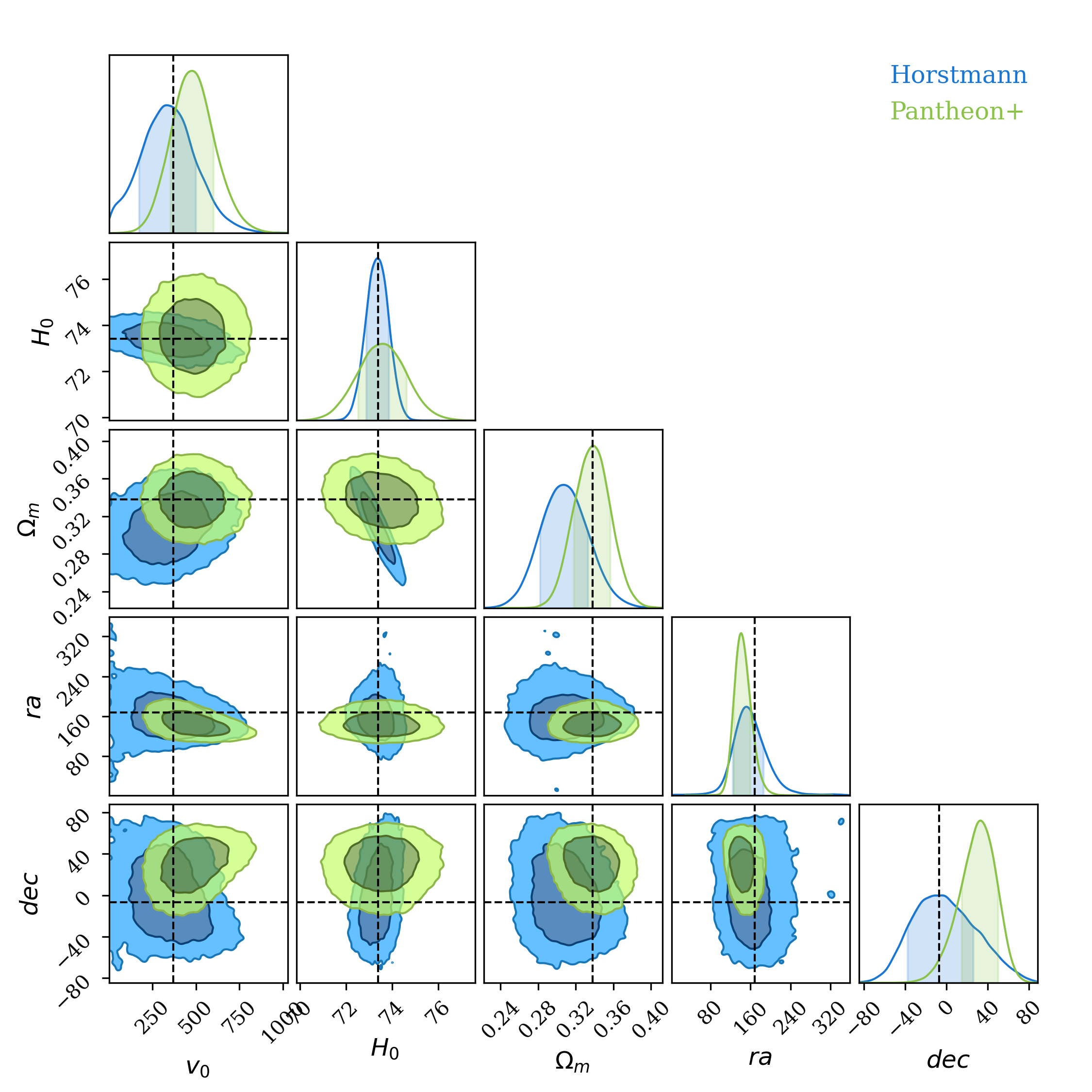}  
		\caption{$z_{\rm cut}$=0.0375}
		\label{fig:sub-comparison-0.0375}
	\end{subfigure}
	\\
	\begin{subfigure}{.5\textwidth}
		\centering
		\includegraphics[width=0.75\linewidth]{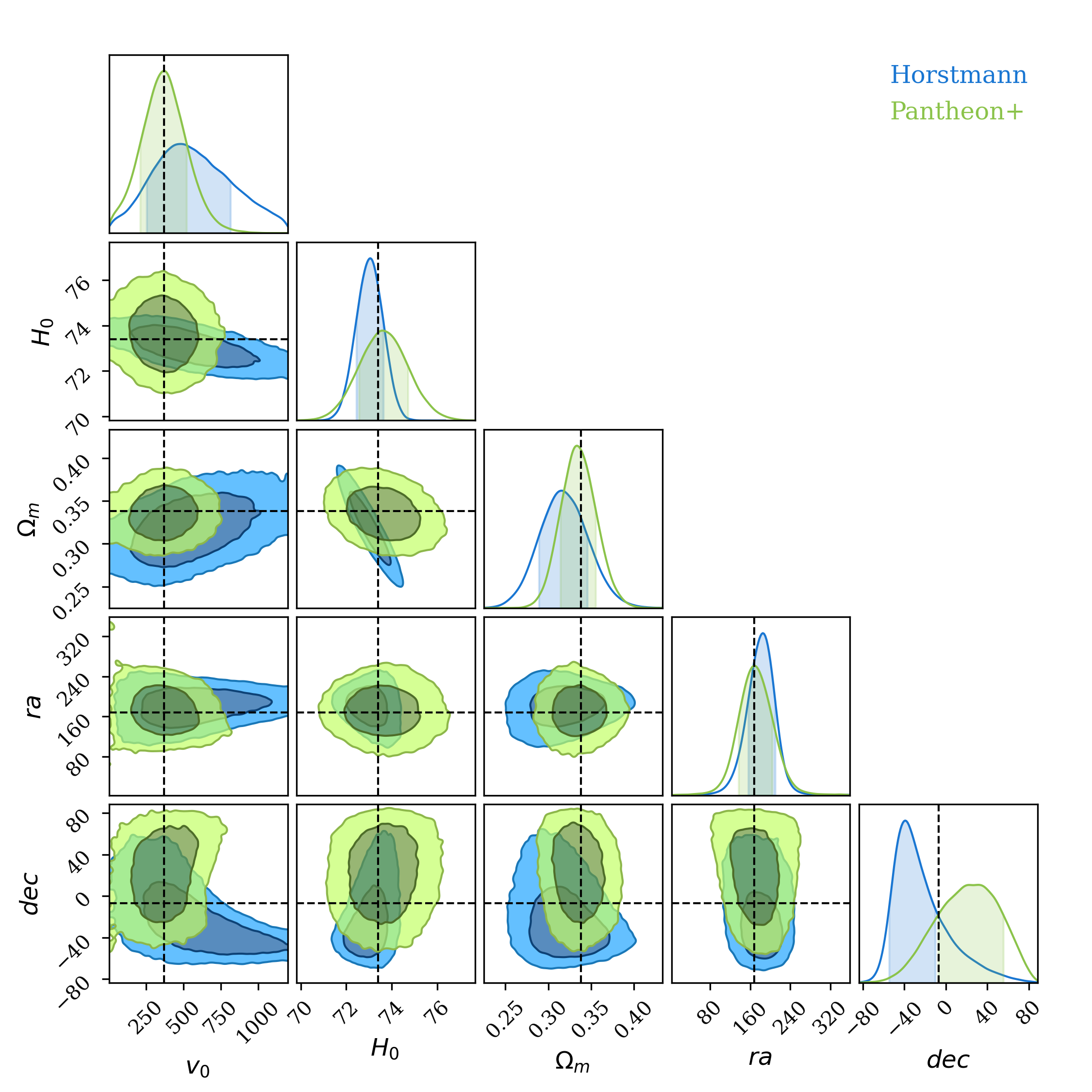}  
		\caption{$z_{\rm cut}$=0.05}
		\label{fig:sub-comparison-0.05}
	\end{subfigure}
	\begin{subfigure}{.5\textwidth}
		\centering
		\includegraphics[width=0.75\linewidth]{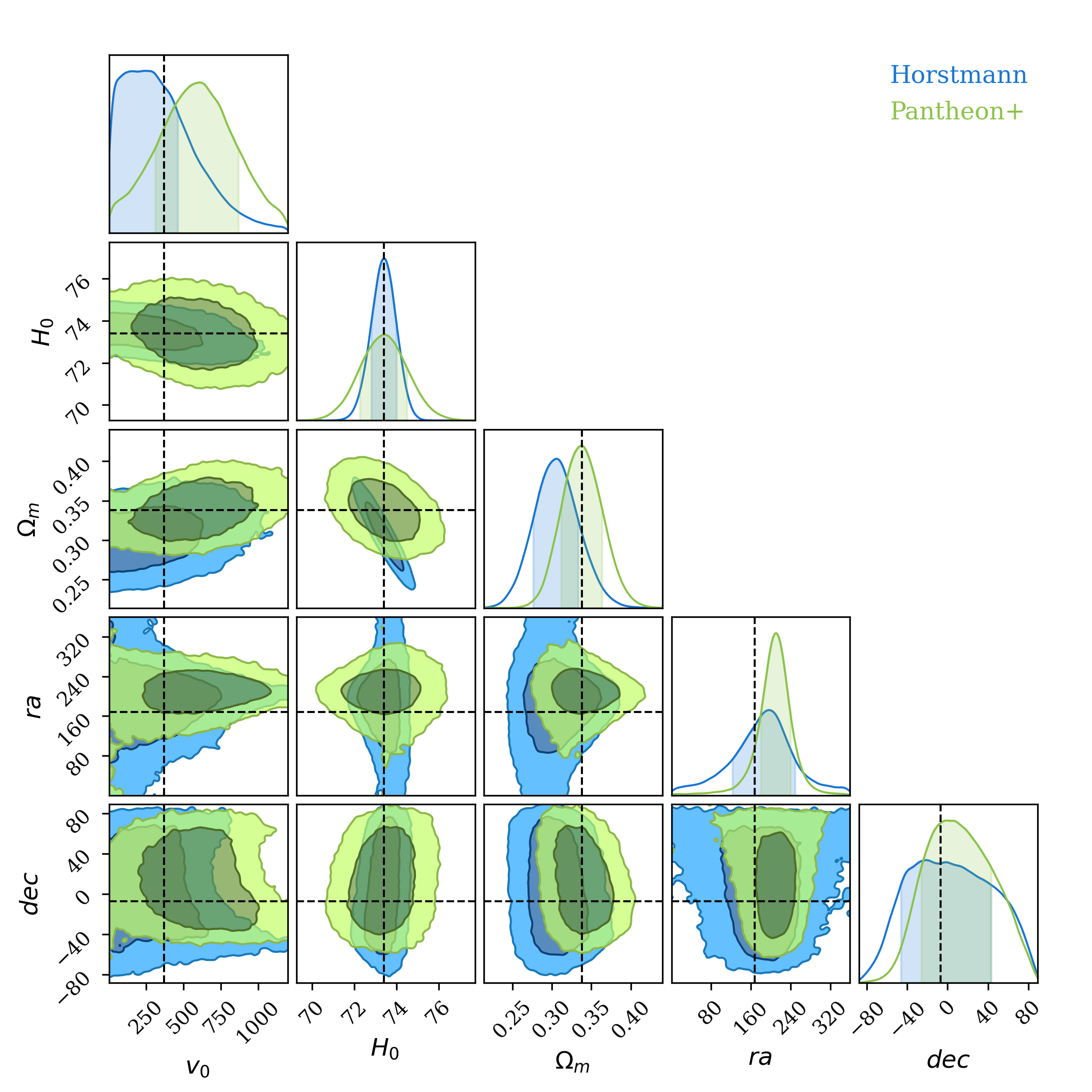}  
		\caption{$z_{\rm cut}$=0.1}
		\label{fig:sub-comparison-0.1}
	\end{subfigure}
	\caption{\label{f:compar}Contour plots for the Pantheon+ data and the Pantheon dataset as provided by Horstmann  with different lower cuts in redshift.}
\end{figure}

In Fig.~\ref{f:compar} we show both, the Pantheon and the Pantheon+ data for different lower cuts in redshift. While the error bars of Pantheon are large enough so that the dipole direction is in agreement with the one found in Planck, this is no longer so for the Pantheon+ data set. Note also that for $z_{\rm cut}>0.025$ the Pantheon data no longer discover a dipole at 95\% confidence; for these cuts, the 95\% confidence contours include $v_0=0$.

\subsection*{Redshift binning for the visual inspection}
In Table~\ref{tab:redshift-bin} we report the redshift bins used for Fig.~\ref{f:mu-residuals}.

\begin{table}[!ht]
	\centering
	\begin{tabular}{cc}
		\toprule
		& Redshift interval \\ 
		\midrule
		Bin 1 \qquad & $[8.2000\e{-2},\, 1.5380\e{-2})$ \vspace{6 pt}  \\ 
		Bin 2 \qquad&$[1.5380\e{-2},\, 2.4380\e{-2})$ \vspace{6 pt}  \\ 
		Bin 3 \qquad&$[2.4380\e{-2},\, 3.6940\e{-2})$ \vspace{6 pt} \\ 
		Bin 4 \qquad&$[3.6940\e{-2},\, 1.1371\e{-1})$ \vspace{6 pt} \\ 
		Bin 5 \qquad&$[1.1371\e{-1},\, 2.0000\e{-1})$  \vspace{6 pt} \\ 
		Bin 6 \qquad&$[2.0000\e{-1},\, 2.6722\e{-1})$  \vspace{6 pt} \\ 
		Bin 7 \qquad&$[2.6722\e{-1},\, 3.4910\e{-1})$  \vspace{6 pt} \\ 
		Bin 8 \qquad&$[3.4910\e{-1},\, 5.2000\e{-1})$ \vspace{6 pt} \\ 
		Bin 9 \qquad&$[5.2000\e{-1},\, 2.2600]$\qquad\quad ~~   \\
		\bottomrule
	\end{tabular}
	\caption{Redshift intervals for the 9 bins containing the same number of supernovae (189) used in the analysis of residuals for Fig.~\ref{f:mu-residuals}.  \label{tab:redshift-bin}}
\end{table}

\subsection*{Study of the redshift cuts in mock data}
\begin{figure}[!ht]
	
	\begin{subfigure}{.49\textwidth}
		
		\centering
		\includegraphics [scale=0.4]{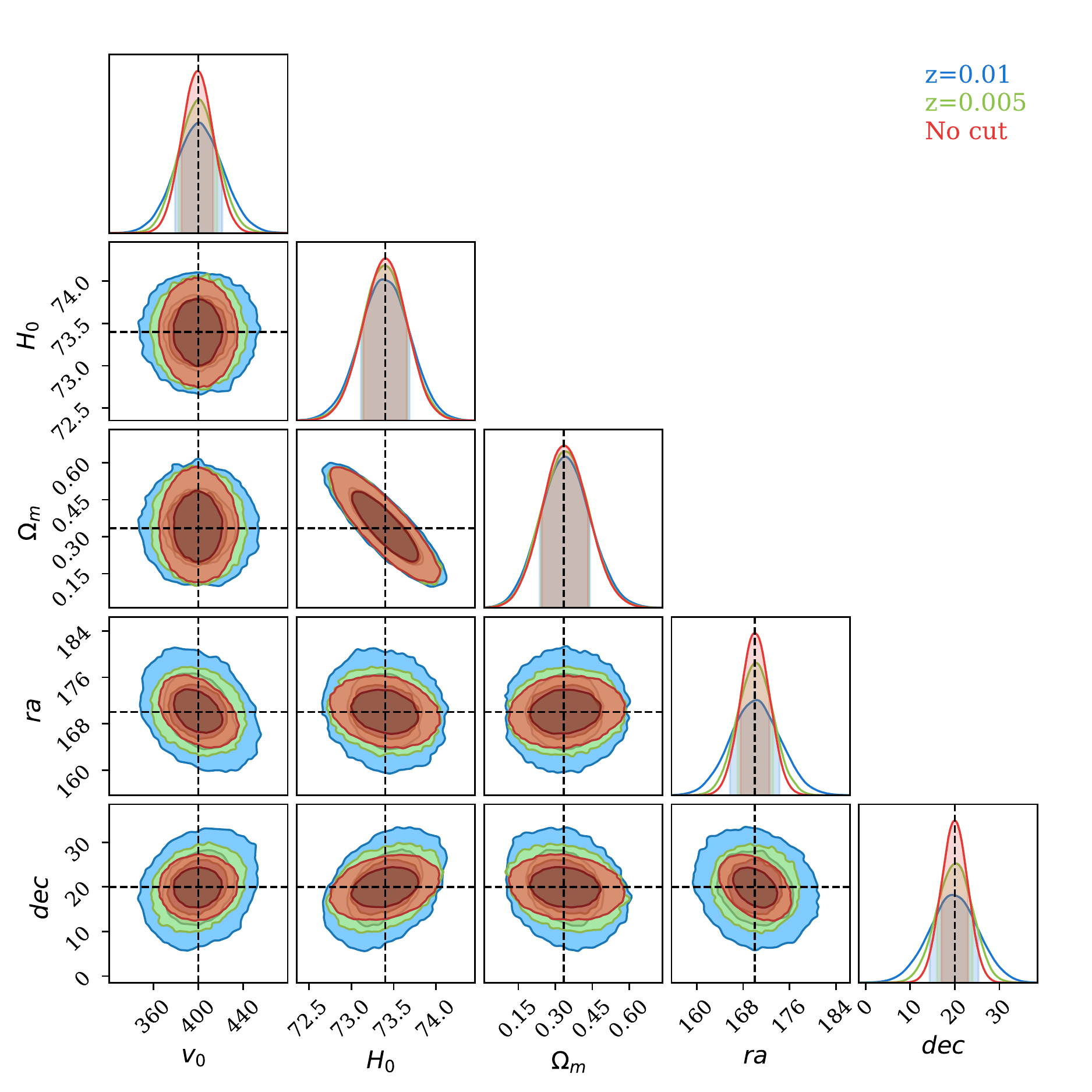}
		\caption{Low $z_{\rm cut}$.} 
		
	\end{subfigure}
	\begin{subfigure}{.49\textwidth}
		\centering
		\includegraphics [scale=0.4]{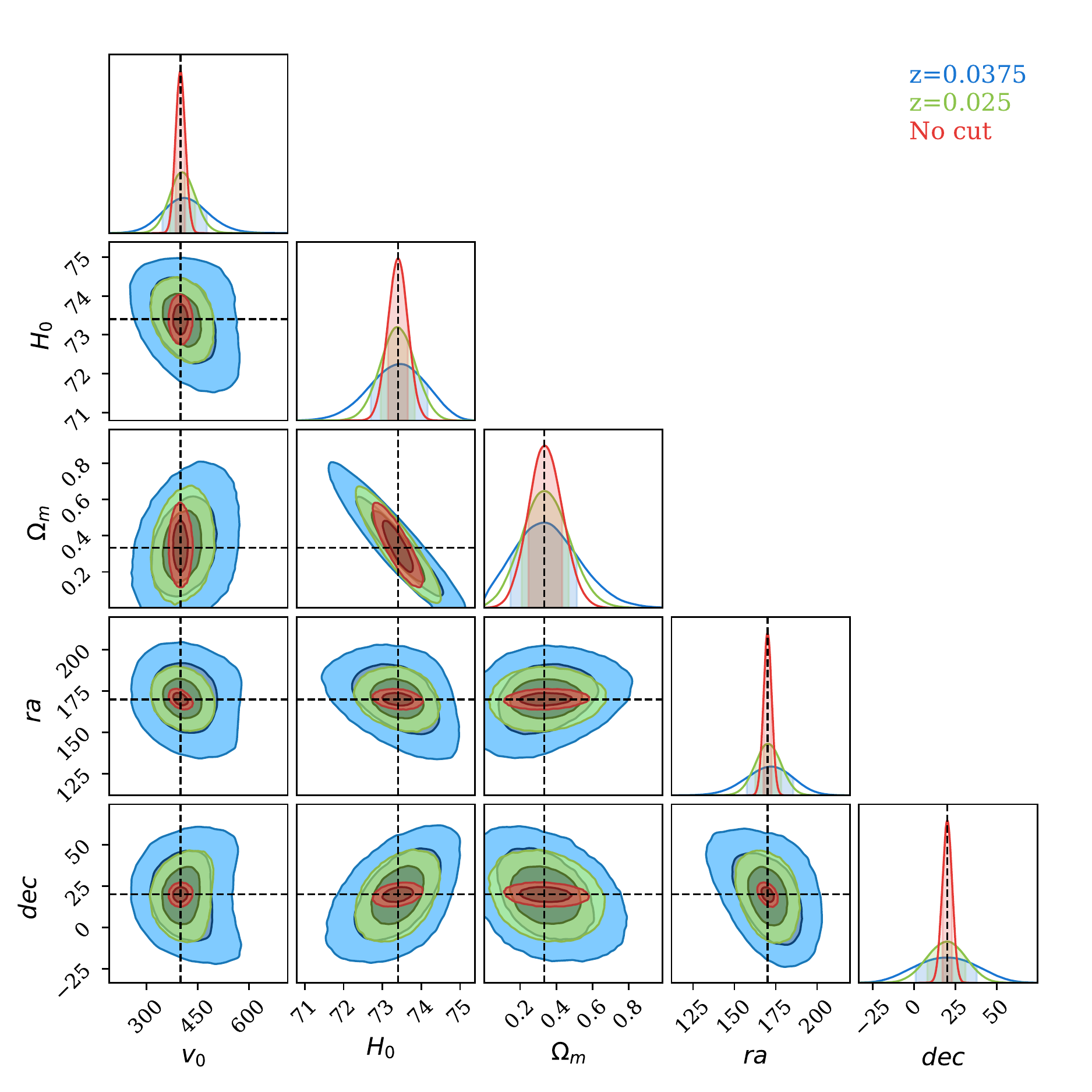}
		\caption{Medium $z_{\rm cut}$.} 
		
	\end{subfigure}
	\\
	\begin{subfigure}{.49\textwidth}
		\centering
		\includegraphics [scale=0.4]{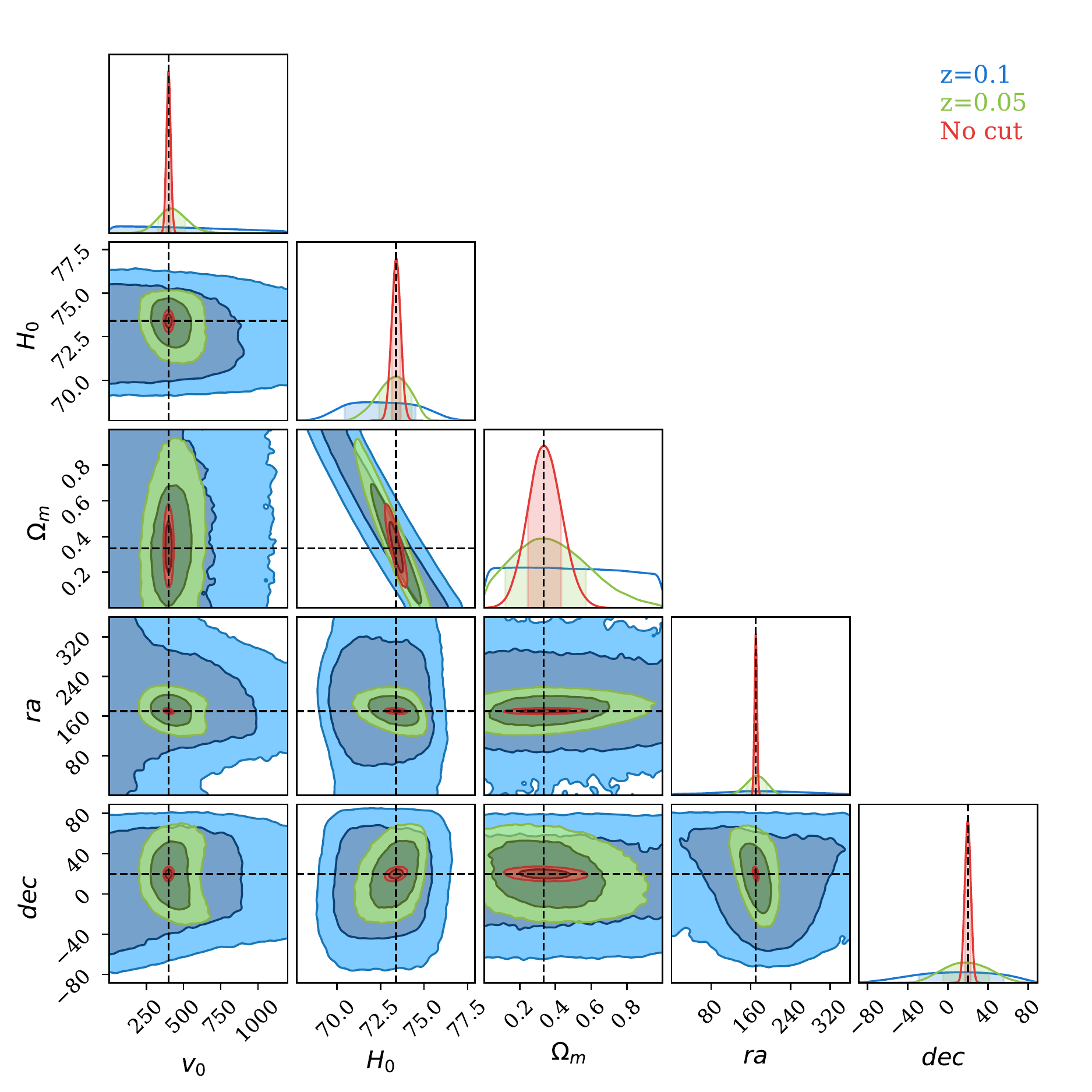}
		\caption{High $z_{\rm cut}$.} 
	\end{subfigure}
	\caption{\label{pic:mock_extra_plots_low_z} The results of our MCMC fitting procedures for the redshift cuts not shown in Fig.~\ref{pic:mock_low_z_175e2_1e2_no_filter}. We are considering the mock dataset at \textbf{low} redshift described in Sec. \ref{sec:redshift_cut}, hence the high redshift cut contains rather few supernovae, see Table~\ref{t:mock}. }
	
\end{figure}

\begin{figure}[!ht]
	
	\begin{subfigure}{.49\textwidth}
		
		\centering
		\includegraphics [scale=0.4]{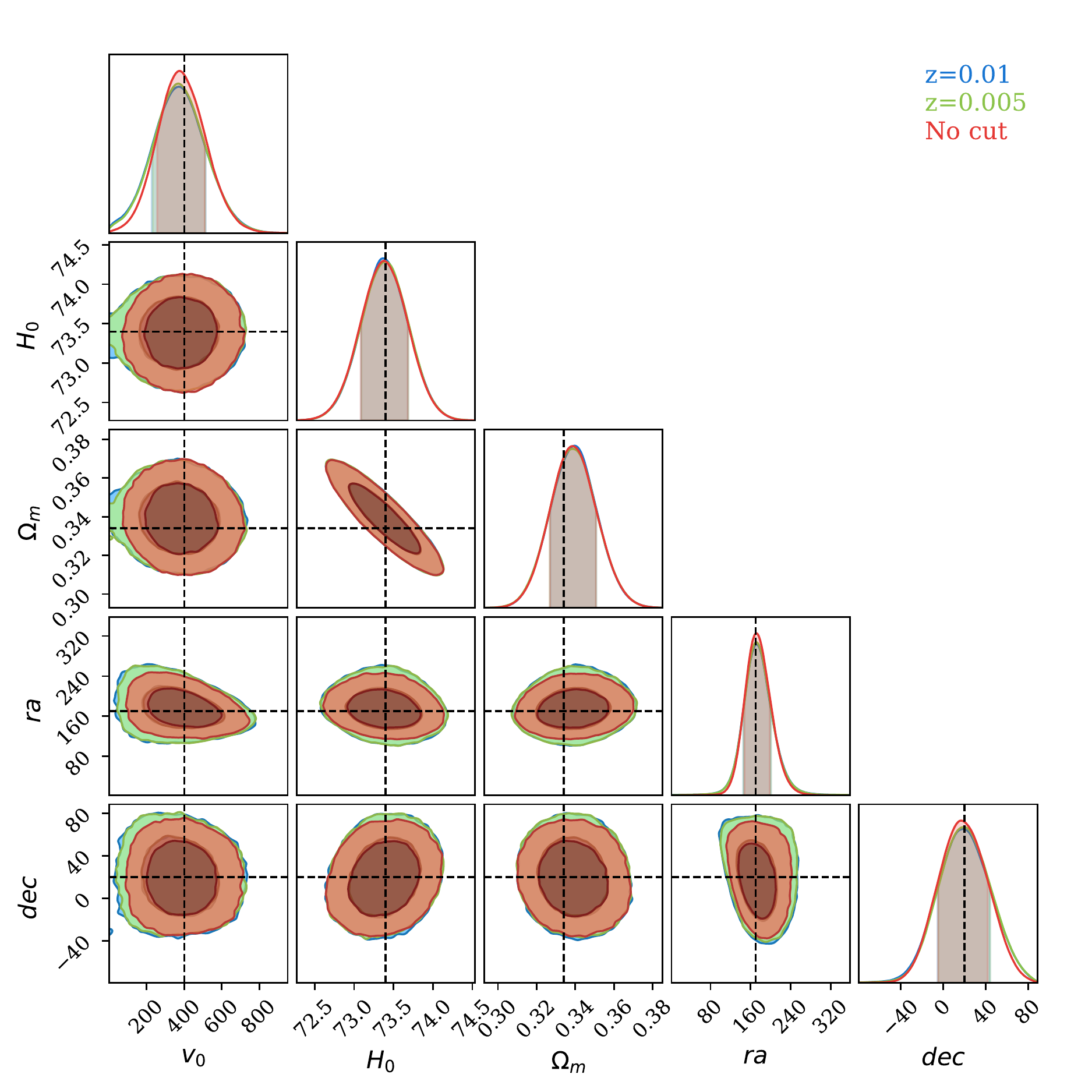}
		\caption{Low $z_{\rm cut}$.} 
		
	\end{subfigure}
	\begin{subfigure}{.49\textwidth}
		\centering
		\includegraphics [scale=0.4]{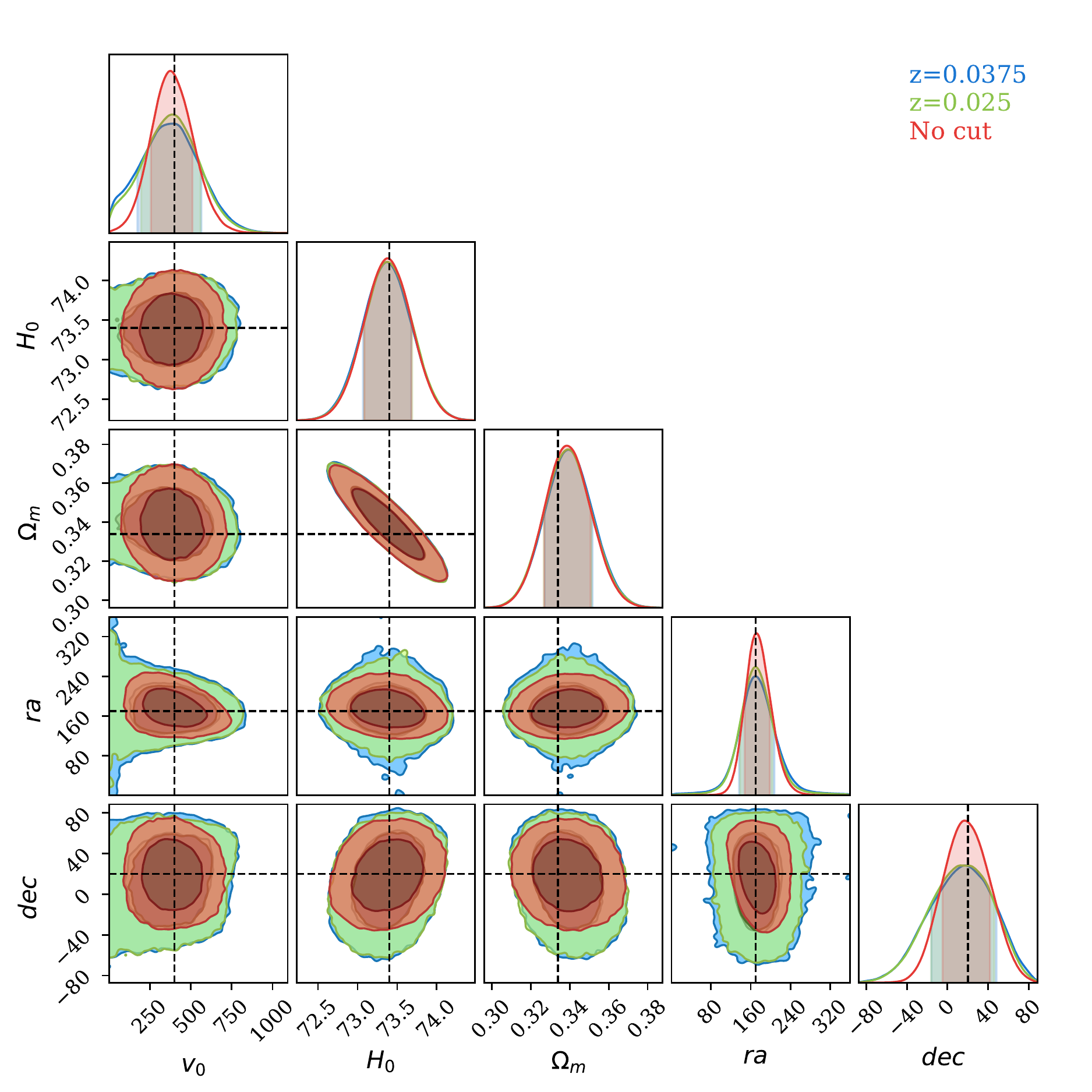}
		\caption{Medium $z_{\rm cut}$.} 
		
	\end{subfigure}
	\\
	\begin{subfigure}{.49\textwidth}
		\centering
		\includegraphics [scale=0.4]{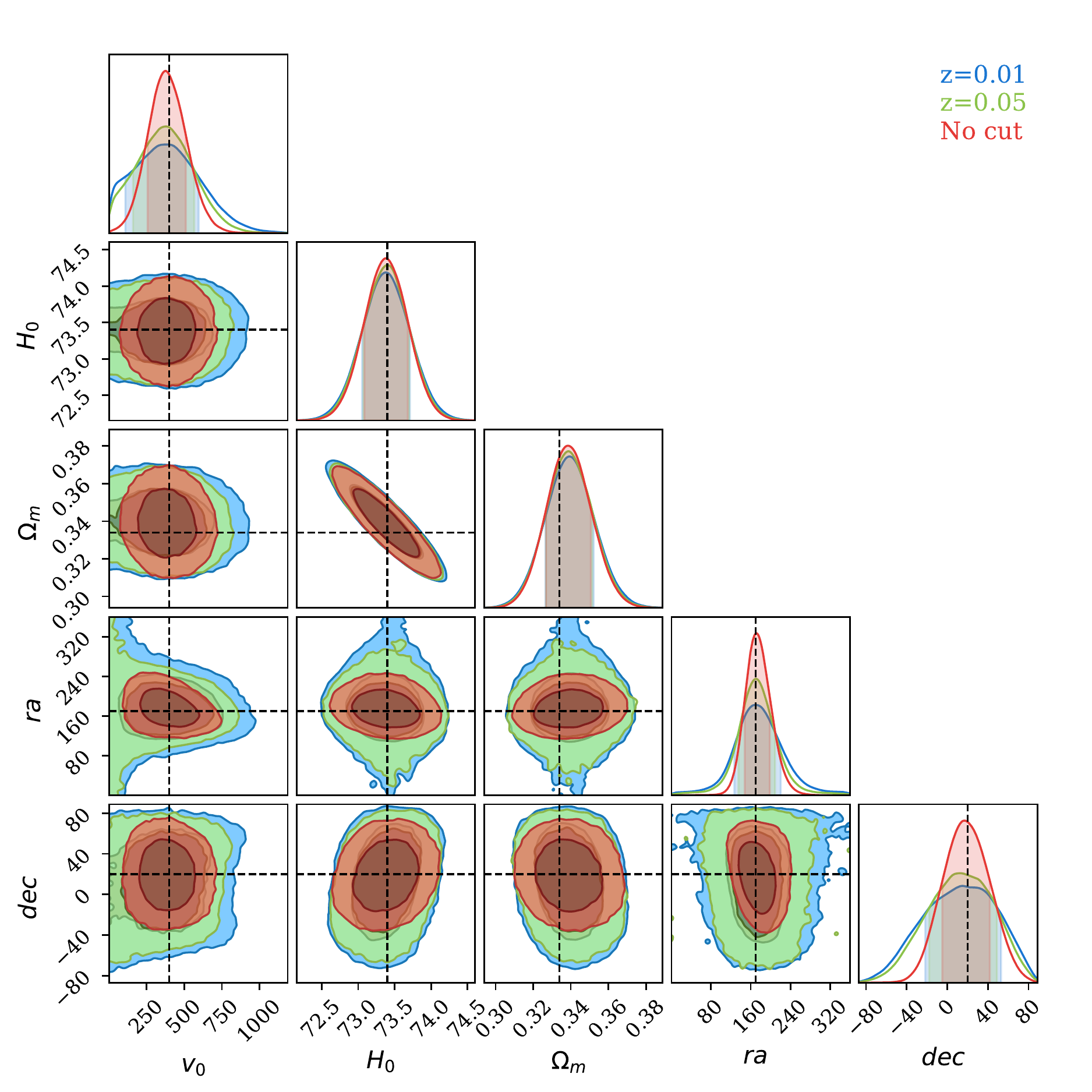}
		\caption{High $z_{\rm cut}$.} 
	\end{subfigure}
	\caption{\label{pic:mock_extra_plots_high_z} The results of our MCMC fitting procedures for the redshift cuts not shown in Fig.~\ref{pic:mock_high_z_175e2_1e2_no_filter}. We are considering the mock dataset at \textbf{higher} redshift described in Sec. \ref{sec:redshift_cut}. }
	
\end{figure}

In Figs.~\ref{pic:mock_extra_plots_low_z} and \ref{pic:mock_extra_plots_high_z} we show the contour plots applying different value of $z_{\rm cut}$ for the mock dataset with the low respectively high redshift distribution
described in Sec.~\ref{sec:redshift_cut}. As also mentioned in the main text, with increasing $z_{\rm cut}$ the dipole detectability is reduced. For the higher cuts in the low-z distribution, Fig.~\ref{pic:mock_extra_plots_low_z}, also the errors in the parameters $H_0$ and $\Om_m$ increase significantly since there are very few supernovae left at these redshifts.

\FloatBarrier

\bibliography{refs}

\providecommand{\href}[2]{#2}\begingroup\raggedright\begin{thebibliography}{10}

\bibitem{Kogut:1993ag}
A.~Kogut et~al., {\it {Dipole anisotropy in the COBE DMR first year sky maps}},
   {\em Astrophys. J.} {\bf 419} (1993) 1,
  [\href{http://arxiv.org/abs/astro-ph/9312056}{{\tt astro-ph/9312056}}].

\bibitem{Planck:2013kqc}
{\bf Planck} Collaboration, N.~Aghanim et~al., {\it {Planck 2013 results.
  XXVII. Doppler boosting of the CMB: Eppur si muove}},  {\em Astron.
  Astrophys.} {\bf 571} (2014) A27, [\href{http://arxiv.org/abs/1303.5087}{{\tt
  arXiv:1303.5087}}].

\bibitem{Planck:2018nkj}
{\bf Planck} Collaboration, N.~Aghanim et~al., {\it {Planck 2018 results. I.
  Overview and the cosmological legacy of Planck}},  {\em Astron. Astrophys.}
  {\bf 641} (2020) A1, [\href{http://arxiv.org/abs/1807.06205}{{\tt
  arXiv:1807.06205}}].

\bibitem{Saha:2021bay}
S.~Saha, S.~Shaikh, S.~Mukherjee, T.~Souradeep, and B.~D. Wandelt, {\it
  {Bayesian estimation of our local motion from the Planck-2018 CMB temperature
  map}},  {\em JCAP} {\bf 10} (2021) 072,
  [\href{http://arxiv.org/abs/2106.07666}{{\tt arXiv:2106.07666}}].

\bibitem{Ellis1984}
G.~F.~R. {Ellis} and J.~E. {Baldwin}, {\it {On the expected anisotropy of radio
  source counts}},  {\em MNRAS} {\bf 206} (Jan., 1984) 377--381.

\bibitem{Tiwari:2016}
P.~Tiwari and A.~Nusser, {\it Revisiting the {NVSS} number count dipole},  {\em
  Journal of Cosmology and Astroparticle Physics} {\bf 2016} (mar, 2016)
  062--062.

\bibitem{Bengaly:2017slg}
C.~A.~P. Bengaly, R.~Maartens, and M.~G. Santos, {\it {Probing the Cosmological
  Principle in the counts of radio galaxies at different frequencies}},  {\em
  JCAP} {\bf 04} (2018) 031, [\href{http://arxiv.org/abs/1710.08804}{{\tt
  arXiv:1710.08804}}].

\bibitem{Colin:2019opb}
J.~Colin, R.~Mohayaee, M.~Rameez, and S.~Sarkar, {\it {Evidence for anisotropy
  of cosmic acceleration}},  {\em Astron. Astrophys.} {\bf 631} (2019) L13,
  [\href{http://arxiv.org/abs/1808.04597}{{\tt arXiv:1808.04597}}].

\bibitem{Siewert:2021}
{Siewert, Thilo M.}, {Schmidt-Rubart, Matthias}, and {Schwarz, Dominik J.},
  {\it Cosmic radio dipole: Estimators and frequency dependence},  {\em A\&A}
  {\bf 653} (2021) A9.

\bibitem{Secrest:2020has}
N.~J. Secrest, S.~v. Hausegger, M.~Rameez, R.~Mohayaee, S.~Sarkar, and
  J.~Colin, {\it A test of the cosmological principle with quasars},  {\em The
  Astrophysical Journal} {\bf 908} (2021), no.~2 L51,
  [\href{http://arxiv.org/abs/2009.14826}{{\tt arXiv:2009.14826}}].

\bibitem{Dam:2022wwh}
L.~Dam, G.~F. Lewis, and B.~J. Brewer, {\it {Testing the Cosmological Principle
  with CatWISE Quasars: A Bayesian Analysis of the Number-Count Dipole}},
  \href{http://arxiv.org/abs/2212.07733}{{\tt arXiv:2212.07733}}.

\bibitem{Migkas:2021zdo}
K.~Migkas, F.~Pacaud, G.~Schellenberger, J.~Erler, N.~T. Nguyen-Dang, T.~H.
  Reiprich, M.~E. Ramos-Ceja, and L.~Lovisari, {\it {Cosmological implications
  of the anisotropy of ten galaxy cluster scaling relations}},  {\em Astron.
  Astrophys.} {\bf 649} (2021) A151,
  [\href{http://arxiv.org/abs/2103.13904}{{\tt arXiv:2103.13904}}].

\bibitem{Secrest:2022uvx}
N.~J. Secrest, S.~von Hausegger, M.~Rameez, R.~Mohayaee, and S.~Sarkar, {\it {A
  Challenge to the Standard Cosmological Model}},  {\em Astrophys. J. Lett.}
  {\bf 937} (2022), no.~2 L31, [\href{http://arxiv.org/abs/2206.05624}{{\tt
  arXiv:2206.05624}}].

\bibitem{Dalang:2021ruy}
C.~Dalang and C.~Bonvin, {\it {On the kinematic cosmic dipole tension}},  {\em
  Mon. Not. Roy. Astron. Soc.} {\bf 512} (2022), no.~3 3895--3905,
  [\href{http://arxiv.org/abs/2111.03616}{{\tt arXiv:2111.03616}}].

\bibitem{Guandalin:2022tyl}
C.~Guandalin, J.~Piat, C.~Clarkson, and R.~Maartens, {\it {Theoretical
  systematics in testing the Cosmological Principle with the kinematic quasar
  dipole}},  \href{http://arxiv.org/abs/2212.04925}{{\tt arXiv:2212.04925}}.

\bibitem{Atrio-Barandela:2014nda}
F.~Atrio-Barandela, A.~Kashlinsky, H.~Ebeling, D.~J. Fixsen, and D.~Kocevski,
  {\it {Probing the Dark Flow Signal in Wmap 9 -year and Planck Cosmic
  Microwave Background Maps}},  {\em Astrophys. J.} {\bf 810} (2015), no.~2
  143, [\href{http://arxiv.org/abs/1411.4180}{{\tt arXiv:1411.4180}}].

\bibitem{Nadolny:2021hti}
T.~Nadolny, R.~Durrer, M.~Kunz, and H.~Padmanabhan, {\it {A new way to test the
  Cosmological Principle: measuring our peculiar velocity and the large-scale
  anisotropy independently}},  {\em JCAP} {\bf 11} (2021) 009,
  [\href{http://arxiv.org/abs/2106.05284}{{\tt arXiv:2106.05284}}].

\bibitem{Bonvin:2006en}
C.~Bonvin, R.~Durrer, and M.~Kunz, {\it {The dipole of the luminosity distance:
  a direct measure of H(z)}},  {\em Phys. Rev. Lett.} {\bf 96} (2006) 191302,
  [\href{http://arxiv.org/abs/astro-ph/0603240}{{\tt astro-ph/0603240}}].

\bibitem{Javanmardi:2015sfa}
B.~Javanmardi, C.~Porciani, P.~Kroupa, and J.~Pflamm-Altenburg, {\it {Probing
  the isotropy of cosmic acceleration traced by Type Ia supernovae}},  {\em
  Astrophys. J.} {\bf 810} (2015), no.~1 47,
  [\href{http://arxiv.org/abs/1507.07560}{{\tt arXiv:1507.07560}}].

\bibitem{Horstmann:2021jjg}
N.~Horstmann, Y.~Pietschke, and D.~J. Schwarz, {\it {Inference of the cosmic
  rest-frame from supernovae Ia}},  \href{http://arxiv.org/abs/2111.03055}{{\tt
  arXiv:2111.03055}}.

\bibitem{Dhawan:2022lze}
S.~Dhawan, A.~Borderies, H.~J. Macpherson, and A.~Heinesen, {\it {The
  quadrupole in the local Hubble parameter: first constraints using Type Ia
  supernova data and forecasts for future surveys}},
  \href{http://arxiv.org/abs/2205.12692}{{\tt arXiv:2205.12692}}.

\bibitem{Kalbouneh:2022tfw}
B.~Kalbouneh, C.~Marinoni, and J.~Bel, {\it {Multipole expansion of the local
  expansion rate}},  {\em Phys. Rev. D} {\bf 107} (2023), no.~2 023507,
  [\href{http://arxiv.org/abs/2210.11333}{{\tt arXiv:2210.11333}}].

\bibitem{Tully:2016ppz}
R.~B. Tully, H.~M. Courtois, and J.~G. Sorce, {\it {Cosmicflows-3}},  {\em
  Astron. J.} {\bf 152} (2016) 50, [\href{http://arxiv.org/abs/1605.01765}{{\tt
  arXiv:1605.01765}}].

\bibitem{Brout:2022vxf}
D.~Brout et~al., {\it {The Pantheon+ Analysis: Cosmological Constraints}},
  {\em Astrophys. J.} {\bf 938} (2022), no.~2 110,
  [\href{http://arxiv.org/abs/2202.04077}{{\tt arXiv:2202.04077}}].

\bibitem{Pan-STARRS1:2017jku}
{\bf Pan-STARRS1} Collaboration, D.~M. Scolnic et~al., {\it {The Complete
  Light-curve Sample of Spectroscopically Confirmed SNe Ia from Pan-STARRS1 and
  Cosmological Constraints from the Combined Pantheon Sample}},  {\em
  Astrophys. J.} {\bf 859} (2018), no.~2 101,
  [\href{http://arxiv.org/abs/1710.00845}{{\tt arXiv:1710.00845}}].

\bibitem{Sasaki:1987ad}
M.~Sasaki, {\it {The Magnitude - Redshift relation in a perturbed Friedmann
  universe}},  {\em Mon. Not. Roy. Astron. Soc.} {\bf 228} (1987) 653--669.

\bibitem{Bonvin:2005ps}
C.~Bonvin, R.~Durrer, and M.~A. Gasparini, {\it {Fluctuations of the luminosity
  distance}},  {\em Phys. Rev. D} {\bf 73} (2006) 023523,
  [\href{http://arxiv.org/abs/astro-ph/0511183}{{\tt astro-ph/0511183}}].
  [Erratum: Phys.Rev.D 85, 029901 (2012)].

\bibitem{Durrer:2020fza}
R.~Durrer, {\em {The Cosmic Microwave Background}}.
\newblock Cambridge University Press, 12, 2020.

\bibitem{Biern:2016kys}
S.~G. Biern and J.~Yoo, {\it {Gauge-Invariance and Infrared Divergences in the
  Luminosity Distance}},  {\em JCAP} {\bf 04} (2017) 045,
  [\href{http://arxiv.org/abs/1606.01910}{{\tt arXiv:1606.01910}}].

\bibitem{Carr:2021lcj}
A.~Carr, T.~M. Davis, D.~Scolnic, D.~Scolnic, K.~Said, D.~Brout, E.~R.
  Peterson, and R.~Kessler, {\it {The Pantheon+ analysis: Improving the
  redshifts and peculiar velocities of Type Ia supernovae used in cosmological
  analyses}},  {\em Publ. Astron. Soc. Austral.} {\bf 39} (2022) e046,
  [\href{http://arxiv.org/abs/2112.01471}{{\tt arXiv:2112.01471}}].

\bibitem{astropy:2013}
{Astropy Collaboration} and {Astropy Project Contributors}, {\it {Astropy: A
  community Python package for astronomy}},  {\em Astron. Astrophys.} {\bf 558}
  (Oct., 2013) A33, [\href{http://arxiv.org/abs/1307.6212}{{\tt
  arXiv:1307.6212}}].

\bibitem{astropy:2018}
{Astropy Collaboration} and {Astropy Project Contributors}, {\it {The Astropy
  Project: Building an Open-science Project and Status of the v2.0 Core
  Package}},  {\em Astron. J.} {\bf 156} (Sept., 2018) 123,
  [\href{http://arxiv.org/abs/1801.02634}{{\tt arXiv:1801.02634}}].

\bibitem{astropy:2022}
{Astropy Collaboration} and {Astropy Project Contributors}, {\it {The Astropy
  Project: Sustaining and Growing a Community-oriented Open-source Project and
  the Latest Major Release (v5.0) of the Core Package}},  {\em Astrophys. J.}
  {\bf 935} (Aug., 2022) 167, [\href{http://arxiv.org/abs/2206.14220}{{\tt
  arXiv:2206.14220}}].

\bibitem{emcee}
D.~{Foreman-Mackey}, D.~W. {Hogg}, D.~{Lang}, and J.~{Goodman}, {\it {emcee:
  The MCMC Hammer}},  {\em Publ. Astron. Soc. Pac.} {\bf 125} (Mar., 2013) 306,
  [\href{http://arxiv.org/abs/1202.3665}{{\tt arXiv:1202.3665}}].

\bibitem{chainconsumer}
S.~R. {Hinton}, {\it {ChainConsumer}},  {\em The Journal of Open Source
  Software} {\bf 1} (Aug., 2016) 00045.

\bibitem{schwimmbad}
A.~M. Price-Whelan and D.~Foreman-Mackey, {\it schwimmbad: A uniform interface
  to parallel processing pools in python},  {\em The Journal of Open Source
  Software} {\bf 2} (sep, 2017).

\bibitem{autocorr}
J.~{Goodman} and J.~{Weare}, {\it {Ensemble samplers with affine invariance}},
  {\em Communications in Applied Mathematics and Computational Science} {\bf 5}
  (Jan., 2010) 65--80.

\bibitem{uncertainties}
E.~O. LEBIGOT, ``Uncertainties: a python package for calculations with
  uncertainties.'' \url{http://pythonhosted.org/uncertainties/}.

\bibitem{Carrick:2015xza}
J.~Carrick, S.~J. Turnbull, G.~Lavaux, and M.~J. Hudson, {\it {Cosmological
  parameters from the comparison of peculiar velocities with predictions from
  the 2M++ density field}},  {\em Mon. Not. Roy. Astron. Soc.} {\bf 450}
  (2015), no.~1 317--332, [\href{http://arxiv.org/abs/1504.04627}{{\tt
  arXiv:1504.04627}}].

\bibitem{Tully_2023}
R.~B. Tully, E.~Kourkchi, H.~M. Courtois, G.~S. Anand, J.~P. Blakeslee,
  D.~Brout, T.~de~Jaeger, A.~Dupuy, D.~Guinet, C.~Howlett, J.~B. Jensen,
  D.~Pomar{\`{e}}de, L.~Rizzi, D.~Rubin, K.~Said, D.~Scolnic, and B.~E. Stahl,
  {\it Cosmicflows-4},  {\em The Astrophysical Journal} {\bf 944} (feb, 2023)
  94.

\bibitem{Watkins_2023}
R.~Watkins, T.~Allen, C.~J. Bradford, A.~Ramon, A.~Walker, H.~A. Feldman,
  R.~Cionitti, Y.~Al-Shorman, E.~Kourkchi, and R.~B. Tully, {\it Analysing the
  large-scale bulk flow using cosmicflows4: increasing tension with the
  standard cosmological model},  {\em Monthly Notices of the Royal Astronomical
  Society} {\bf 524} (jul, 2023) 1885--1892.

\bibitem{Steinhardt:2020kul}
C.~L. Steinhardt, A.~Sneppen, and B.~Sen, {\it {Effects of Supernova Redshift
  Uncertainties on the Determination of Cosmological Parameters}},  {\em
  Astrophys. J.} {\bf 902} (2020), no.~1 14,
  [\href{http://arxiv.org/abs/2005.07707}{{\tt arXiv:2005.07707}}].

\bibitem{Scolnic_2022}
D.~Scolnic, D.~Brout, A.~Carr, A.~G. Riess, T.~M. Davis, A.~Dwomoh, D.~O.
  Jones, N.~Ali, P.~Charvu, R.~Chen, E.~R. Peterson, B.~Popovic, B.~M. Rose,
  C.~M. Wood, P.~J. Brown, K.~Chambers, D.~A. Coulter, K.~G. Dettman,
  G.~Dimitriadis, A.~V. Filippenko, R.~J. Foley, S.~W. Jha, C.~D. Kilpatrick,
  R.~P. Kirshner, Y.-C. Pan, A.~Rest, C.~Rojas-Bravo, M.~R. Siebert, B.~E.
  Stahl, and W.~Zheng, {\it The pantheon+ analysis: The full data set and
  light-curve release},  {\em The Astrophysical Journal} {\bf 938} (oct, 2022)
  113.

\bibitem{Peterson:2021hel}
E.~R. Peterson et~al., {\it {The Pantheon+ Analysis: Evaluating Peculiar
  Velocity Corrections in Cosmological Analyses with Nearby Type Ia
  Supernovae}},  {\em Astrophys. J.} {\bf 938} (2022), no.~2 112,
  [\href{http://arxiv.org/abs/2110.03487}{{\tt arXiv:2110.03487}}].

\bibitem{Jelic-Cizmek:2018gdp}
G.~Jelic-Cizmek, F.~Lepori, J.~Adamek, and R.~Durrer, {\it {The generation of
  vorticity in cosmological N-body simulations}},  {\em JCAP} {\bf 09} (2018)
  006, [\href{http://arxiv.org/abs/1806.05146}{{\tt arXiv:1806.05146}}].

\end{thebibliography}\endgroup
\bibliographystyle{JHEP}
\end{document}